\documentclass[12pt]{iopart}
\usepackage{iopams}  
\usepackage{color}
\usepackage{graphicx}
\usepackage{subfig}

\newcommand{\tauD}{{\tau_{\rm D}}}
\newcommand{\tauB}{{\tau_{\rm B}}}
\newcommand{\tauM}{{\tau_{\rm M}}}

\newcommand{\tauDD}{{\tau_{\rm BnG}}}

\newcommand{\mD}{{\mathcal{D}}}
\newcommand{\R}{\mathbb{R}}
\newcommand{\N}{\mathbb{N}}
\newcommand{\E}{\mathbb{E}}
\newcommand{\C}{\mathbb{C}}

\newcommand{\be}{\begin{equation}}
\newcommand{\ee}{\end{equation}}

\begin{document}

\title[Anomalous diffusion originated by 
two Markovian hopping-trap mechanisms]
{Anomalous diffusion originated by two Markovian 
hopping-trap mechanisms}

\author{S Vitali$^1$, P Paradisi$^{1,2}$, G Pagnini$^{1,3}$}

\address{$^1$ BCAM--Basque Center for Applied Mathematics,
Alameda de Mazarredo 14, E-48009 Bilbao, Basque Country - Spain}
\address{$^2$ ISTI--CNR, Institute of Information Science and Technologies 
"A. Faedo", Via Moruzzi 1, I-56124 Pisa, Italy}
\address{$^3$ Ikerbasque--Basque Foundation for Science,
Plaza Euskadi 5, E-48009 Bilbao, Basque Country - Spain}
\ead{gpagnini@bcamath.org}
\vspace{10pt}
\begin{indented}
\item[]Original Submission: 27 October 2021\\
Revised version: 11 April 2022 
\end{indented}

\begin{abstract}
We show through intensive simulations that the paradigmatic 
features of anomalous diffusion are indeed the features of
a (continuous-time) random walk 
driven by two different Markovian hopping-trap mechanisms.
If $p \in (0,1/2)$ and $1-p$ are the probabilities 
of occurrence of each 
Markovian mechanism,
then the anomalousness parameter $\beta \in (0,1)$ results to be   
$\beta \simeq 1 - 1/\{1 + \log[(1-p)/p]\}$.
Ensemble and single-particle observables of this model have been studied
and they match the main characteristics of anomalous diffusion 
as they are typically measured in living systems.
In particular, the celebrated transition of the walker's distribution
from exponential to stretched-exponential and finally
to Gaussian distribution is displayed by including also 
the Brownian yet non-Gaussian interval. 
\end{abstract}

%
%
\submitto{\jpa}  
%
\maketitle
%
%

\section{Introduction}
\label{sec:introduction}

We show that anomalous diffusion emerges from a  
process that goes through the action of two {\it co-existing}
Markovian mechanisms acting with different statistical frequency. 
In other words,
anomalous diffusion emerges from standard diffusion
when very seldomly the process switches to another standard diffusion with
a different set-up: the probability of occurrence of this switch
originates and fully characterizes the anomalous diffusion such that
anomalous diffusion is indeed not originated 
by a broad distribution of relaxation times 
\cite{shlesinger-arpc-1988,pagnini-pa-2014},
or by a crowded environment \cite{sokolov-sm-2012}, 
or by other mechanisms linking it to complexity \cite{west-rmp-2014}.

The motion of a random walker is called diffusion 
when it goes through a dissipative evolution and 
the ensemble statistics of the process 
are characterized by the convergence of 
the walker's probability density function (PDF) 
to the Gaussian distribution and also
by a mean-square displacement (MSD) that is linear in time,
namely the Brownian motion (Bm).
An offspring of the Gaussianity and of the Bm  
is that the governing equation of the walker's PDF,
i.e., the Fokker--Planck (FP) equation,
is an equation with a single and constant coefficient, 
that is the diffusion coefficient.
Since the Gaussian distribution is also named Normal distribution,
we have that the term diffusion turns into normal diffusion and
whenever one, or both, of the
characteristic features of the ensemble statistics 
of the normal diffusion are not fulfilled then
the corresponding process falls into the class of the anomalous diffusion.

We study here anomalous diffusion as it emerges 
from over-damped processes only.
In this respect, we report in this introductory section, as an overview,
that the random walk for normal diffusion goes through 
the Galton board setting, 
namely, at each fixed time-step the walker performs 
a jump drawn from a symmetric distribution with finite variance 
\cite{klafter_sokolov-2011}.
But theories of random walks could be even very refined.
So when, few decades ago,
anomalous diffusion catched the attention of the scientific community
"Random walks were an old topic that seemed fully understood and explored, 
belonging to textbooks and not having novel research directions" 
\cite{shlesinger-epjb-2017}. 
But, if the simple setting of the Galton board works well for normal
diffusion with the assumption of indepedence between consecutive states,
i.e., the Markovianity property,
a fundamental feature of anomalous diffusion is embodied
indeed by memory effects between consecutive states, 
i.e., the non-Markovianity property. 
Therefore the best candidate for modelling anomalous diffusion
emerged to be the continuous-time random walk (CTRW),
first introduced by Montroll and Weiss in 1965 
\cite{montroll_etal-jmp-1965}.
Namely, the CTRW is a random walk which allows for 
random waiting-times between consecutive jumps 
and so there is no more a fixed time-step for time evolution
but a time-step drawn by a distribution.
As a matter of fact, this is a procedure for introducing non-Markovianity 
into the settings of the random walk.
Later, many successes of the CTRW have been reported,
see, e.g., \cite{
shlesinger_etal-prl-1987,
bouchaud_etal-pr-1990,
metzler_etal-pr-2000,
metzler_etal-jpa-2004}.

Anomalous diffusion took its place in 1973 when
Scher and Lax discussed
\cite{scher_etal-prb-1973a,scher_etal-prb-1973b},
in general, transport processes in disordered systems,
and, in particular, the diffusion of carriers 
in amorphous semiconductor films for photocopying machines.
In 1975 a successful model on the basis of the CTRW was
proposed by Scher and Montroll \cite{scher_etal-prb-1975}
by using recent calculations in 1973 by 
Montroll and Scher \cite{montroll_etal-jsp-1973} and in 1974
by Shlesinger \cite{shlesinger-jsp-1974}.
1973--1975 were anni mirabiles for the anomalous diffusion. 
After this, new applications of diffusion theory 
started to call for new modelling approaches and,
by passing through a number of other applications in physics
\cite{anomaloustransport},
anomalous diffusion landed nowadays in living systems 
\cite{barkai_etal-pt-2012,sabri_etal-prl-2020}.

At the same time, the field of fractional calculus
found its glorious application in modelling anomalous diffusion
through the time-fractional generalisation of the diffusion equation, 
i.e., by replacing the first-order time-derivative with a non-local
derivative operator of a fractional (actually a positive real) order.
The link between anomalous diffusion and fractional calculus
is embodied by the so-called memory effect that governs the diffusion
and is encoded into the power-law kernel of the operators
of fractional calculus.  
The story of fractional models for anomalous diffusion 
started in 1986 with Nigmatullin 
who modelled diffusion in porous medium by using fractal comb-like structures
and come down to derive a time-fractional diffusion equation 
\cite{nigmatullin-pssb-1986}, see also reference 
\cite{duarte_etal-fcaa-2014} for other
pioneering applications of fractional calculus. 
The same year, the solution to such equation was provided by Wyss 
\cite{wyss-jmp-1986}
for a time fractional-order less than $1$, that was extended to 
be less than $2$ in 1989 by Schneider and Wyss \cite{schneider_etal-jmp-1989}.
Following Schneider and Wyss \cite{schneider_etal-jmp-1989},
in 1989, Nonnenmacher and Nonnenmacher applied fractional calculus  
to the Boltzmann equation for deriving a fractional extended 
irreversible thermodynamics \cite{nonnenmacher_etal-cph-1989},
and in the same year Nonnenmacher \cite{nonnenmacher-ebj-1989}
applied fractional calculus to lateral diffusion processes 
in biomembranes predicting the measured values more accurately 
than when being compared with the predictions of
standard diffusion.

Later, in 1995--1996, Mainardi published noteworthy papers 
\cite{mainardi-rqe-1995,mainardi-csf-1996,mainardi-aml-1996}
where the time-fractional diffusion and its solution were 
put in an easy-to-understand setting that widely popularised the topic, 
for an historical summary we refer the reader to reference 
\cite{pagnini_etal-caim-2015},
and such popularisation continued with other noteworthy 
papers co-authored with Gorenflo,
see, e.g., \cite{
gorenflo_etal-cism-1997,
scalas_etal-pa-2000,
mainardi_etal-pa-2000,
gorenflo_etal-cp-2002}.
Fractional diffusion were definitively legitimised in 2002 by 
Sokolov, Klafter and Blumen
\cite{sokolov_etal-pt-2002}.
The meaning of time fractional-derivative in physical models
was investigated since 1995 by Hilfer,
see, among many, the papers 
\cite{hilfer-f-1995,hilfer-csf-1995,hilfer-a-2016,kleiner_etal-amp-2021}, 
and see also two critical analysis 
about the relation between fractional and fractality
by Rutman dated 1994 and 1995 \cite{rutman-tmp-1994,rutman-tmp-1995}.
The reader interested on the success of fractional calculus in anomalous diffusion
can pass through a number of edited books, e.g.,
\cite{hilfer-book,klm,tarasov_volume4,tarasov_volume5}.
CTRW and fractional diffusion
emerged to be linked {\it de facto} under certain mild conditions in 1985
when, without referring to fractional operators,
Balakrishnan showed for the first time a similar integral representation
\cite{balakrishnan-pa-1985}, but unfortunately without well-posing 
the problem with respect to the initial condition. 
The link between CTRW and fractional diffusion was indeed
derived on rigorous basis only 
in 1995 by Hilfer and Anton \cite{hilfer_etal-pre-1995},
after that in 1993 the CTRW were linked to fractional relaxation phenomena
by Gl\"ockle and Nonnenmacher \cite{glockle_etal-jsp-1993}.
So, the correct setting of the CTRW in the framework of
non-local fractional operators was derived quite late
in spite of the fact that the relation between the CTRW and the
generalised Master equation was already known since the 70s 
\cite{bedeaux_etal-jmp-1971,
kenkre_etal-jsp-1973,kenkre_etal-prb-1974,
kehr_etal-pa-1978},
as well as the use of Fourier and Laplace multipliers 
in the framework of the CTRW
\cite{tunaley-jsp-1974,shlesinger-jsp-1974},
and very close results were obtained in the 80s 
\cite{klafter_etal-prl-1980,
shlesinger_etal-jsp-1982,
zwanzig-jsp-1983,klafter_etal-pra-1987}.
For a critical review about the link between the 
CTRW and fractional calculus, the reader is referred to 
the introductory section in reference \cite{hilfer-epjb-2017} 
and to references \cite{hilfer-pa-2003,barkai_etal-pa-2007}.
However, the link CTRW--fractional diffusion is only an oversimplified picture 
that is unable to cover the rich phenomenology 
that has a place behind the label of anomalous diffusion, 
but for sure it was the most successful way for the scientific community
to become acquainted with anomalous diffusion and
to bring out the most important observables.

During the years, 
anomalous diffusion was slowly established both theoretically,
see, e.g., the interpretation of fractional calculus 
as a macroscopic manifestation of randomness \cite{grigolini_etal-pre-1999} or
the relation with Hamiltonian chaos \cite{zaslavsky-2005},
and experimentally, see, e.g., 
\cite{tolicnorrelykke_etal-prl-2004,klafter_etal-pw-2005,
golding_etal-prl-2006,bronstein_etal-prl-2009,regner_etal-bj-2013},
up to the recent confident exhortation by Metzler:
"Experimentalists, keep reporting unexpected behaviors!"
\cite{metzler-bj-2017}.

In the last decades, 
a plethora of models were proposed and investigated, 
each one for fixing and explaining some observables,
see, e.g.,
\cite{baeumer_etal-fcaa-2001,
beghin-csf-2012,
jeon_etal-pccp-2014,metzler_etal-pccp-2014,
yuste_etal-pre-2016,molina_etal-pre-2016,
vitali_etal-jrsi-2018,
dossantos-csf-2019,sliusarenko_etal-jpa-2019,lanoiselee_etal-jpa-2019,
dossantos_etal-csf-2021,itto_etal-jrsi-2021,chechkin_etal-pre-2021}.
Among these recent models, the so-called diffusing-diffusivity
(DD) approach \cite{chechkin_etal-prx-2017} 
resulted to be well performing with respect to some relevant features.
The DD approach is based on two stochastic differential 
equations (SDEs):
the over-damped Langevin equation for driving the walker's trajectory
and a SDE for the time-dependent diffusion coefficient.
This two-equation model resembles the subordination approach 
and it leads to a superstatistical solution 
at elapsed time shorter than the correlation time-scale of
the diffusion coefficient \cite{chechkin_etal-prx-2017}.
In particular, the DD approach allows for a transition 
from an exponential walker's PDF
to a Gaussian PDF displaying also the Brownian yet non-Gaussian (BynG) interval
\cite{chechkin_etal-prx-2017,postnikov_etal-njp-2020}. 

To conclude this section,
in this paper we show, by adopting the CTRW formalism, that
anomalous diffusion can emerge from a co-existing pair of well-set Markovian 
hopping-trap mechanisms with only two different time-scales.
Therefore, it is not needed indeed to introduce a broad distribution 
of relaxation times \cite{shlesinger-arpc-1988,pagnini-pa-2014}
and neither cumbersome random-walk models.
Actually, this model meets many ensemble and also single-particle statistics 
that define anomalous diffusion: in particular, 
the transitions of the walker's PDF and the BynG interval 
like the DD model does. The present model has a number of analogies with the 
DD-like model recently studied by Hidalgo--Soria, Barkai and Burov
\cite{hidalgosoria_etal-e-2021}. That model is based on an 
over-damped Langevin equation for a Gaussian process 
with a dichotomous diffusion coefficient that switches after
a random time. The main difference with the present research lays
on the fact that any DD-like approach describes through the Langevin
equation a pure Lagrangian point-of-view of the continuos walker's wandering 
by a fully characterisation in terms of the elapsed time.
In our CTRW setting, 
the Eulerian point-of-view of the hopping-trap mechanism is previleged,
and the adoption of two Markovian mechanisms allows for 
fulfilling the Onsager principle \cite{allegrini_etal-pre-2003}. 
Moreover, recent e-prints appeared
\cite{doerries_etal-arxiv-2022,metzler_etal-arxiv-2022}
where the authors, 
motivated by experimental parameters for tau proteins in neuronal cells, 
through a simple Markovian mobile-immobile transport of particles,
which has some similarities with the present model, 
unveil certain features of anomalous diffusion as the 
transitions of the walker's PDF by including also the BynG interval. 

The rest of the paper is organised as follows.
In section \ref{sec:paradigmatic} 
we report the features of the paradigmatic anomalous diffusion.
In section \ref{sec:definition} we introduce the model 
and in section \ref{sec:simulations} 
we present and discuss the results of an intense study by simulations.
Conclusions are reported in final section \ref{sec:conclusions}.

\section{The paradigmatic anomalous diffusion}
\label{sec:paradigmatic}
With the advent of techniques
for single-particle tracking in living systems, anomalous diffusion
found a paradigmatic setting on the basis of experimental data,
see, e.g., \cite{metzler_etal-pccp-2014,manzo_etal-rpp-2015}.
Let $t \ge 0$ be the time-parameter and $\Omega$ be the sample space,
then we denote a stochastic process in unbounded domain by 
$X^\omega_t:[0,\infty) \times \Omega \to \R$ where 
$\omega \in \Omega$ indexes each independent realization
(namely, each single-particle trajectory).
What we call {\it paradigmatic anomalous diffusion} is a 
generic one-dimensional random walk diffusing in an unbounded domain
that meets the followings features.

At the level of ensemble statistics, anomalous diffusion displays a regime
characterized by a MSD that grows in time
according to a sub-linear power-law, namely
\be 
\E[X_t^2] \sim t^\beta \,, \quad 0 < \beta < 1 \,,
\ee
and a stretched-exponential distribution that is related to the
anomalousness parameter $\beta$ by  
\be
\rho(z) \sim |z|^{(\beta-1)/(2-\beta)} \exp\{- |z|^{2/(2-\beta)}\}
\,, \quad |z| \to + \infty \,.
\label{TFDE}
\ee
Here, we refer to (\ref{TFDE}) as the time-fractional diffusion law.

In terms of FP equation, this phenomenology is 
modelled by the time-fractional diffusion equation 
\cite{schneider_etal-jmp-1989,mainardi-csf-1996,mainardi-aml-1996},
that is the governing equation, for example,
of the walker's distribution in the case of 
the CTRW with infinite-mean waiting times 
\cite{hilfer_etal-pre-1995}
or of the gray Brownian motion (gBm) by Schneider 
\cite{schneider-1990,schneider-1992}.
This last was originally based
on the over-damped fractional Brownian motion (fBm) but 
it could be extended to the under-damped Langevin equation as well 
\cite{vitali_etal-jrsi-2018}.
In terms of physical interpretation, 
the approach based on the
CTRW describes a diffusion process 
in an inhomogeneous medium \cite{berkowitz_etal-wrs-2002,sposini_etal-nfo-2021}
while the approach based on the gBm describes a diffusion process by an
heterogeneous ensemble of walkers \cite{vitali_etal-jrsi-2018}.
Another stochastic modelling, 
whose walker's PDF is governed by the time-fractional diffusion equation,
is the subordination approach \cite{baeumer_etal-fcaa-2001} and 
it is somehow related to the CTRW \cite{baeumer_etal-tams-2009}.
In this approach, 
the Bm evolves with respect to an operational time that is a random variable
driven by the physical time. Such randomness can indeed be re-phrased as
a randomess of the time-scale of the physical time \cite{ditullio_etal-fp-2019}.

At the same time, the anomalous diffusion regime is indeed 
an intermediate regime,
and further features emerged 
as well as prototypical features of the anomalous diffusion
from the improved experimental capacities 
\cite{barkai_etal-pt-2012,metzler_etal-pccp-2014,
manzo_etal-rpp-2015,manzo_etal-prx-2015}.
So, nowadays, the paradigmatic anomalous diffusion includes also 
a walker's distribution $\rho(x;t)$ with exponential tails 
at short elapsed time, $t \ll \tauB$,  
i.e., $\rho(z) \sim \rme^{- |z|}$ as $|z| \to + \infty$,
where $\tauB$ stays for the Barkai--Burov time-scale 
who proved that such exponential tails 
(up to a logarithmic correction:
$\rho(x;t) \sim \rme^{- |x|\log(|x|/t)^\gamma - C t}$ as $|x|/t \to + \infty$
and $\gamma, C > 0$) 
are indeed universal for diffusing walkers 
\cite{barkai_etal-prl-2020,wang_etal-e-2020},
and also a distribution with Gaussian tails 
at large elapsed times, $t \gg \tauD$,  
i.e., $\rho(z) \sim \rme^{- z^2}$ as $|z| \to + \infty$,
where $\tauD$ stays for the normal diffusion time-scale when,
as a matter of fact, 
walkers go through a Brownian and Gaussian diffusion 
(anomalous-to-normal transition)
\cite{sandev_etal-jpa-2018,
molina_etal-njp-2018,
sliusarenko_etal-jpa-2019}.
We remark that here the universal exponential-tailed distribution
by Barkai \& Burov \cite{barkai_etal-prl-2020} is intended
as small-time universality rather than as large-space universality:
this exchange is done on the basis of the limit $|x|/t \to \infty$.
The ensemble phenomenology is finally enriched by the
mentioned BynG interval \cite{chechkin_etal-prx-2017,postnikov_etal-njp-2020}, 
when the MSD starts to grow linearly in time before 
than the anomalous-to-normal transition occurs,
namely at $\tauB \ll \tauDD \ll t \ll \tauD$. 
Within these PDF-transitions, 
the DD approach emerged to be the higher flexible formulation 
\cite{chechkin_etal-prx-2017,hidalgosoria_etal-e-2021}.
In its minimal-model scheme,
the diffusion coefficent in the DD approach is determined by the square of
an Ornstein--Uhlenbeck process and this 
introduces an extra time-scale that allows for a transition of the
walker's PDF through the mentioned regimes and also 
for the appearing of the BynG interval.
The PDF-transition is the strength point of the DD model
with respect to other superstatistical-like approaches as:
the gBm \cite{schneider-1990,schneider-1992}
and the generalized gray Brownian motion (ggBm) 
\cite{molina_etal-pre-2016},
which are both superstatistical-like fBm \cite{mackala_etal-pre-2019},
that indeed do not display a transition of the walker's PDF 
between different shapes.
The CTRW approach allows indeed for the anomalous-to-normal transition
\cite{saichev_etal-jetp-2004}, that can be observed also in 
the under-damped ggBm as a consequence of the finite statistical sampling 
of the time-scales \cite{sliusarenko_etal-jpa-2019}.

At the level of single-particle statistics, 
the observables that characterize anomalous diffusion are 
the p-variation test \cite{magdziarz_etal-prl-2009,magdziarz_etal-jpa-2013}, 
the time-averaged MSD (TAMSD) and the ensemble-averaged TAMSD (ETAMSD).
The p-variation test provides information on the stochastic origins
of the data allowing for discriminating among processes. 
The TAMSD and the ETAMSD allow for observing 
the dependence of the statistics on the time-lag between 
the start of the process and the start of the measurement 
\cite{weigel_etal-pnas-2011,tabei_etal-pnas-2013,manzo_etal-prx-2015},
which is called aging.
TAMSD and ETAMSD also allow
for estimating the degree of ergodicity breaking 
\cite{bouchaud-jpf-1992,schulz_etal-prl-2013},
namely when the walkers need an infinite time for exploring an infinite system 
but they can access to the whole domain because it is not 
split in mutually inaccessible regions \cite{bouchaud-jpf-1992}.
In formulae they are as follows.

Let $T$ be the measurement time, i.e., $t\in [0,T]$,
with time-step $h$ such that $T=N h$ and $\Delta = m h$, 
where $N \,, m \in \N$ and $N > m$,  
then the p-variation test $V^{(p)}(t)$ is defined as
\begin{equation}
 V^{(p)}(t)=\lim_{n \to \infty}
\sum_{j=0}^{2^n-1}
\left\vert 
X_{t_{j+1}\wedge t} - X_{t_{j}\wedge t}\right\vert^p \,,
\quad {\rm with} \quad 2^n=N \,,
\label{eq:p-variation}
\end{equation}
where $t_j= j T/2^n$ and $a \wedge b = \min\{a \,, b\}$.
The TAMSD is calculated by the formula
\begin{equation}
 \overline{\delta^2}(T,\Delta) =
\frac{1}{N-m+1}\sum_{k=0}^{N-m} \left[
X_{kh + \Delta} - X_{kh} \right]^2 \,, 
\label{eq:TAMSD}
\end{equation}
as a function of $\Delta$, 
while when the ETAMSD $\E\left[ \overline{\delta^2}^2 \right]$ is calculated
as a function of $T$ then aging is observed.
Finally, the degree of ergodicity breaking is estimated by
the parameter
\begin{equation}
E_B(T,\Delta) = \lim_{T \to \infty} \frac{\E\left[ \overline{\delta^2}^2 \right]}
{\E^2\left[\overline{\delta^2} \right]} - 1 \,.
\label{eq:EB}
\end{equation}
Actually, $E_B$ is an indicator of the  
inequality between time-averaged and ensemble-averaged statistics.

Paradigmatic anomalous diffusion displays a p-variation consistent with
Gaussian processes, in particular with the fBm \cite{magdziarz_etal-prl-2009}.
Moreover, 
TAMSD displays a linear growing in time,
suggesting an underlying Bm, but the diffusion coefficient differs
among single-trajectories, see, e.g., reference \cite{manzo_etal-prx-2015}.
The distribution of the diffusion coefficients among
the trajectories causes the weak ergodicity breaking
\cite{he_etal-prl-2008,manzo_etal-prx-2015,molina_etal-pre-2016}.
The CTRW \cite{he_etal-prl-2008}, 
the subordinated fBm \cite{thiel_etal-pre-2014} and
the ggBm \cite{molina_etal-pre-2016} have 
the same degree of ergodicity breaking, 
i.e., they provide the same value of $E_B$.
Furthermore, anomalous diffusion in living systems displays aging 
with an ETAMSD that decreases as $T^{-\lambda}$ with $\lambda > 0$  
\cite{weigel_etal-pnas-2011,tabei_etal-pnas-2013,manzo_etal-prx-2015}.
All these properties are reproduced by the ggBm \cite{molina_etal-pre-2016}, 
while the CTRW cannot reproduce the p-variation test \cite{magdziarz_etal-prl-2009}
together with other failures \cite{manzo_etal-prx-2015}.

\section{The model}
\label{sec:definition}
\subsection{Definition}
We propose a model based on the theory of the CTRW,
see references \cite{
shlesinger-epjb-2017,
weiss-1994,zaburdaev_etal-rmp-2015,
scher-epjb-2017,kutner_etal-epjb-2017}
for technical and historical reviews.
Let $\Omega$ be the sample space,
then $\omega \in \Omega$ indexes 
each independent realization of the walker's trajectory.
In the CTRW approach, each $\omega$-realization of the walker's 
trajectory goes through the pair of iterative processes 
\be
X_N^\omega-X_{N-1}^\omega = R_N \,, 
\quad 
t_N^{\omega}-t_{N-1}^{\omega} = \tau_N \,, 
\quad N=1 \,, 2 \,, \dots \,, 
\label{CTRWrule}
\ee
where 
the displacements $R$ are i.i.d. random variables distributed
according to the jump-size distribution $\lambda(x)$
and
the positive time-increments $\tau$ between consecutive jumps
are i.i.d. random variables distributed according to the 
waiting-time distribution $\psi(t)$
such that, after $N$ iterations, 
the walker of the $\omega$-realization 
is located in $X_N^{\omega}=x_0 + \sum_{i=1}^N R_i$
at the elapsed time 
$t_N^{\omega}=t_0 + \sum_{i=1}^N \tau_i$.
It holds that, for the same number of iterations $N$,
the elapsed times of 
two different $\omega$-realizations are different.
We can compress this notation in 
$(X_N^{\omega},t_N^{\omega})=X^\omega_{t_N}$. 
We set $x_0=0$ and $t_0=0$.

Without loss in generality, 
we consider the case of a random walk
in continuous space, rather than in a lattice,
and then $X^\omega_{t_N}:[0,+\infty) \times \Omega \to \R$ with 
$\lambda(x):\R \to \R_+$ and $\psi(t):\R_+ \to \R_+$.
Within the formalism of the uncoupled CTRW
\cite{germano_etal-pre-2009},
if $\rho(x;t):\R \times (0,+\infty) \to \R_+$ 
is the walker's PDF at site $x \in \R$ and time $t > 0$
with initial datum $\rho(x;0)=\delta(x)$,
such that $\displaystyle{\int_{\R}\rho(x;t) \, \rmd x=1}$ 
and $\rho(x;t) > 0$ for all
$(x,t) \in \R \times (0,+\infty)$,
then in the Fourier--Laplace domain it holds
\be
{\widehat{\widetilde{\rho}}}(\kappa;s)
=\int_0^{+\infty} \e^{-st}\int_{-\infty}^{+\infty} \e^{-i \kappa x}
\rho(x;t) \, \rmd x \, \rmd t
=\frac{1-\widetilde{\psi}(s)}{s [1-\widehat{\lambda}(\kappa)
\widetilde{\psi}(s)]} \,, 
\label{CTRW}
\ee
where 
$\widetilde{\psi}(s)$ and $\widehat{\lambda}(\kappa)$
are the Laplace and the Fourier transforms
of the waiting-time distribution $\psi(t)$
and of the jump-length distribution $\lambda(x)$,
respectively. Since from the normalization condition of
$\psi(t)$ and $\lambda(x)$ it follows that
$\widetilde{\psi}(0)=\widehat{\lambda}(0)=1$,
then it holds $\widehat{\widetilde{\rho}}(0;s) = 1/s$.

The memory of the process is provided by the
waiting-time distribution $\psi(t)$. In fact,
by inverting (\ref{CTRW}) we have that the FP equation is
\cite{mainardi_etal-pa-2000}
\be
\int_0^t \Phi(t-\tau) \frac{\partial \rho}{\partial \tau} d\tau=
-\rho(x;t) + \int_{-\infty}^{+\infty}
\lambda(x-\xi)\rho(\xi;t) \, \rmd \xi \,,
\label{FPmemory}
\ee 
where the memory kernel $\Phi(t)$ is defined by
\be
\widetilde{\Phi}(s)=\frac{1-\widetilde{\psi}(s)}{s \widetilde{\psi}(s)} \,.
\label{memory}
\ee
Therefore, the process is memory-less, i.e., Markovian,
when $\Phi(t)=\delta(t/\tauM)=\tauM \delta(t)$, 
where $\tauM$ is the time-scale of the Markovian process,
which means $\widetilde{\Phi}(s)=\tauM$ and then
$\psi(t)=\rme^{-t/\tauM}/\tauM$ \cite{zwanzig-jsp-1983,mainardi_etal-pa-2000}
such that the mean waiting-time is:
$\displaystyle{
\langle t \rangle = \int_0^{+\infty} t \, \psi(t) \, \rmd t =
\tauM}$.

We consider a Gaussian jump-length distribution, i.e.,
\be
\lambda(x)=\frac{\rme^{-x^2/(2\sigma^2)}}{\sqrt{2\pi\sigma^2}} \,, \quad
\langle x^2 \rangle = 
\int_{-\infty}^{+\infty} 
x^2 \, \frac{\rme^{-x^2/(2\sigma^2)}}{\sqrt{2\pi\sigma^2}} \, \rmd x 
= \sigma^2 \,,
\label{modelsetting}
\ee
and the following model for the waiting-time distribution
\be
\psi(t)= 
p \, \frac{\rme^{-t/\tauD}}{\tauD} + (1-p) \, \frac{\rme^{-t/\tauB}}{\tauB} \,, 
\quad
\tauB \le \tauD \,,
\label{model}
\ee
where $p \in [0,1]$ is the probability of occurrence
of each family of Markovian mechanisms: 
$\psi_{\rm D}(t)=\rme^{-t/\tauD}/\tauD$ and
$\psi_{\rm B}(t)=\rme^{-t/\tauB}/\tauB$,
$\tauD$ is the time-scale of the diffusion-limit and  
$\tauB$ is the Barkai--Burov time-scale 
\cite{barkai_etal-prl-2020,wang_etal-e-2020}, 
both have been discussed in section \ref{sec:paradigmatic},
and the diffusion coefficient results to be  
$\mD=\sigma^2/(2 \, \langle t \rangle)$
with $\langle t \rangle = p \, \tauD + (1-p) \, \tauB$.

\subsection{Non-Markovianity}
By using the Laplace transform of the memory kernel (\ref{memory}), 
we can study the non-Markovianity of model (\ref{model}). 
Actually, we obtain
\be
\widetilde{\Phi}(s)=\frac
{\tauB (1+\tauD s) + p (\tauD - \tauB)}
{p (1+\tauB s) + (1-p)(1+\tauD s)}
\,,
\label{modelmemory}
\ee 
and it can be checked that model (\ref{model}) 
is Markovian when $\tauB=\tauD=\tauM$,
such that $\widetilde{\Phi}(s)=\tauM$ for all $p$, 
and when $p=0$, such that $\widetilde{\Phi}(s)=\tauB=\tauM$ 
for arbitraty $\tauD$. 
At large time,
the memory fades away and then, when $\tauD \gg \tauB$, we have
that in the Laplace domain 
the large-time limit corresponds to the limit $s \ll 1/\tauD \ll 1/\tauB$
and it holds
\be
\widetilde{\Phi}(s) 
\simeq p \, \tauD + (1-p) \, \tauB = \tauM \,,
\quad \tauB s \ll \tauD s \ll 1 \,,
\label{modelmemory2}
\ee 
that is an estimation of $\tauM$ because in this limit
$\widetilde{\Phi}(s)$ is independent of $s$. 
Therefore we have that
\be
\tauM
= \tauB \left(1 - p + p \, \frac{\tauD}{\tauB}\right) \,,
\label{tauM}
\ee 
and by assuming that $\tauM$ exists and is bounded 
for all $p$ and whatever is the inequality $\tauD \gg \tauB$, 
it means that $p \, \tauD/\tauB$ is always bounded for all $p$
and whatever is the inequality $\tauD \gg \tauB$, too, 
such that $\tauB \simeq p \, \tauD$ when $p \to 0$.
In fact, if $\tauD < \infty$
then the quantity $p \, \tauD/\tauB$ is not bounded for all $p$ 
when for example $\tauB \propto p^2$ and $p \to 0$.  
To conclude, if we plug $\tauB \propto p \, \tauD$ 
into (\ref{tauM}),
together with $p \to 0$ such that $\tauD \gg \tauB$,
we have that $\tauM \simeq \tauB \, (2 - p) \propto 2 \, p \, \tauD$
and, by plugging this last into (\ref{modelmemory2}), 
we finally have the following constraints for non-Markovianity
\be
\tauB=\tauD \, \frac{p}{1-p} \,, \quad p \in (0,1/2) \,,
\label{model2}
\ee
when $p=1/2$ then $\tauB=\tauD$ and, according to (\ref{modelmemory}),
model (\ref{model}) results to be Markovian.
In particular, constraint (\ref{model2}) provides an estimation of 
the Barkai--Burov time-scale $\tauB$ 
for defining the small elapsed-times, i.e., $t \ll \tauB$,
and states the time-scale $\tauD$ for defining the large elapsed-times, 
i.e., $t \gg \tauD$. 

Hence, from (\ref{memory}) we have that model 
(\ref{model}, \ref{model2}) is non-Markovian, 
in fact it holds
\be
\widetilde{\Phi}(s)= \tauD \,
\frac{p (1+\tauB s) + p (1+\tauD s)}
{p (1+\tauB s) + (1-p) (1+\tauD s)} 
\,, \quad p \in (0,1/2) \,,
\label{Phimodel}
\ee
that is dependent on $s$ when $\tauD s \gg \tauB s \gg 1$
and $p \in (0,1/2)$, and it goes to 
$\widetilde{\Phi}(s)=2 \, p \, \tauD = \tauM$
when $\tauB s \ll \tauD s \ll 1$. 
Note that in (\ref{model2}) and (\ref{Phimodel}) $p \ne 0$, 
see also below (\ref{modelmemory}),
otherwise for model (\ref{model}, \ref{model2}) it holds 
$\tauB=0$ for $\tauD < \infty$
and $\widetilde{\Phi}(s)=0$ for all $s$,
namely $\psi(t)=0$ for all $t$.

Moreover, in opposition to literature on CTRW models for 
anomalous diffusion, see, among many, the reviews 
\cite{metzler_etal-pr-2000,metzler_etal-jpa-2004,zaburdaev_etal-rmp-2015}, 
the mean waiting-time of model (\ref{model}, \ref{model2}) is finite, i.e.,
\be
\langle t \rangle = \int_0^{+\infty} t \, \psi(t) \, \rmd t = 
2 \, p \, \tauD = \tauM < \infty \,.
\label{meanwt}
\ee 
By setting $\tauD < +\infty$,  
a finite-mean waiting-times exists (\ref{meanwt})
which guarantees a transition to normal diffusion.

\subsection{Power-law memory fading and anomalousness parameter $\beta$}
Together with non-Markovianity, model (\ref{model})
displays also a power-law memory fading.
We introduce the so-called survival probability, 
see, e.g., \cite{weiss-1994}, 
namely the probability for the walker to remain at the initial position:
\be
\Psi(t)=1-\int_0^t \psi(\xi) d\xi \,, \quad
\frac{d\Psi}{dt}=-\psi(t) \,, 
\quad \widetilde{\Psi}(s)= \frac{1-\widetilde{\psi}(s)}{s} \,,
\ee
that, for the present model (\ref{model}), is
\be
\Psi(t)= 
p \, \rme^{-t/\tauD} + (1-p) \, \rme^{-t/\tauB} \,, 
\quad
\widetilde{\Psi}(s)= 
\frac{p \, \tauD}{1+\tauD s} + \frac{(1-p) \, \tauB}{1+\tauB s}  
\,. 
\label{Psi}
\ee

It is known that anomalous diffusion is obtained 
within the CTRW formalism when 
$\widetilde{\psi}(s) \sim 1 - s^\beta$ with $s \to 0$,
e.g., \cite{metzler_etal-pr-2000},
and it is embodied in the CTRW/fractional diffusion formalism
by \cite{hilfer_etal-pre-1995}
\be
\Psi_{\rm ML}(t)=E_\beta(-t^\beta) \,, \quad 
\widetilde{\Psi}_{\rm ML}(s)= \frac{s^{\beta-1}}{s^\beta + 1} \,,
\label{PsiML}
\ee
with
\be
\widetilde{\Psi}_{\rm ML}(s) \sim \frac{1}{s^{1-\beta}} \,,
\quad s \to 0 \,,
\label{PsiMLpower}
\ee
where $E_\beta(-t^\beta)$, with $0 < \beta < 1$, is the Mittag--Leffler
function \cite[Appendix E]{mainardi-2010}
\be
E_\alpha(z)=\sum_{n=0}^\infty \frac{z^n}{\Gamma(1 + \alpha n)} \,,
\quad \alpha > 0 \,, \quad z \in \C \,.
\ee

The Laplace transforms of the two survival probabilities 
(\ref{Psi}) and (\ref{PsiML}) are related by the formula
\cite[formula (E.51)]{mainardi-2010}
\be
\frac{1}{1-s}=\int_0^\infty \e^{-u} \, E_\alpha(u^\alpha s) \, \rmd u \,,
\quad \alpha > 0 \,,
\label{integralML}
\ee
and then
\begin{eqnarray}
\widetilde{\Psi}(s) &=&
p\tauD \, 
\int_0^\infty \e^{-u} \, E_\alpha(- u^\alpha \tauD s) \, \rmd u 
\nonumber \\
& & \hspace{2truecm}
+ (1-p) \tauB \, \int_0^\infty \e^{-u} \, 
\, E_\alpha(-u^\alpha \tauB s) \, \rmd u \,.
\label{Psipower}
\end{eqnarray}
If we consider the interval $1/\tauD \ll s \ll 1/\tauB$ then
it holds $\tauD s \gg 1$ and $\tauB s \ll 1$ such that
$u^\alpha \tauD s \to 0$ for $u \le u_{\rm m}(s) \ll (\tauD s)^{-1/\alpha}$ and
$u^\alpha \tauB s \to 0$ for all $u$.
Therefore, by noting that $E_\alpha(0)=1$, 
in the interval $1/\tauD \ll s \ll 1/\tauB$,
formula (\ref{Psipower}) becomes
\be
\widetilde{\Psi}(s) \simeq
\tauM - p \tauD \, \e^{-u_{\rm m}(s)}\, 
+ p\tauD \, 
\int_{u_{\rm m}(s)}^\infty \e^{-u} \, E_\alpha(- u^\alpha \tauD s) \, \rmd u 
\,,
\label{Psipower2}
\ee
where definition (\ref{modelmemory2}) of $\tauM$ has been used.
Through the change of variable $\xi=(\tauD s)^{1/\alpha} u$,
we have that 
\be
\widetilde{\Psi}(s) \simeq
\tauM - p \tauD \, \e^{-u_{\rm m}(s)}\, 
+ \frac{p\tauD}{(\tauD s)^{1/\alpha}} \, 
\int_{\xi_{\rm m}}^\infty \e^{-\xi/(\tauD s)^{1/\alpha}} 
\, E_\alpha(- \xi^\alpha) \, \rmd \xi 
\,,
\label{Psipower3}
\ee
where $\xi_{\rm m}=(\tauD s)^{1/\alpha} \, u_{\rm m}(s)$ 
is independent of $s$. By looking at the integral term
and by remembering that $\tauD s \gg 1$,
we observe that
for an arranged interval of $s$ such that $\xi_{\rm m} \sim \mathcal{O}(1)$ 
we finally obtain the following power-law scaling
\be
\widetilde{\Psi}(s) \simeq
\tauM  
+ \frac{p \, \tauD^{1-1/\alpha}}{s^{1/\alpha}} \, 
\int_{\xi_{\rm m}}^\infty E_\alpha(- \xi^\alpha) \, \rmd \xi 
\,, \quad \frac{1}{\tauD} \ll s \ll \frac{1}{\tauB} \,, \quad \alpha > 0 \,.
\label{Psipower4}
\ee
From this derivation the value of $\alpha$ is indetermined, as a consequence
of the indetermination of $\alpha$ in formula (\ref{integralML}).
However, the value of $\alpha$
is expected to be determined as function of $p$, i.e., $\alpha=\alpha(p)$.
Moreover, since we are focused on the resulting anomalous diffusion,
we can establish a relation between the value of $\alpha$,
which corresponds to the model (\ref{model}, \ref{model2}), 
and the exponent $\beta$ as follows from the anomalous diffusion 
displayed by a CTRW model equipped with (\ref{PsiML}). 
Hence, by comparing (\ref{Psipower4}) and (\ref{PsiMLpower}),
we have the relation
\be
\beta=1-\frac{1}{\alpha(p)} \,, 
\label{alphabeta}
\ee
and we can fix $\alpha(p)$ by comparing plots of the power-law interval
of (\ref{Psi}) for different values of $p$ against plots of (\ref{PsiML})
for different values of $\beta$.

\begin{figure}
\begin{center}  
\subfloat[][]{
\includegraphics[height=5cm,width=7cm]{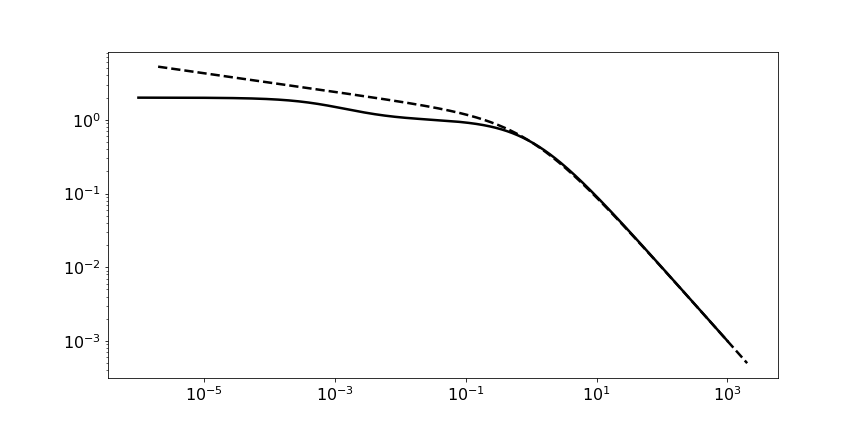}}
\subfloat[][]{
\includegraphics[height=5cm,width=7cm]{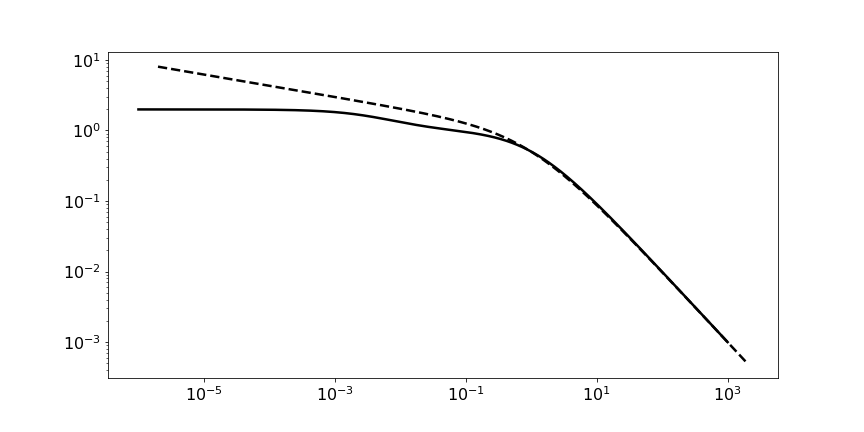}}
\\
\subfloat[][]{
\includegraphics[height=5cm,width=7cm]{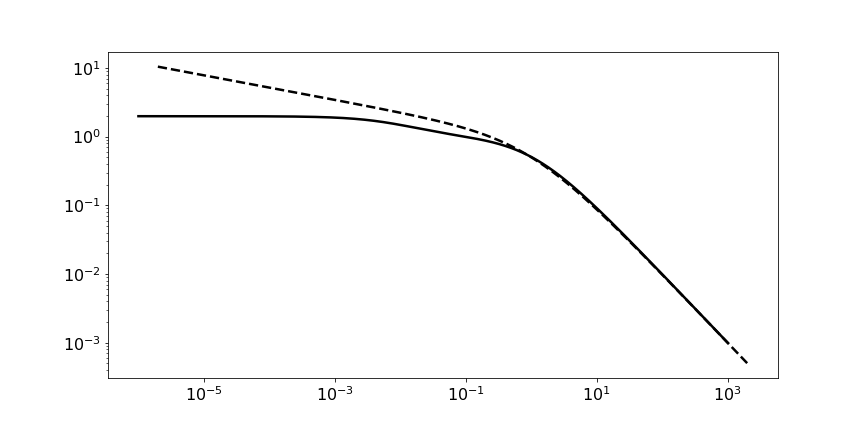}}
\subfloat[][]{
\includegraphics[height=5cm,width=7cm]{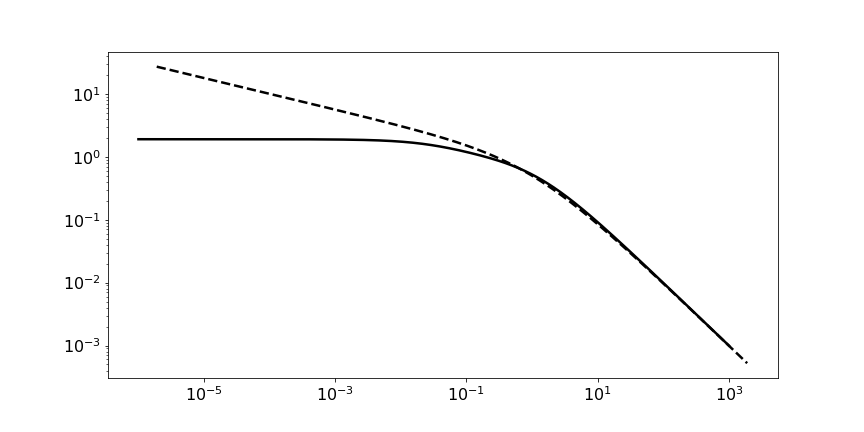}}
\end{center}
\caption{Plots of the Laplace transforms of the 
survival probabilities 
$\widetilde{\Psi}_{\rm ML}(s)/\tau_1 \, vs \, \tau_2 s$ (dashed line)
and $\widetilde{\Psi}(s)/\tauB \, vs \, \tauB s$ (continuos line)
for different values of $p \in (0,1/2)$,
from (a) to (d): $p=0.001 \,, 0.005 \,, 0.01 \,, 0.05$.
The universal short-time behaviour is evident by the overlapping
of the two curves for large values of the corresponding argouments and
starting at $\tau_2 s=1$ and at $\tauB s=1$, respectively.}
\label{fig:Psi1}
\end{figure}

\begin{figure}
\begin{center}
(a) \includegraphics[height=5cm,width=7cm]{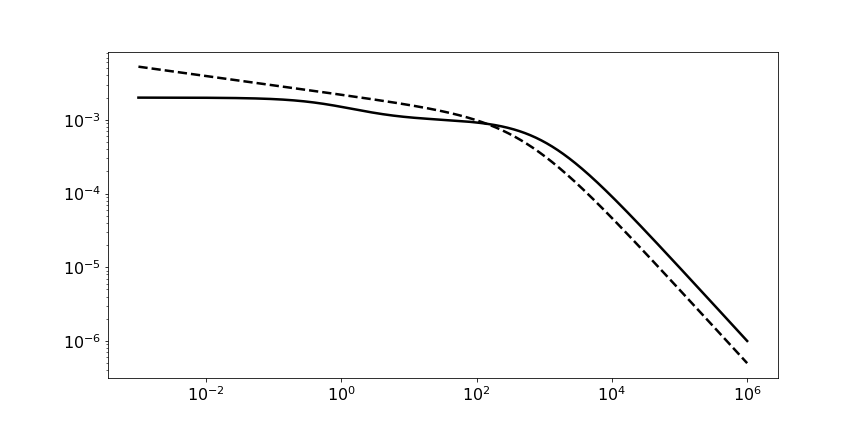}
\includegraphics[height=5cm,width=7cm]{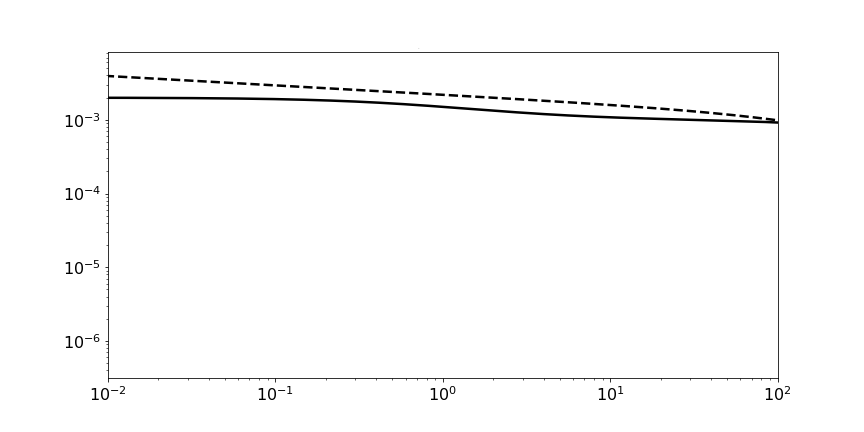}
\\
(b) \includegraphics[height=5cm,width=7cm]{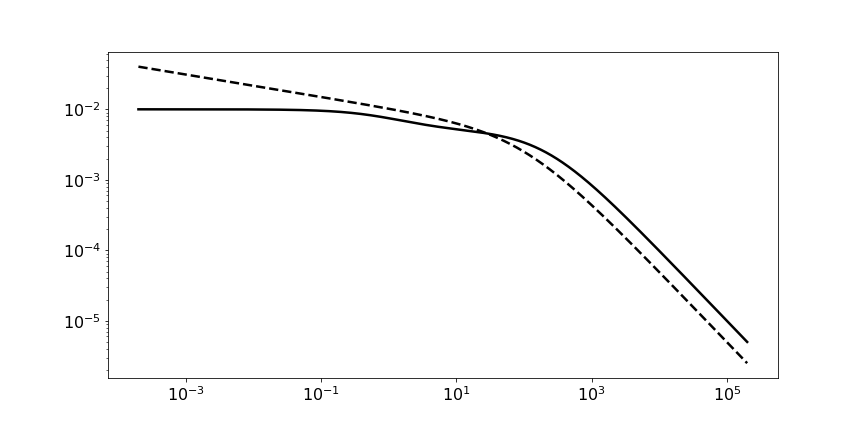}
\includegraphics[height=5cm,width=7cm]{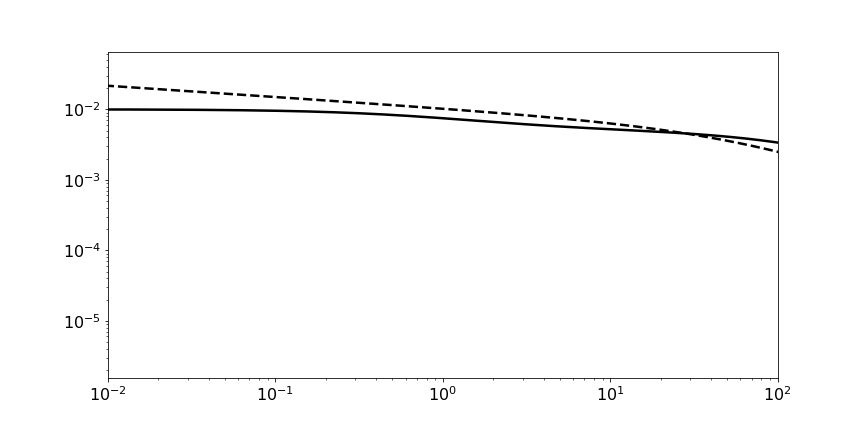}
\\
(c) \includegraphics[height=5cm,width=7cm]{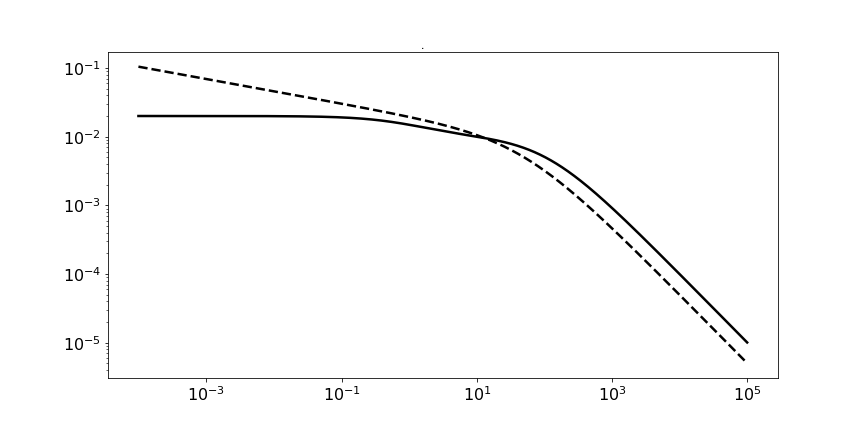}
\includegraphics[height=5cm,width=7cm]{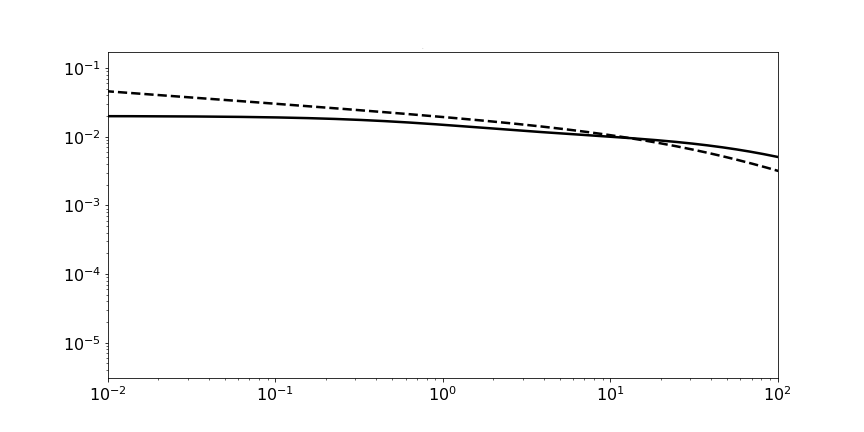}
\\
(d) \includegraphics[height=5cm,width=7cm]{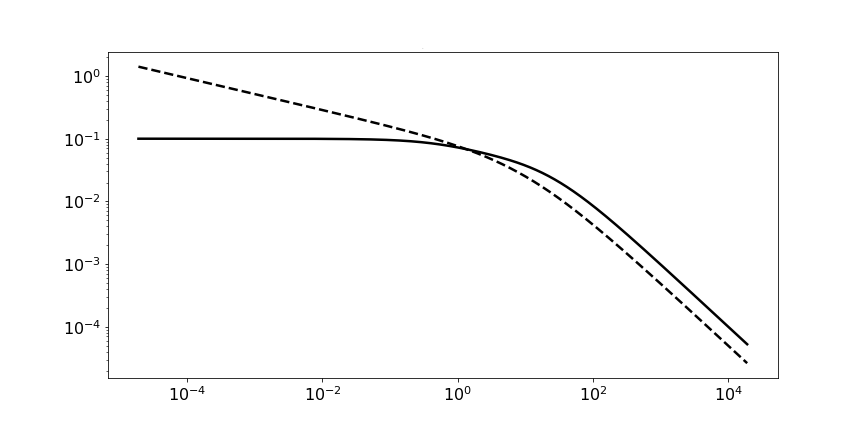}
\includegraphics[height=5cm,width=7cm]{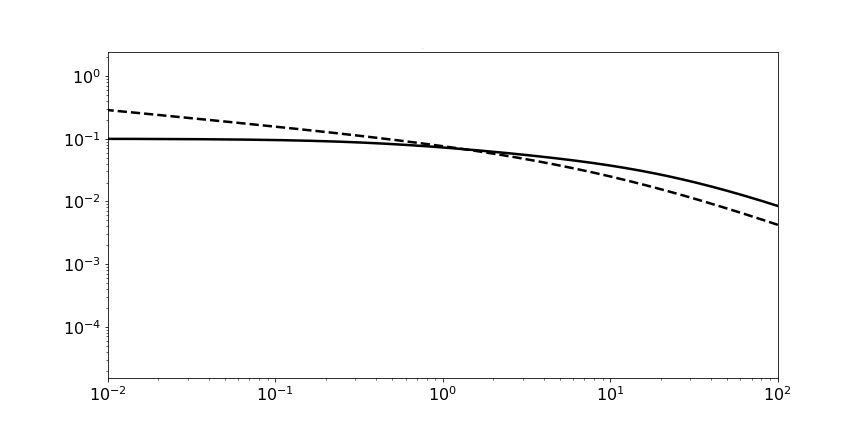}
\end{center}
\caption{
Plots (left column) of the Laplace transforms of the 
survival probabilities 
$\widetilde{\Psi}_{\rm ML}(s)/\tauD \, vs \, \tauD s$ (dashed line)
and $\widetilde{\Psi}(s)/\tauD \, vs \, \tauD s$ (continuos line)
for different values of $p \in (0,1/2)$,
from (a) to (d): $p=0.001 \,, 0.005 \,, 0.01 \,, 0.05$.
It is evident that for small enough $p$ there exists an intermediate interval 
of a couple of decades located around $\tauD s =1$ 
where both Laplace transforms of the
survival probabilities display a power-law decaying, see the zoom in
the right column.}
\label{fig:Psi2}
\end{figure}

In particular, by writing $\widetilde{\Psi}_{\rm ML}(s)$ in dimensional form:
\be
\widetilde{\Psi}_{\rm ML}(s)=
\tau_1 \, \frac{(\tau_2 s)^{\beta-1}}{(\tau_2 s)^\beta + 1} \,,
\label{PsiMLdim}
\ee
and by searching for the time-scales $\tau_1$ and $\tau_2$
by comparing (\ref{PsiMLdim}) against (\ref{Psi}), 
in the limit $\tauB/\tauD \to 0$ we get
\be
\widetilde{\Psi}_{\rm ML}(s) \simeq \widetilde{\Psi}(s) \,,
\quad s \gg \frac{1}{\tau_2} \,, 
\quad {\rm with} 
\quad \tau_1=p \, \tauD \,, 
\quad \tau_2 = 2 \, \frac{\tauD \tauB}{\tauD + \tauB} \,.
\ee
The Laplace transforms of the two survival probabilities, 
i.e., $\widetilde{\Psi}_{\rm ML}(s)$
in (\ref{PsiML}) and $\widetilde{\Psi}(s)$ in (\ref{Psi}), 
are compared in figures \ref{fig:Psi1} and \ref{fig:Psi2}.

In figure \ref{fig:Psi1} we observe that the two survival probabilities 
display an universal behaviour for large values of $s$ with respect
to the associated short-time time-scales: this corresponds to the
universal behaviour for short elapsed-times related to the Barkai--Burov 
time-scale $\tauB$, i.e., $s \gg 1/\tauB$.

In figure \ref{fig:Psi2} we observe that, for small enough $p$,
the two survival probabilities 
display a parallel power-law decreasing for intermediate values of $s$ 
for a couple of decades around the large-time time-scale $\tauD$.
At very large-time, i.e., $s \ll 1/\tauD$, the present model 
transits to the normal
diffusion while the CTRW equipped with the survival probability
$\Psi_{\rm ML}(t)$ continues with the anomalous diffusion.

\begin{figure}
\begin{center}
\includegraphics[height=10cm,width=15cm]{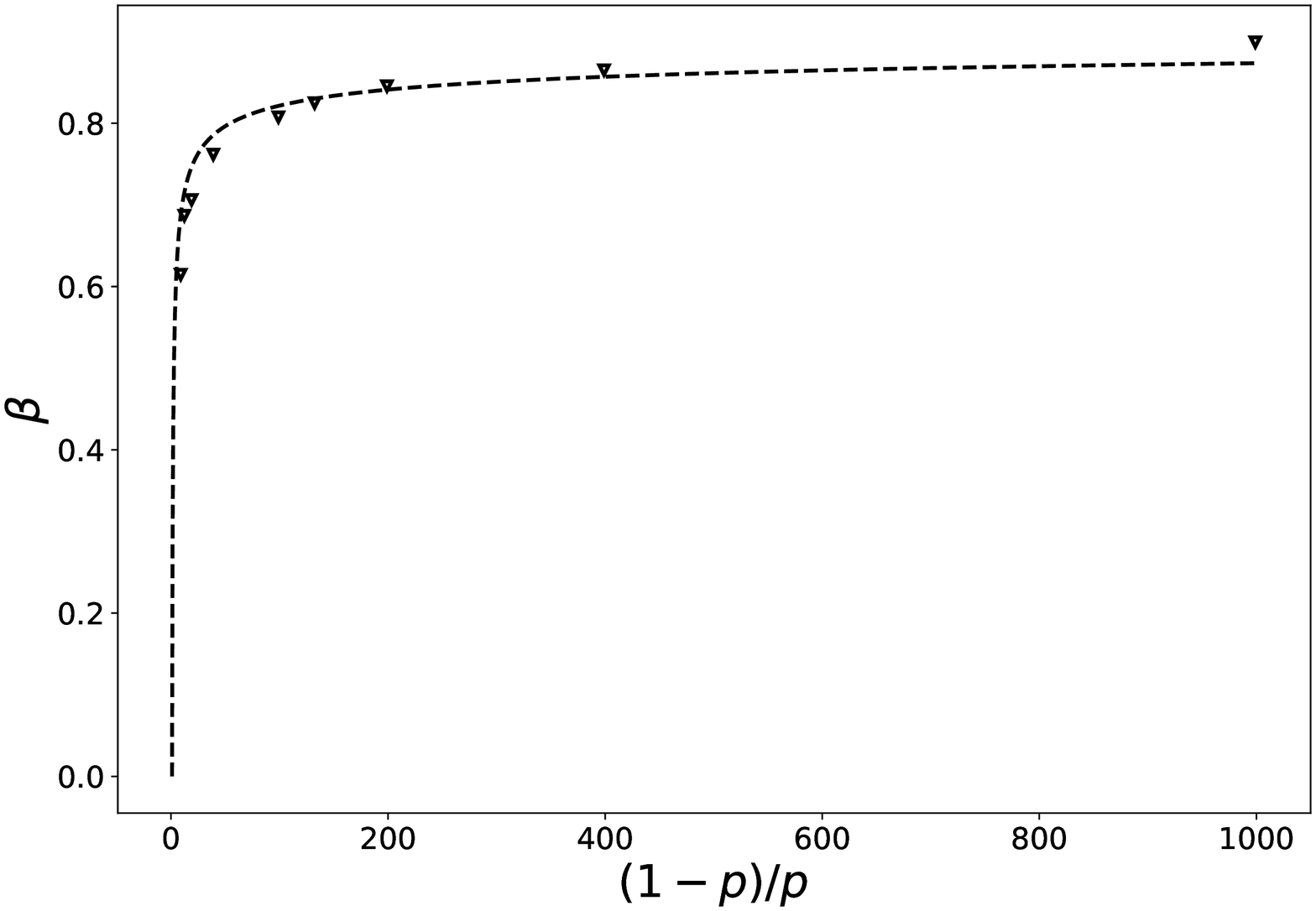}
\end{center}
\caption{Determination of the anomalousness parameter $\beta$ by plotting 
the parameter $\beta$ of $\widetilde{\Psi}_{\rm ML}(s)$ 
and the ratio $(1-p)/p$ of $\widetilde{\Psi}(s)$ 
as they emerge in the equal power-law decaying interval
in figure \ref{fig:Psi2}. 
The fit is given in formula (\ref{beta}).}
\label{fig:beta}
\end{figure}

Moreover, searching for the common power-law between
$\widetilde{\Psi}_{\rm ML}(s)$ and $\widetilde{\Psi}(s)$
in the interval around $\tauD s =1$ leads to a determination
of the anomalousness parameter $\beta$, see figure \ref{fig:beta},
that emerges to be approximated by the formula
\be
\beta \simeq 1- \frac{1}{1 + \displaystyle{\log \frac{1-p}{p}}} \,,
\quad p \in (0,1/2) \,,
\label{beta}
\ee 
and then we have from (\ref{alphabeta}) that
\be
\alpha(p)=1 + \displaystyle{\log \frac{1-p}{p}} \,,
\quad p \in (0,1/2) \,.
\label{alpha}
\ee 

Again we have that $p \ne 0$, otherwise $\beta=1$ and anomalous diffusion 
is lost. On the contrary, we have that when $p=1/2$ it holds $\beta=0$
such that diffusion stops as an extreme consequence of sub-diffusion.
However, when $p=1/2$ the model (\ref{model}, \ref{model2}) reduces
to Markovian standard diffusion and anomalous diffusion is lost again.
An explanation follows from the
estimation of the extension of the anomalous diffusion interval by 
the difference
\be
\tauD - \tauB = \tauD \, \left(1 - \frac{p}{1-p}\right) \,,
\ee 
where (\ref{model2}) has been used. From formula (\ref{beta})
we finally obtain that the extension of the anomalous diffusion interval is
\be
\tauD - \tauB = \tauD \, \left[1 - \e^{-\beta/(1-\beta)}\right] \,,
\label{ADinterval}
\ee 
that decreases when $\beta$ decreases and we have the ratio
\be
\frac{\tauB}{\tauD}=\e^{-\beta/(1-\beta)} \,.
\ee

\subsection{Further to model (\ref{model}, \ref{model2})}
Present model (\ref{model}, \ref{model2}) is fully determined by 
a pair of dimensional parameters, i.e., $[\tauB]=Time$ 
and $[\sigma]=Length$,
and one single adimensional parameter, i.e., $p \in (0,1/2)$.
When $p=1/2$ the model reduces to the Markovian normal diffusion
with $\tauD=\tauB=\tauM$ and $\widetilde{\Phi}(s)= \tauM$
such that the mean waiting-time is $\langle t \rangle = \tauM$.

We highlight that the present model (\ref{model}) 
is not a two-state CTRW \cite{weiss-jsp-1976}.
It is indeed much more close to the model studied by Hilfer
in his objection to the relation between the CTRW and time-fractional 
diffusion equations \cite{hilfer-pa-2003} and 
to the models studied by Barkai and Sokolov in their subsequent reply 
\cite{barkai_etal-pa-2007}.
Those models \cite{hilfer-pa-2003,barkai_etal-pa-2007} 
were based on a waiting-times distribution
builted by the combination of a power-law and an exponential-law:
a recent analog study showed that for an inhomogeneous version
of that model a weak form of the objection still holds 
\cite{carnaffan_etal-c-2020} 
because normal diffusion can be observed for a particular
interval of the exponent of the power-law. 
Similarly, model (\ref{model}) relates also with
the double-order time-fractional diffusion equation in the Caputo sense
\cite{chechkin_etal-pre-2002,
mainardi_etal-jvc-2008,sandev_etal-pre-2015} 
by plugging into (\ref{FPmemory}) 
the memory kernel $\Phi(t)$ as defined in (\ref{memory}) 
and split according to (\ref{model}).
Furthermore, model (\ref{model}, \ref{model2}) shares with the
Weistrass random-walk for L\'evy flights 
\cite{hughes_etal-pnas-1981,klafter_etal-pt-1996}
the fact that it allows for the estimation of the anomalousness parameter
from the dynamics of the process, 
and it is not indeed plugged into the formulation. 

Beside all of this, 
we have that two families of Markovian mechanisms are enough
for generating an anomalous diffusive regime, 
in opposition to the Mittag--Leffler formalism that claims for a
large spectrum of families of Markovian mechanisms 
\cite{pagnini-pa-2014}. 
Moreover, much more unbalanced is the occurrence of each family of 
Markovian mechanisms, i.e., $p \to 0^+$, much more solid is the
intermediate interval where anomalous diffusion is generated
by a power-law memory fading. 
Actually, the approximation of 
power-law functions through exponential sums 
has been already investigated 
\cite{anderson_etal-mc-1997,bochud_etal-qf-2007,mclean-2018}.
Theoretically, this is the main result of the present paper
and, with respect to the existing literature,
we want to stress that 
the present model (\ref{model}, \ref{model2}) 
liberates research on anomalous diffusion
from the shackles of power-law or infinite mean.

In the following we show through intensive simulations that this
two Markovian hopping-trap mechanisms are 
enough for obtaining a random walk that meets the features
that define the here-called paradigmatic anomalous diffusion.

\begin{figure}
\begin{center}
\includegraphics[height=10cm,width=15cm]{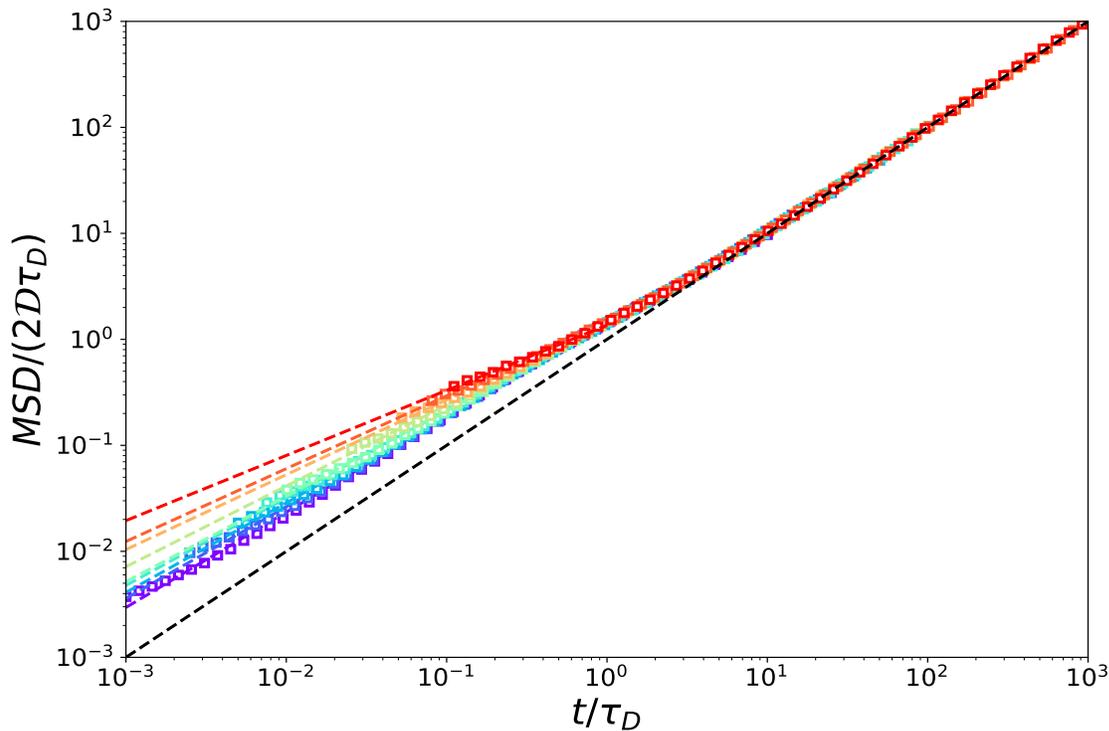}
\caption{Plots of MSD for different values of $p \in (0,1/2)$.
The plots show that at large-times $t > \tauD$ 
the model (\ref{model}, \ref{model2}) diffuses according
to standard diffusion while at intermediate-times $\tauB < t < \tauD$
diffuses anomalously with parameter $\beta$ dependent on
$p \in (0,1/2)$. 
The values of $p$ adopted for this figure are
$p = 0.001 \,, 0.0025 \,, 0.005 \,, 0.0075 \,, 0.01 \,, 0.025 \,,
0.05 \,, 0.075 \,, 0.1$.
Coloured dashed-lines correspond to the estimation of the 
anomalousness parameter $\beta$ as plotted in figure \ref{fig:beta}
and fitted by formula (\ref{beta}).
}
\label{fig:MSD}
\end{center}
\end{figure}

\section{Simulations}
\label{sec:simulations}
In spite of its simple definition,
we computed no further analytical result, yet, and we present
here an intensive study based on simulations. 
Model (\ref{model}, \ref{model2}) is investigated with respect to its 
regime-transitions with focus on the intermediate regime.
Such intermediate regime can be long enough to characterize the
process.
A similar study on L\'evy flights showed that,
in some special settings, 
an intermediate regime could extend so much 
to making the large-time limit unattainable 
from measurements in real systems \cite{pagnini_etal-fcaa-2021}. 

The setting of simulations is the following:
the number of independent realizations is $\max\{\omega \in \Omega\}=10^4$,
the two dimensional parameters are $\tauB=1$ and $\sigma=1$,
and different values of $p \in (0,1/2)$. 

\begin{figure}
\begin{center}
\subfloat[][]{
\includegraphics[height=5cm,width=7cm]
{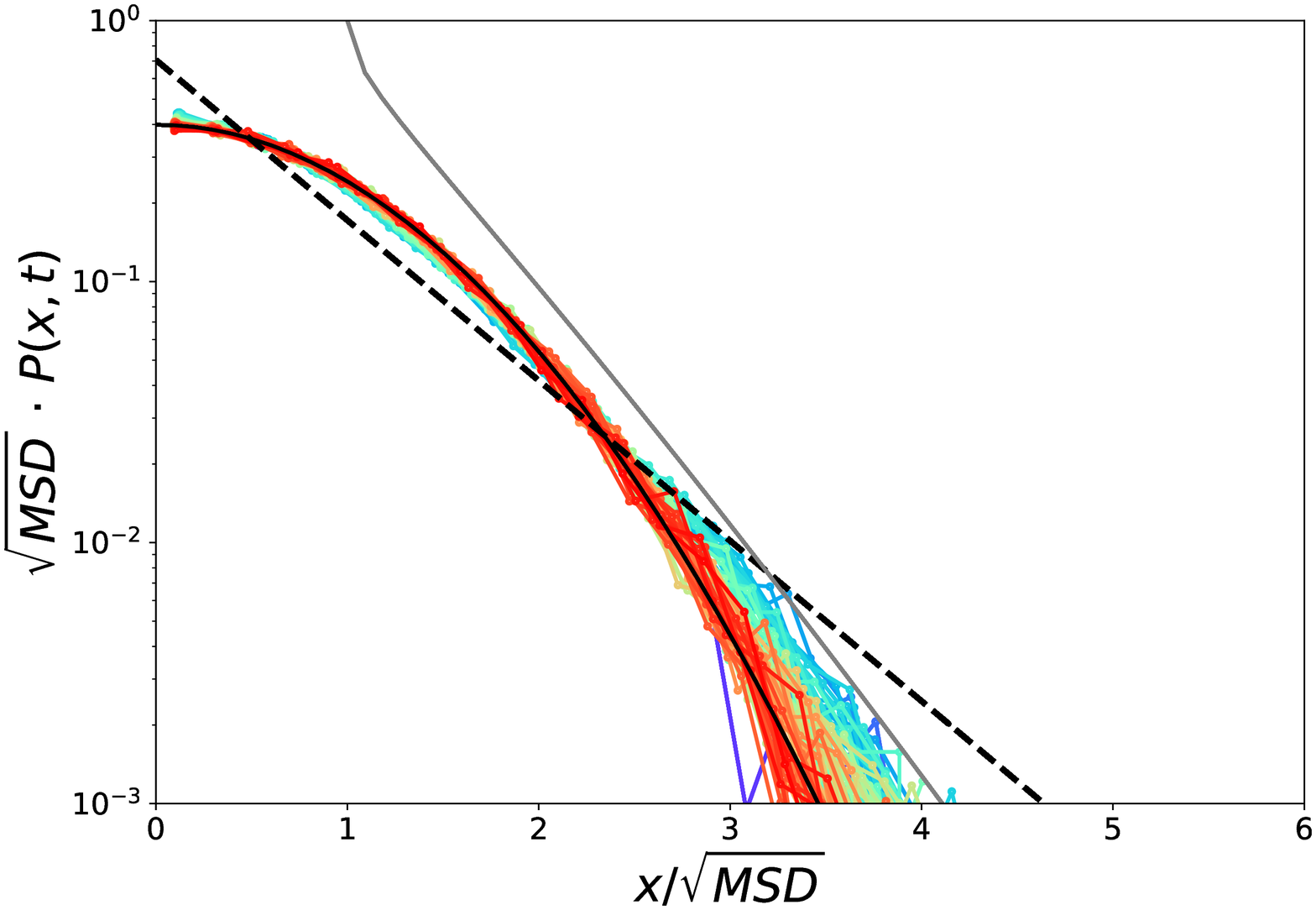}}
\subfloat[][]{
\includegraphics[height=5cm,width=7cm]
{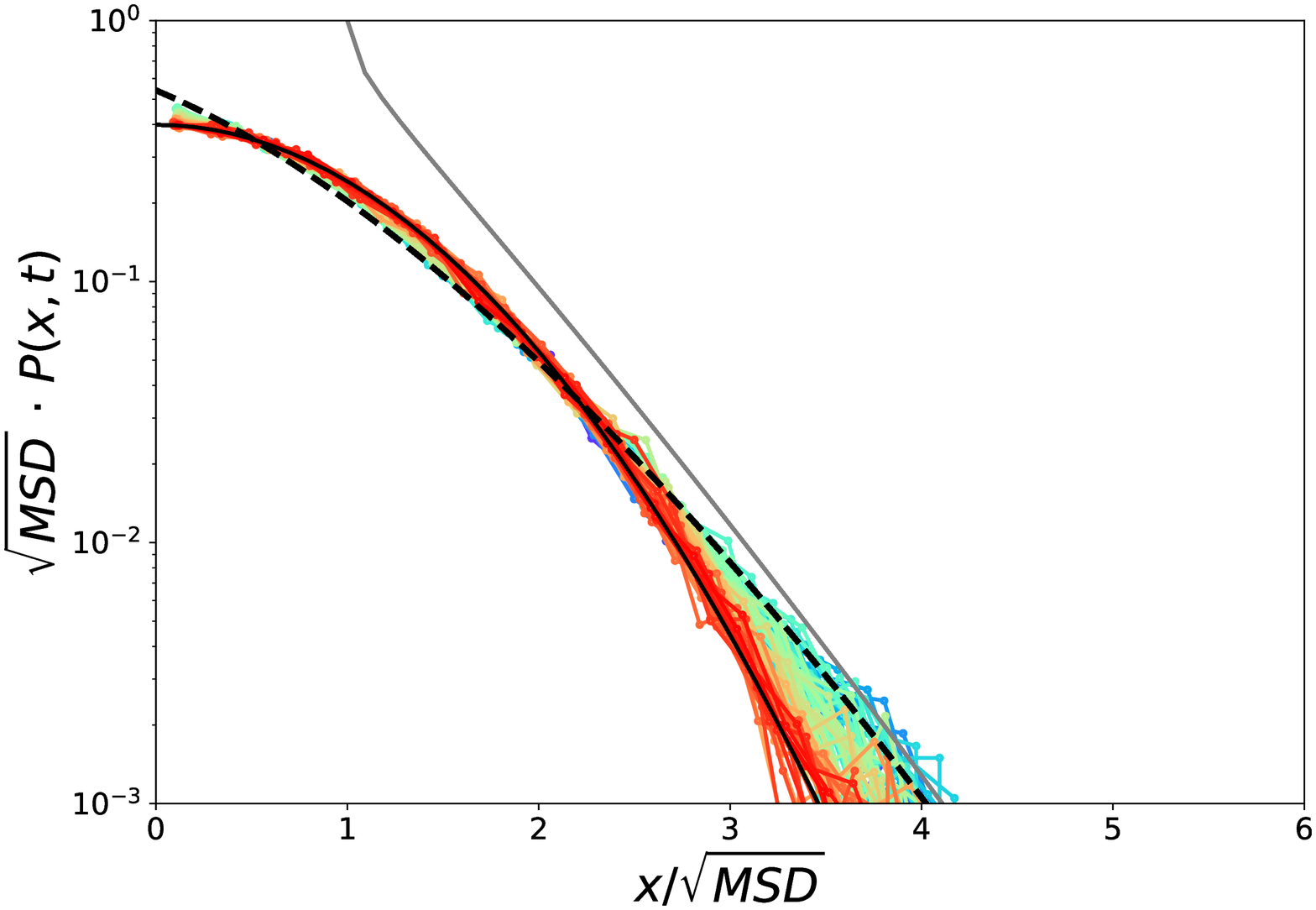}}
\\
\subfloat[][]{
\includegraphics[height=5cm,width=7cm]
{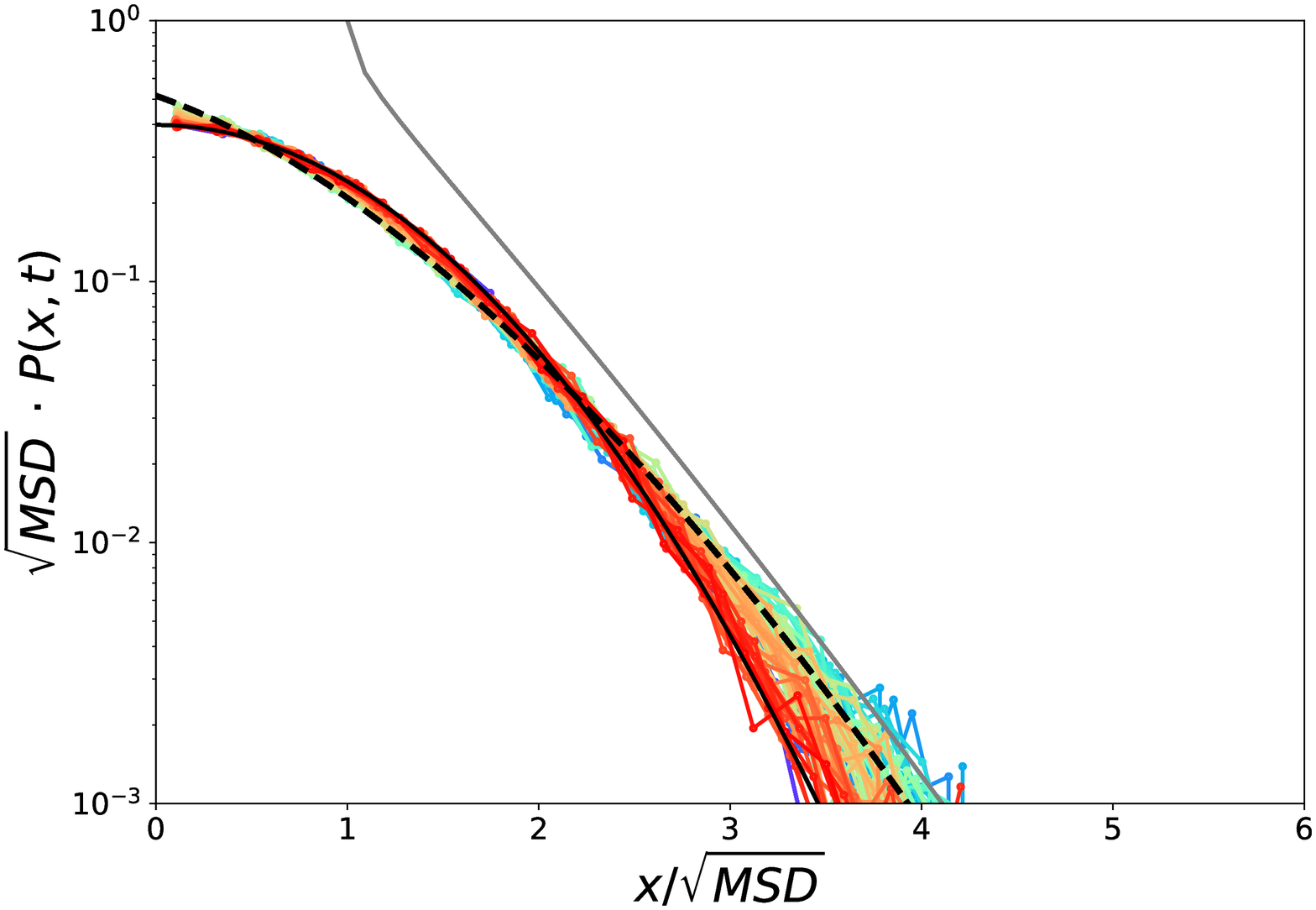}}
\subfloat[][]{
\includegraphics[height=5cm,width=7cm]
{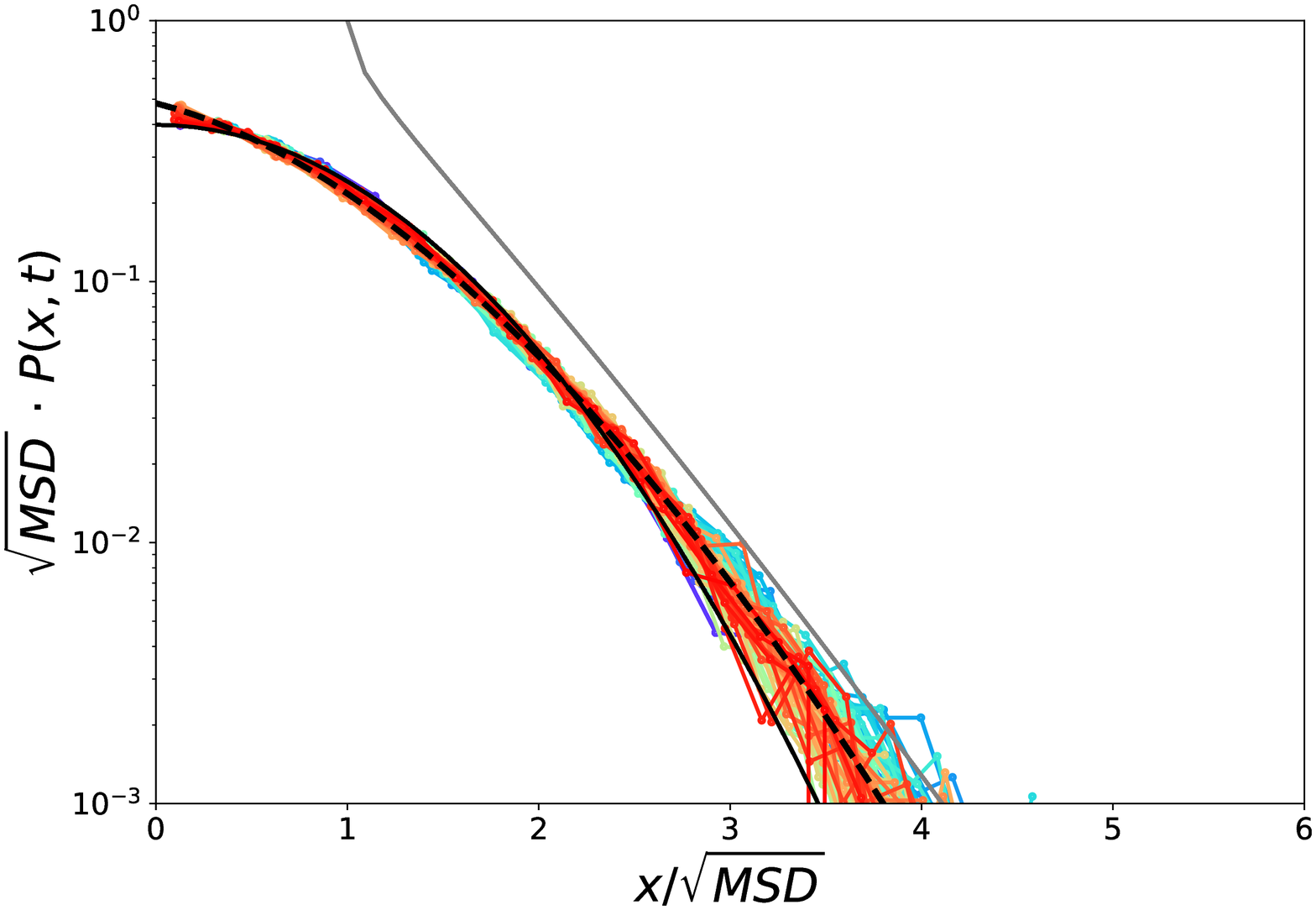}}
\\
\subfloat[][]{
\includegraphics[height=5cm,width=7cm]
{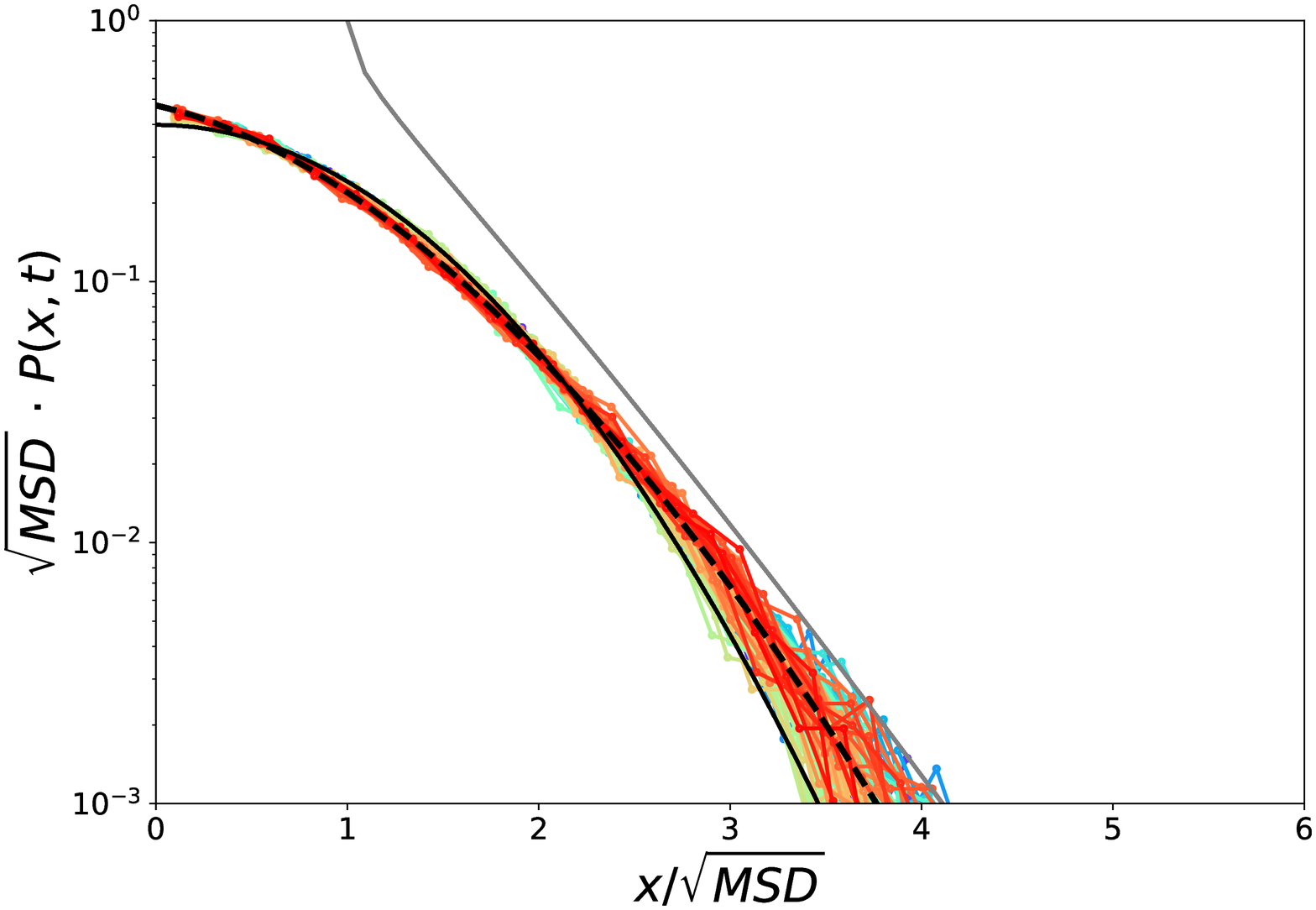}}
\subfloat[][]{
\includegraphics[height=5cm,width=7cm]
{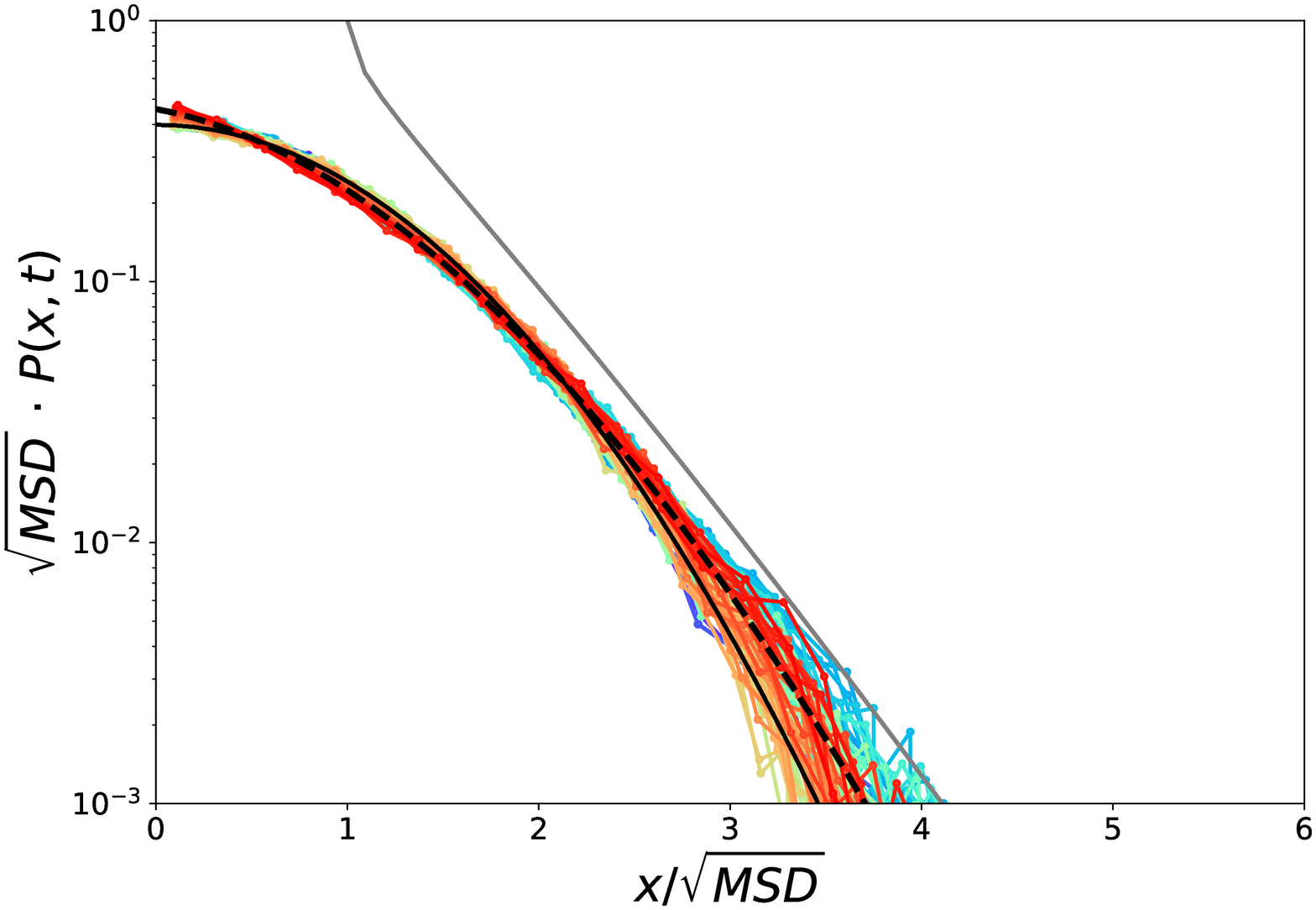}}
\caption{
Plots of walker's PDF at different elapsed-times for different
values of $p \in (0,1/2]$,
from (a) to (f): $p=0.5 \,, 0.1 \,, 0.05 \,, 0.01 \,, 0.005 \,, 0.001$
that is $\beta=0 \,, 0.6872 \,, 0.7465 \,, 0.8213 \,, 0.8411 \,, 0.8735$. 
The PDF goes through two regime-transitions from
short-time $t < \tauB$ (violet), 
to intermediate-time $\tauB < t < \tauD$ (green), and to large-time 
$t > \tauD$ (red).
The reference lines for PDF tails corresponds 
to the exponential (gray solid-line), 
to the stretched-exponential (black dashed-line),  
and to the Gaussian (black solid-line).
The stretched-exponential PDF
follows the time-fractional diffusion law (\ref{TFDE}).
It is important to observe that when $p=0.5$ the tails of the PDF 
transit from exponential to Gaussian by skipping the stretched-exponential
that is in agreement with the Barkai--Burov theory \cite{barkai_etal-prl-2020}.
}
\label{fig:PDF}
\end{center}
\end{figure}

We report the outputs of model (\ref{model}, \ref{model2}) 
regarding the observables proper of
the paradigmatic anomalous diffusion presented in section \ref{sec:paradigmatic}.
We start this numerical study with ensemble statistics.
The MSD is shown in figure \ref{fig:MSD}
and we observe that the MSD
displays an anomalous sub-linear regime with exponent
$\beta$ when $\tauB < t < \tauD$
and Brownian linear diffusion when $t > \tauD$. 
For different values of $p$ different values
of $\beta$ follow according to formula (\ref{beta}). 
Furthermore, consistently with the paradigmatic anomalous 
diffusion discussed in section \ref{sec:paradigmatic}, 
the walker's PDF of model (\ref{model}, \ref{model2}) 
goes through regime-transitions. 
In particular, in figure \ref{fig:PDF} it is shown
that when the walker's PDF is re-scaled with the corresponding variance,
we can distinguish three different tail-behaviours:
exponential ($t < \tauB$), 
stretched-exponential ($\tauB < t < \tauD$)
and Gaussian ($t > \tauD$). 
In particular, 
we can see that the stretched-exponential PDF follows
the time-fractional diffusion law (\ref{TFDE}) 
and the anomalousness parameter $\beta$ is dependent on $p$ according to
formula (\ref{beta}). 
Moreover, we observe that when $p=1/2$ the walker's PDF transits
from the exponential to the Gaussian by skipping the stretched-exponential,
that is in agreement with the Barkai--Burov theory \cite{barkai_etal-prl-2020}.
This is consistent also with the previous analysis where, 
from formula (\ref{beta}), we have that $\beta=0$ when $p=1/2$ 
and then the duration of the 
anomalous diffusion interval is $\tauD-\tauB=0$ 
according to formula (\ref{ADinterval}).

\begin{figure}
\begin{center}
\subfloat[][]{
\includegraphics[height=5cm,width=7cm]{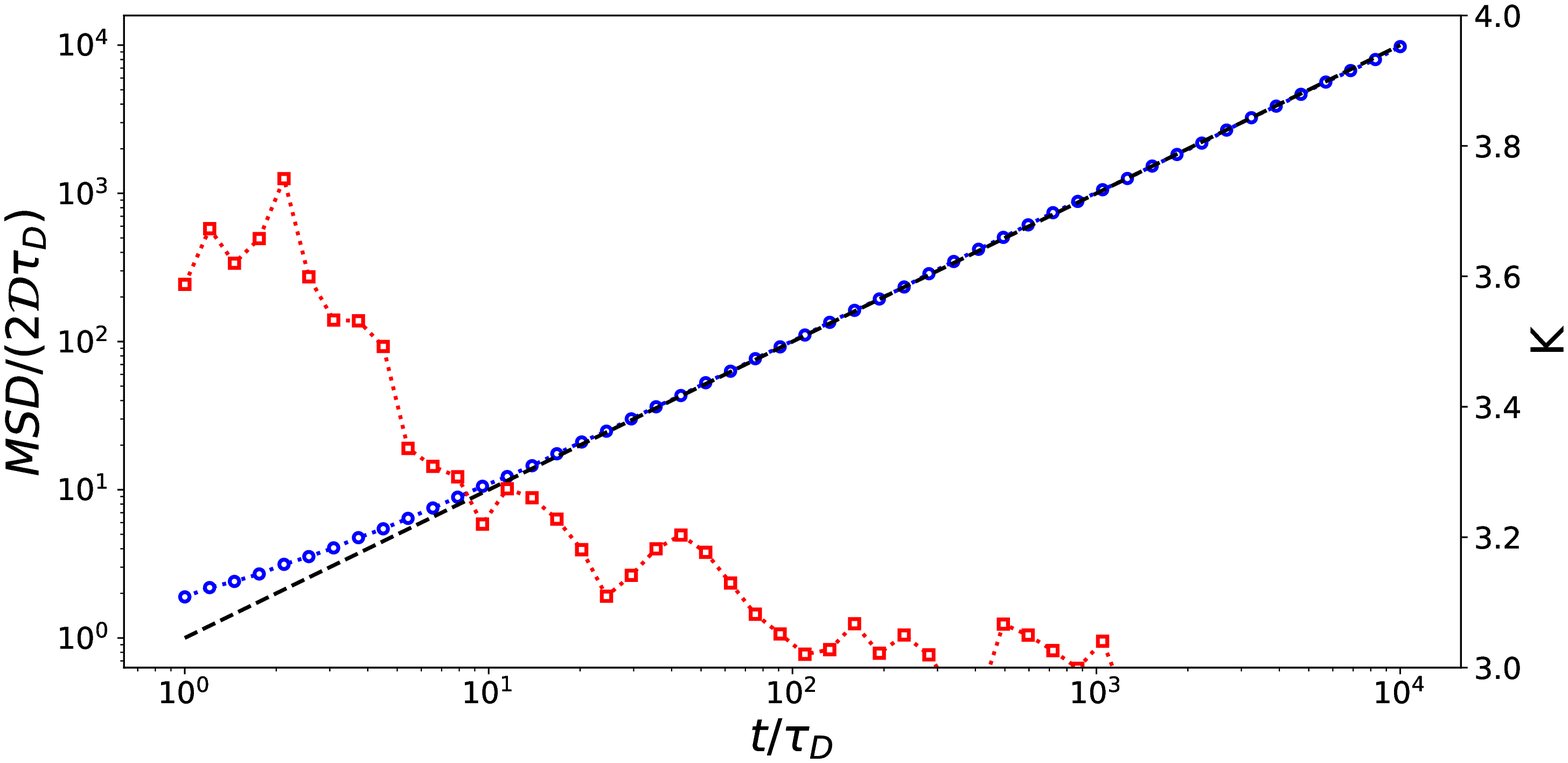}} 
\subfloat[][]{
\includegraphics[height=5cm,width=7cm]{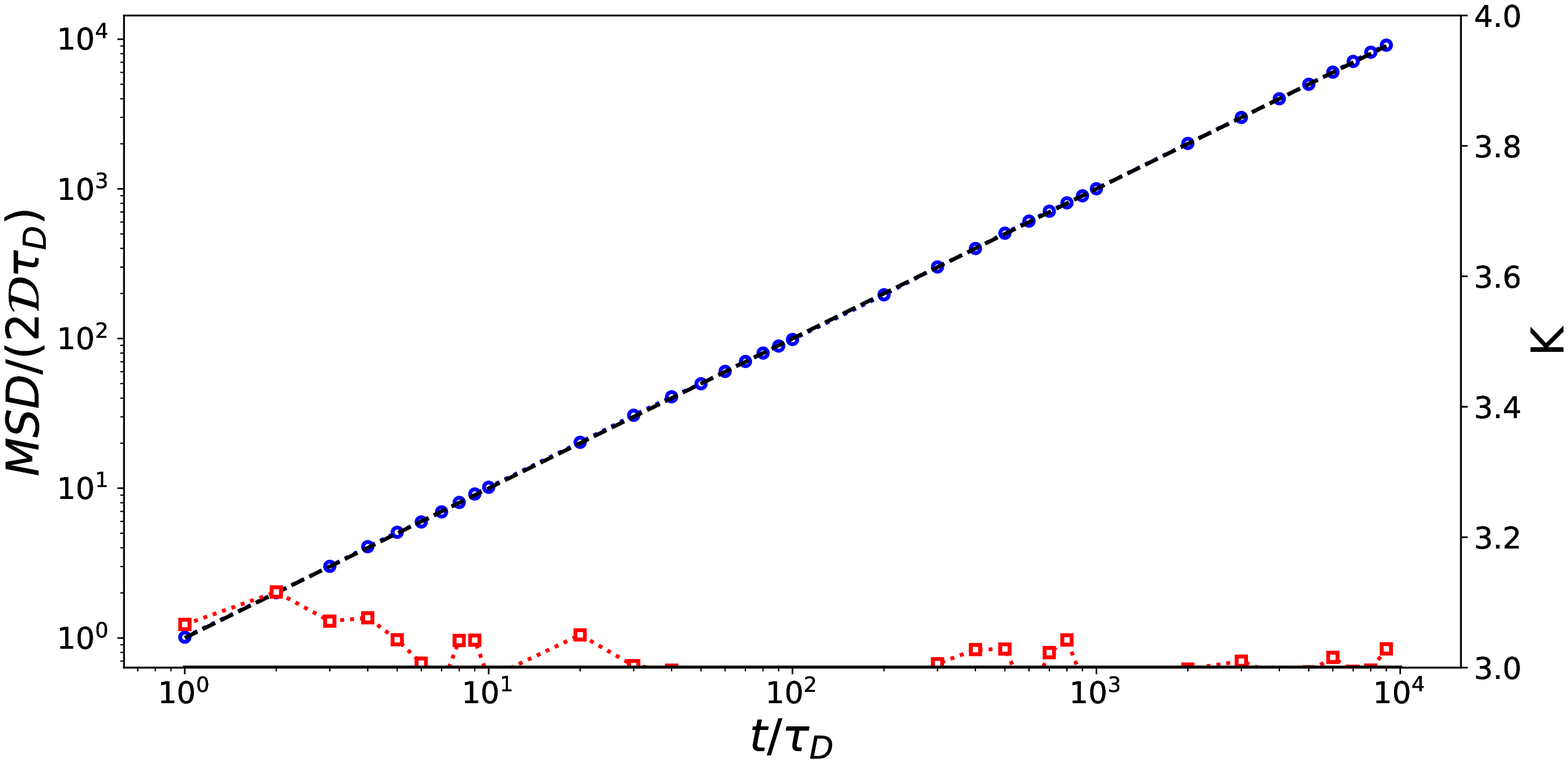}}
\caption{
Comparison of the starting of the 
Brownian linear scaling of the MSD (blue) against the kurtosis (red), 
this last expresses through its value $K=3$ the Gaussianity of the PDF. 
In panel (a) it is reported the behaviour of the 
basic Markovian CTRW ($p=1/2$) and in panel (b) that of 
the over-damped Langevin equation.
An intrinsic BynG delay of one decade is displayed by an hopping-trap mechanism
with respect to a continuous stochastic process.
}
\label{fig:BynG-Wiener}
\end{center}
\end{figure}

We end the analysis of the ensemble statistics by discussing the occurrence
of the BynG interval.
In particular we plot in the same figures the MSD and the kurtosis 
to compare the starting of the Brownian linear scaling of the MSD against 
the Gaussianity of the PDF, 
this last is expressed through its kurtusis value $K=3$.
First of all we observe that the hopping-trap mechanism 
intrinsically displays a delay in the spirit of the BynG with respect
to continuous processes. This is shown in figure \ref{fig:BynG-Wiener}
where the basic Markovian CTRW model ($p=1/2$) 
displays a one-decade of delay while 
this delay is indeed not displayed at all by the over-damped
Langevin equation, i.e., 
$\rmd Y^\omega_t=\sqrt{2 \mD} \, \rmd W^\omega_t$, with $\omega \in \Omega$,
where $Y_t^\omega:[0,+\infty)\times \Omega \to \R$
and $\rmd W_t^\omega$ is the delta-correlated Wiener process
with variance $\E[(\rmd W_t)^2]=\rmd t$.

\begin{figure}
\begin{center}
\subfloat[][]{
\includegraphics[height=5cm,width=7cm]{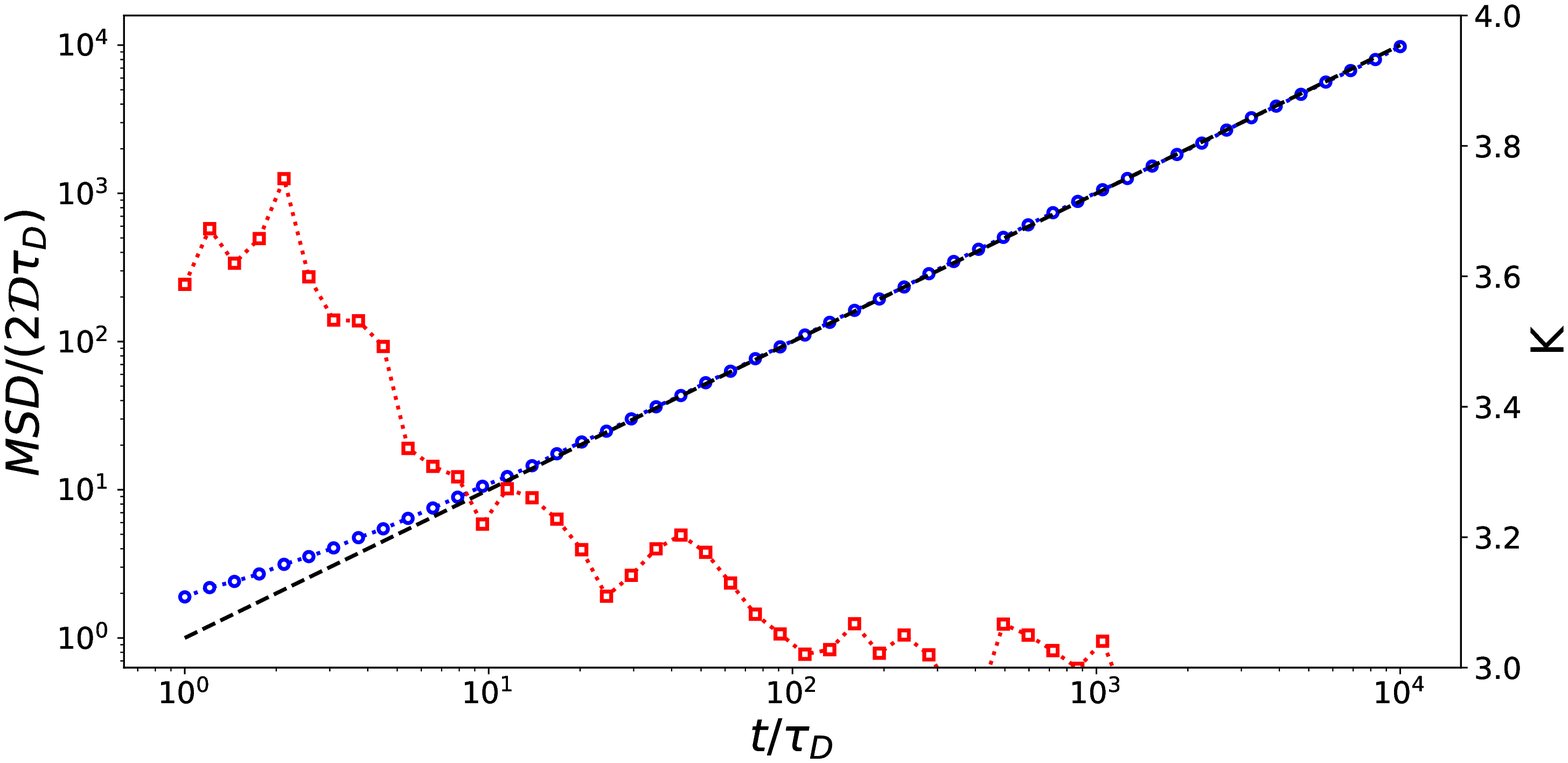}}
\subfloat[][]{
\includegraphics[height=5cm,width=7cm]{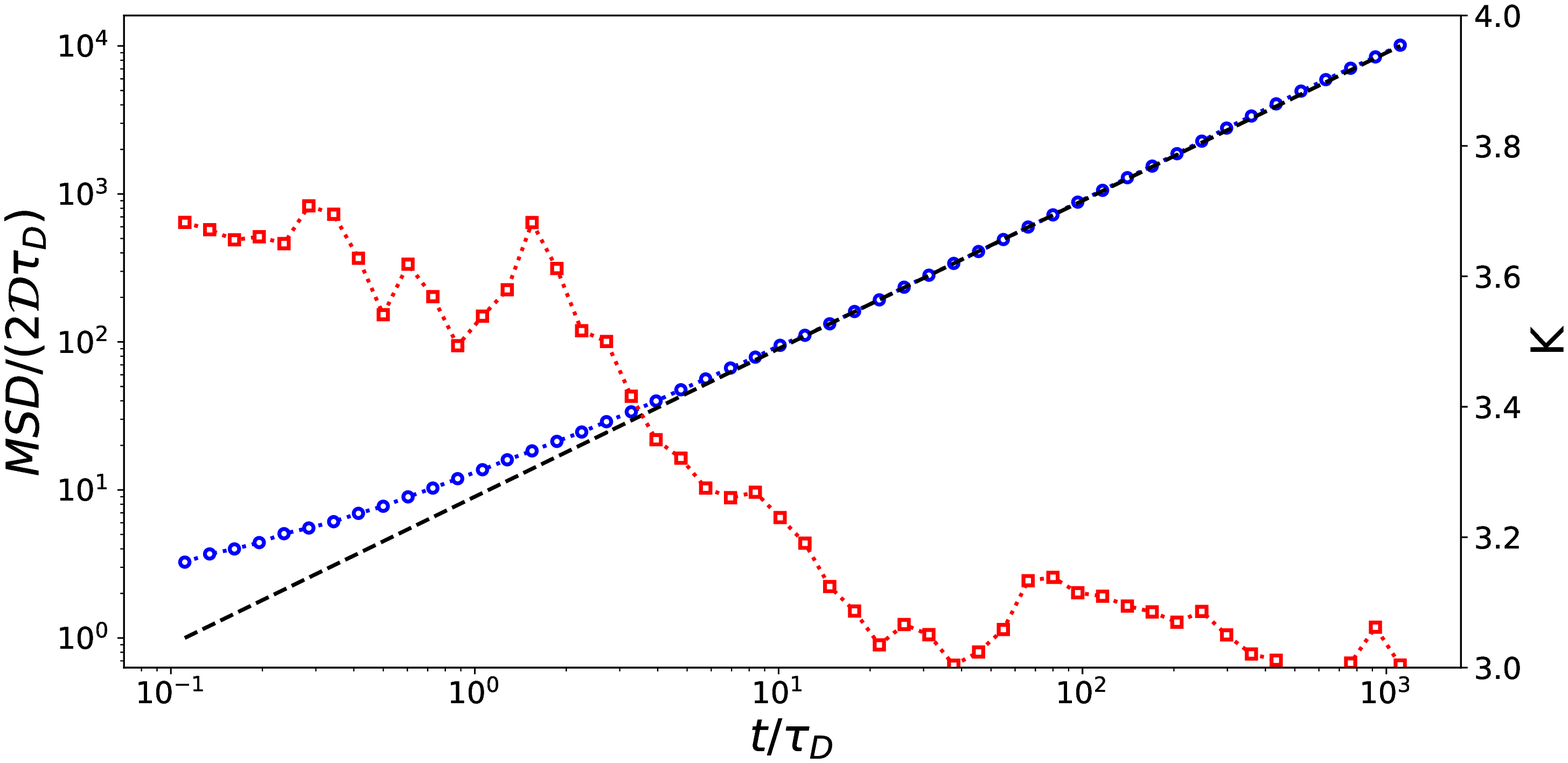}}\\
\subfloat[][]{
\includegraphics[height=5cm,width=7cm]{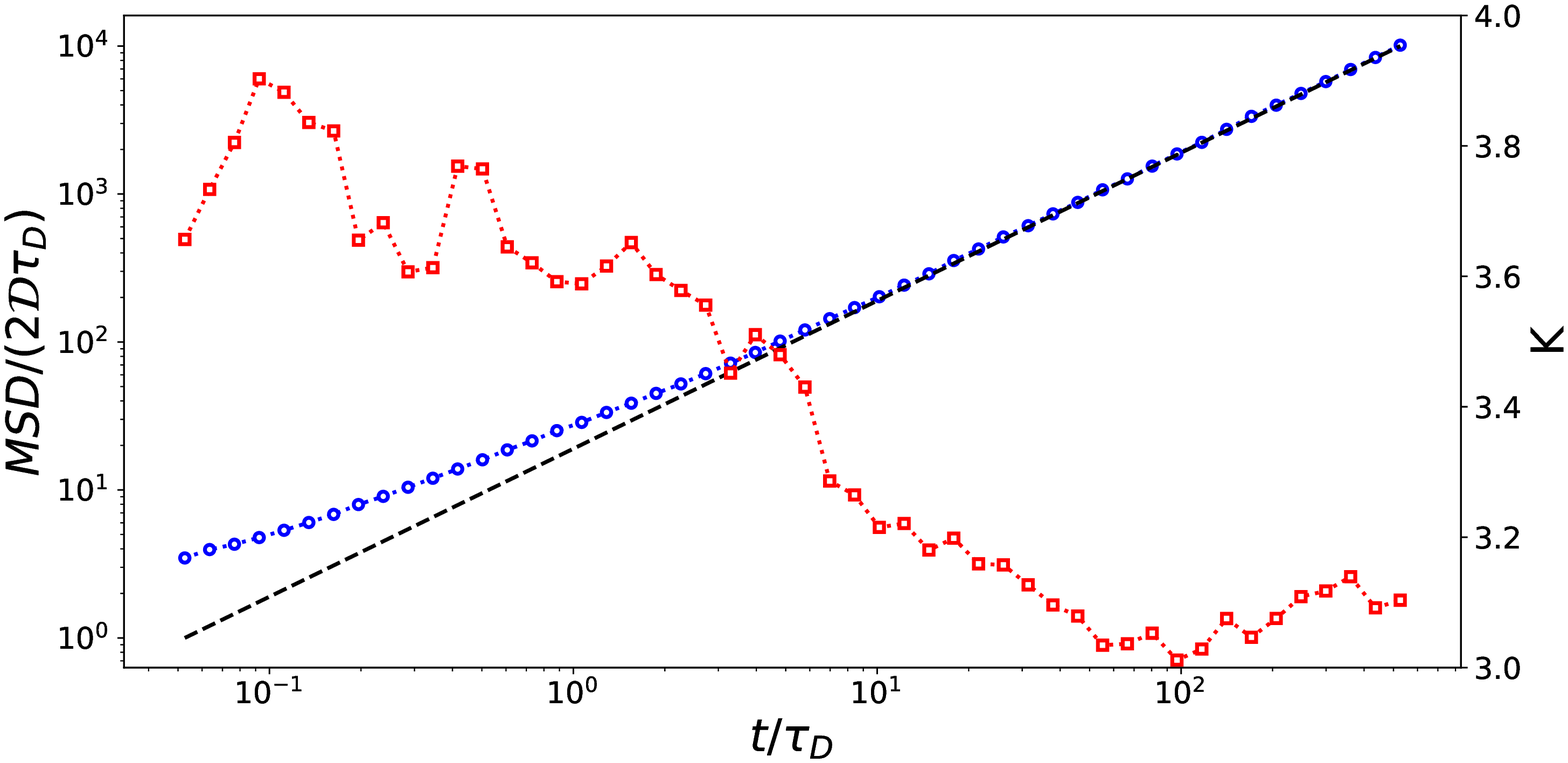}}
\subfloat[][]{
\includegraphics[height=5cm,width=7cm]{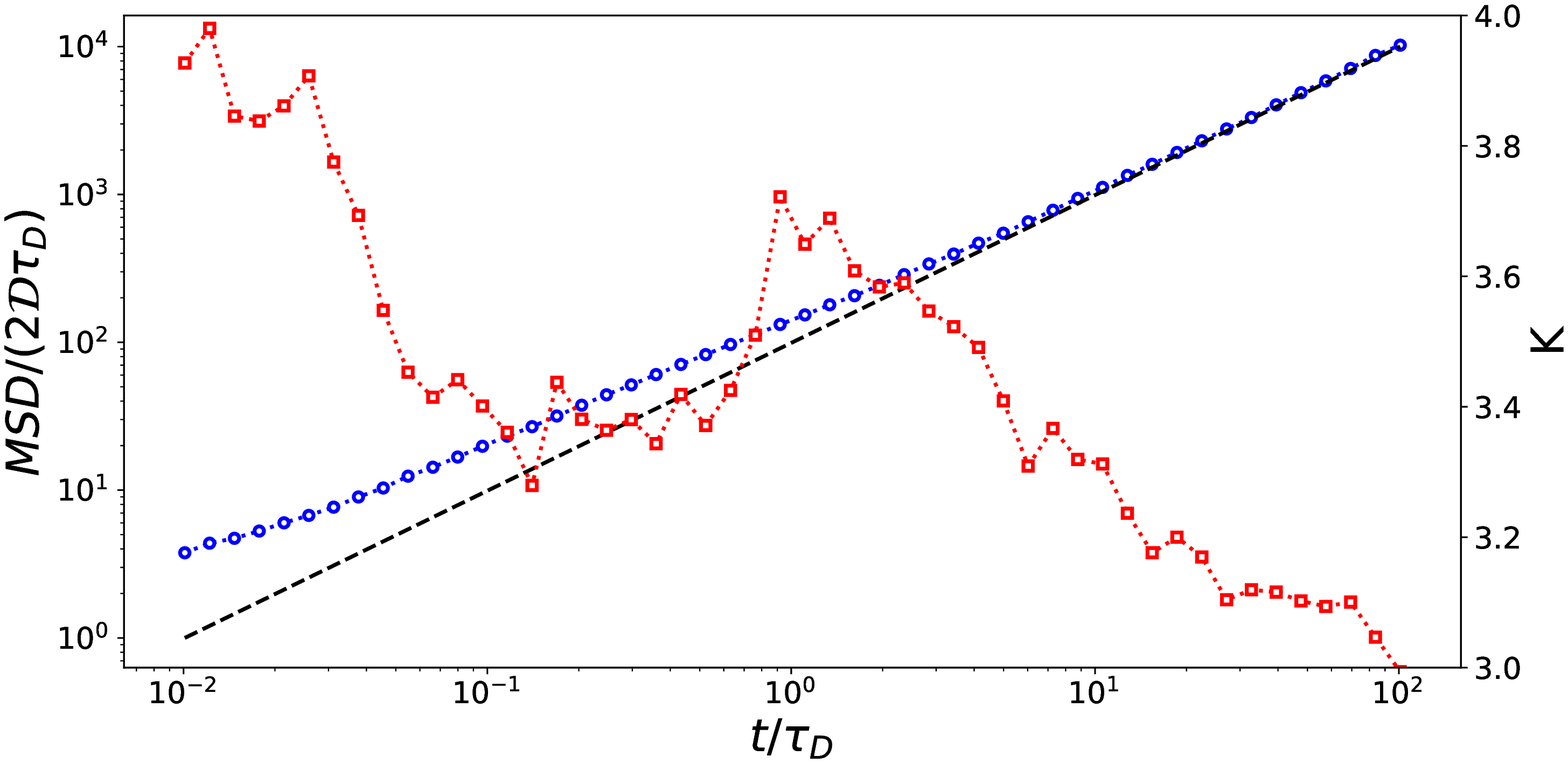}}\\
\subfloat[][]{
\includegraphics[height=5cm,width=7cm]{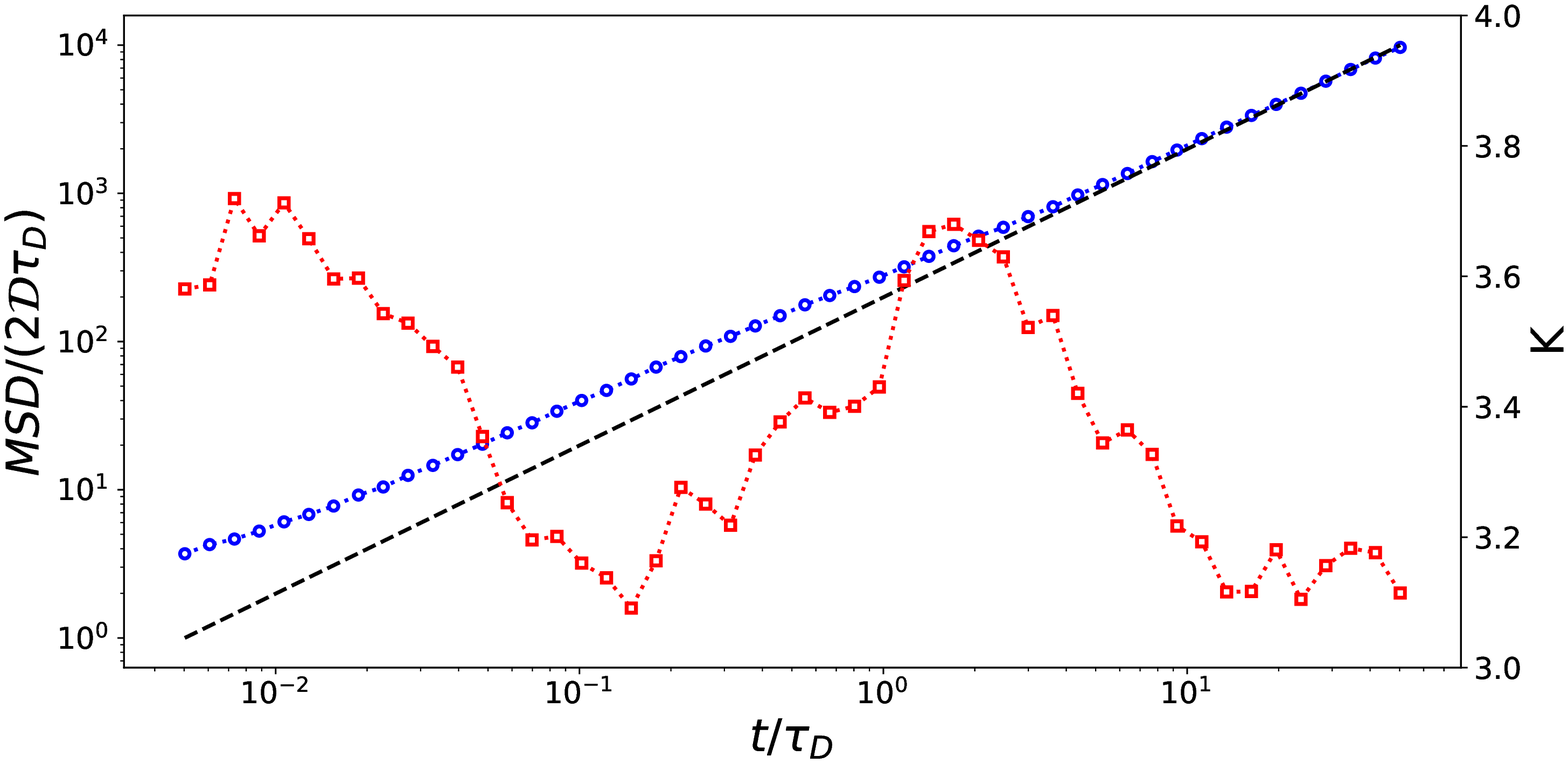}}
\subfloat[][]{
\includegraphics[height=5cm,width=7cm]{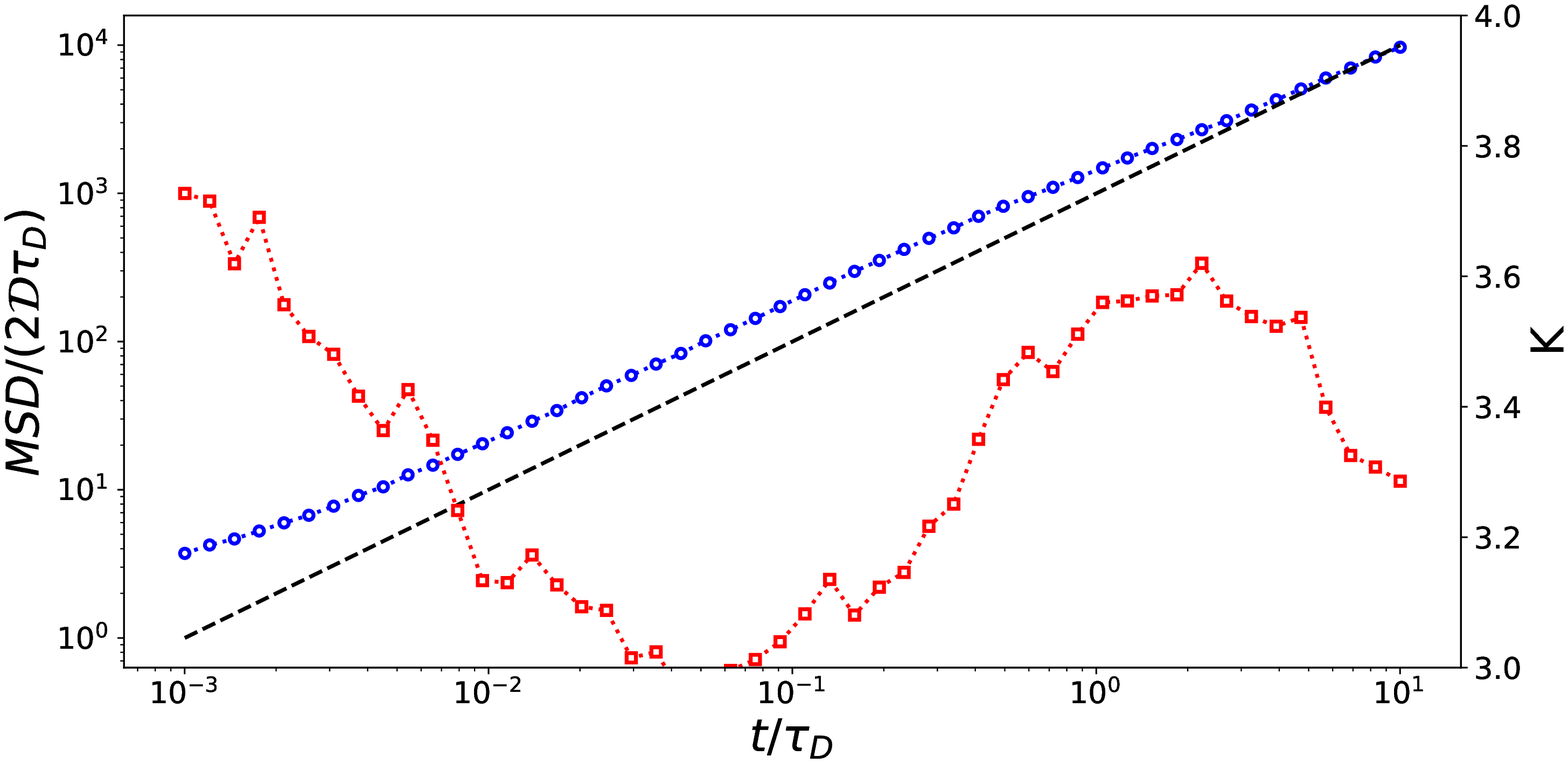}}
\caption{
The same as in figure \ref{fig:BynG-Wiener} for model (\ref{model}, \ref{model2})
only with different values of $p\in(0,1/2]$, from (a) to (f): 
$p=0.5 \,, 0.1 \,, 0.05 \,, 0.01 \,, 0.005 \,, 0.001$.
Panel (a) shows the basic Markovian CTRW model ($p=1/2$) 
as in panel (a) of figure \ref{fig:BynG-Wiener},
it is shown again for convenience for comparison 
with model (\ref{model}, \ref{model2}).
The oscillating behaviour of the kurtosis reflects the
action of two co-existing Markovian mechanisms 
acting on the walker with different statistical frequency (\ref{model}).
The intrinsic one-decade delay of an hopping-trap mechanism displayed
in panel (a) increases for decreasing values of $p$ up to an extention
of two decades. 
}
\label{fig:BynG}
\end{center}
\end{figure}

The BynG interval for model (\ref{model}, \ref{model2}) is
shown in figure (\ref{fig:BynG}). 
Qualitatively, we observe that the kurtosis has an oscillating behaviour,
as a consequence of the action of two co-existing Markovian mechanisms 
acting on the walker with different statistical frequency (\ref{model}).
Quantitatively, we observe that 
the intrinsic one-decade delay of an hopping-trap mechanism,
see panel (a) of both figures \ref{fig:BynG-Wiener} and \ref{fig:BynG}, 
increases for decreasing values of $p$ up to an extention of two decades
of the relaxation time for the diffusive limit $\tauD$. 
This behaviour can be compared against
the behaviour of the minimal DD model displayed 
in \cite[figure 3]{chechkin_etal-prx-2017}.
In particular, 
the duration of the BynG interval in the minimal DD model 
is two decades with respect to the relaxation 
time of the stochastic diffusion coefficient.
Therefore, the BynG interval of the present model (\ref{model}, \ref{model2})
is comparable with that of the prototypical DD model \cite{chechkin_etal-prx-2017}.

\begin{figure}
\begin{center}
(a)
\includegraphics[height=4cm,width=7cm]{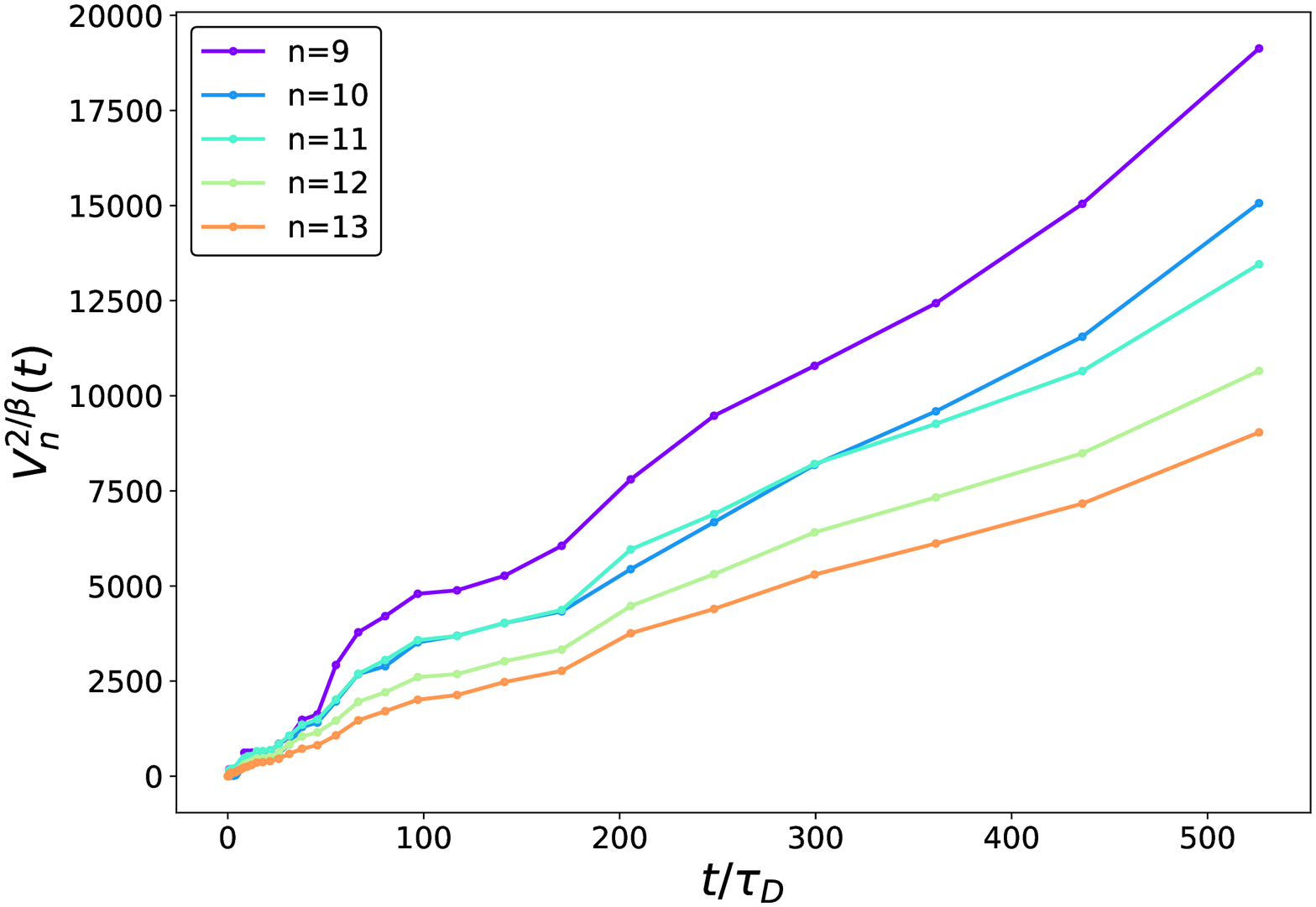}
\includegraphics[height=4cm,width=7cm]{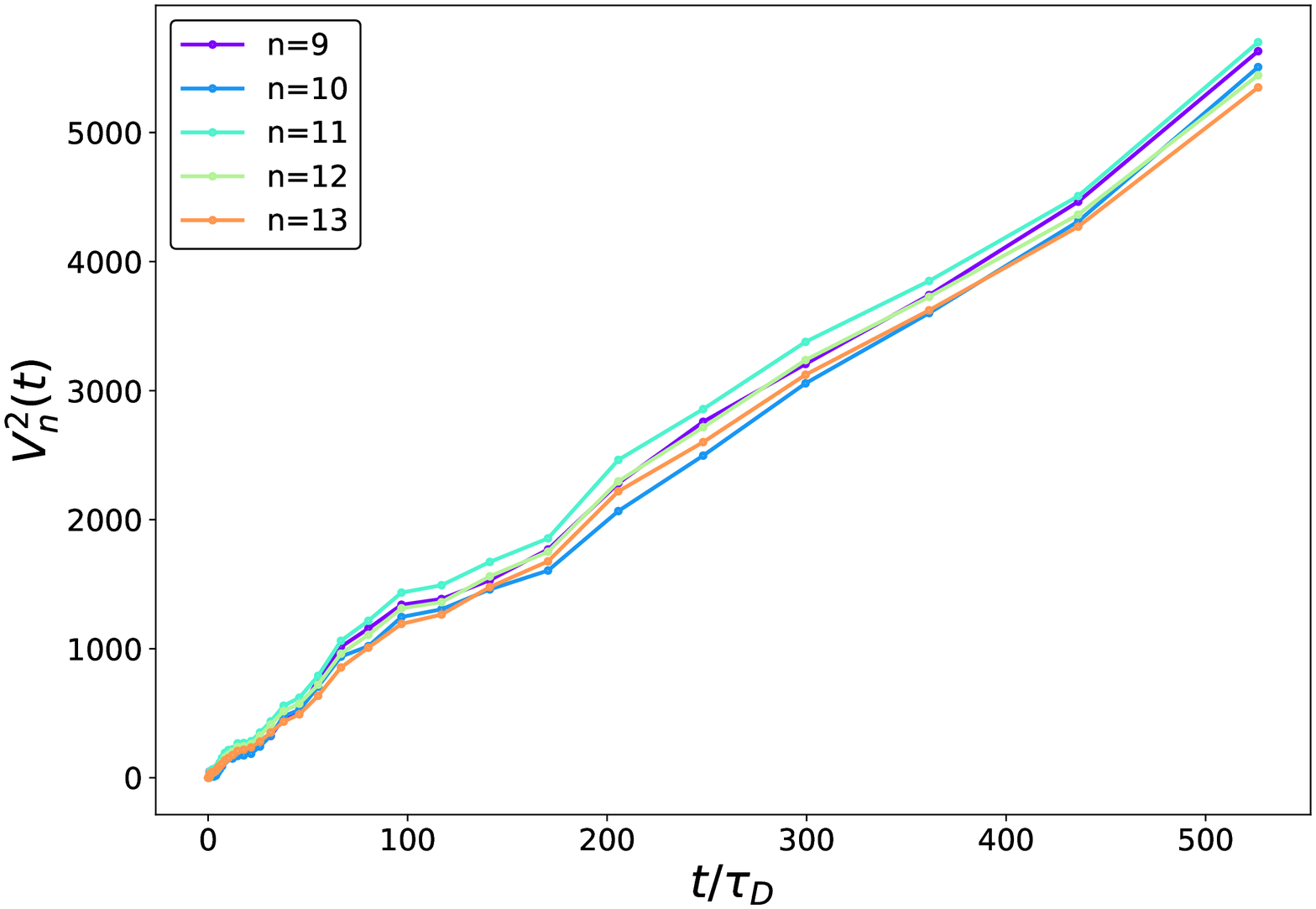}
\\
(b)
\includegraphics[height=4cm,width=7cm]{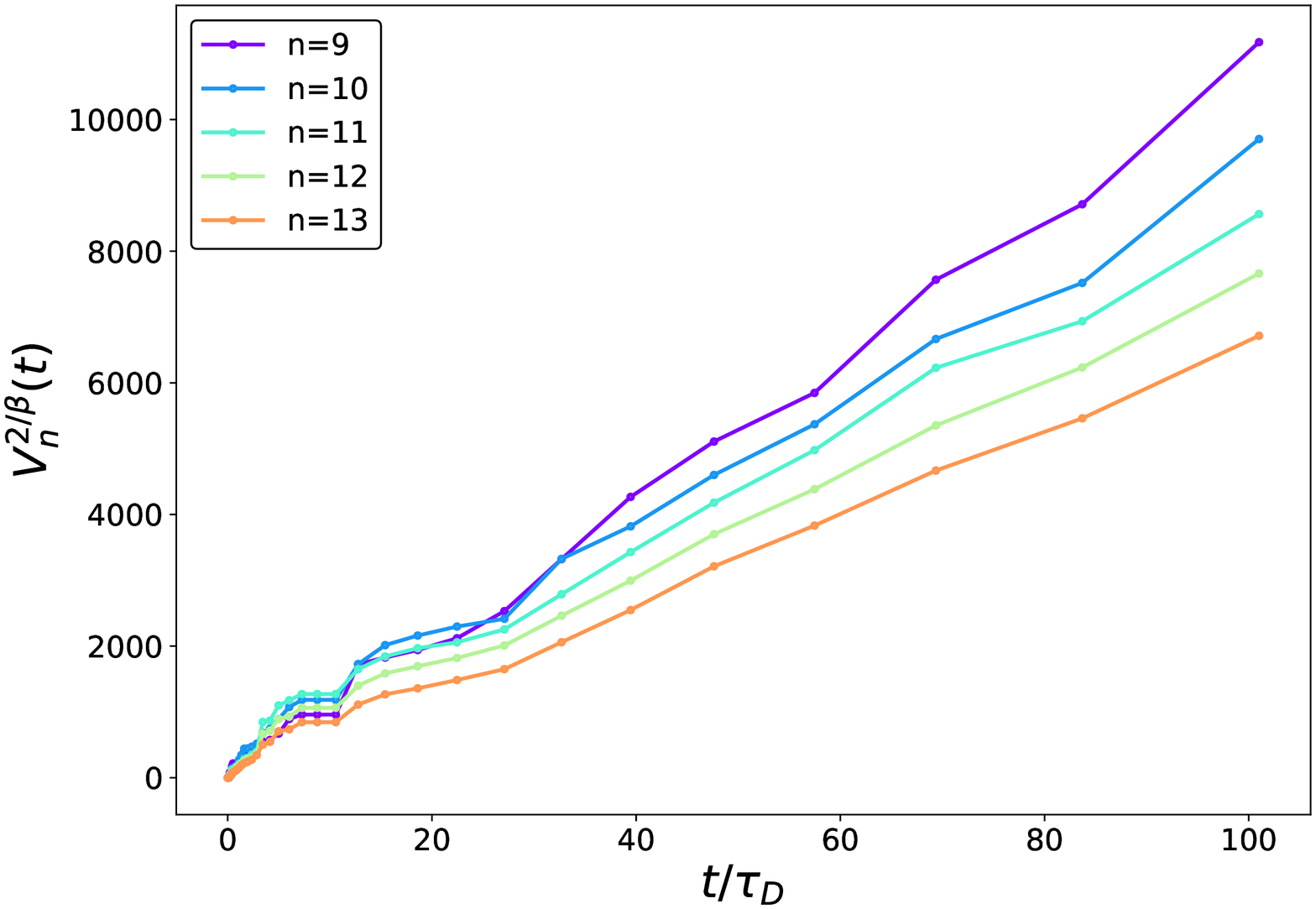}
\includegraphics[height=4cm,width=7cm]{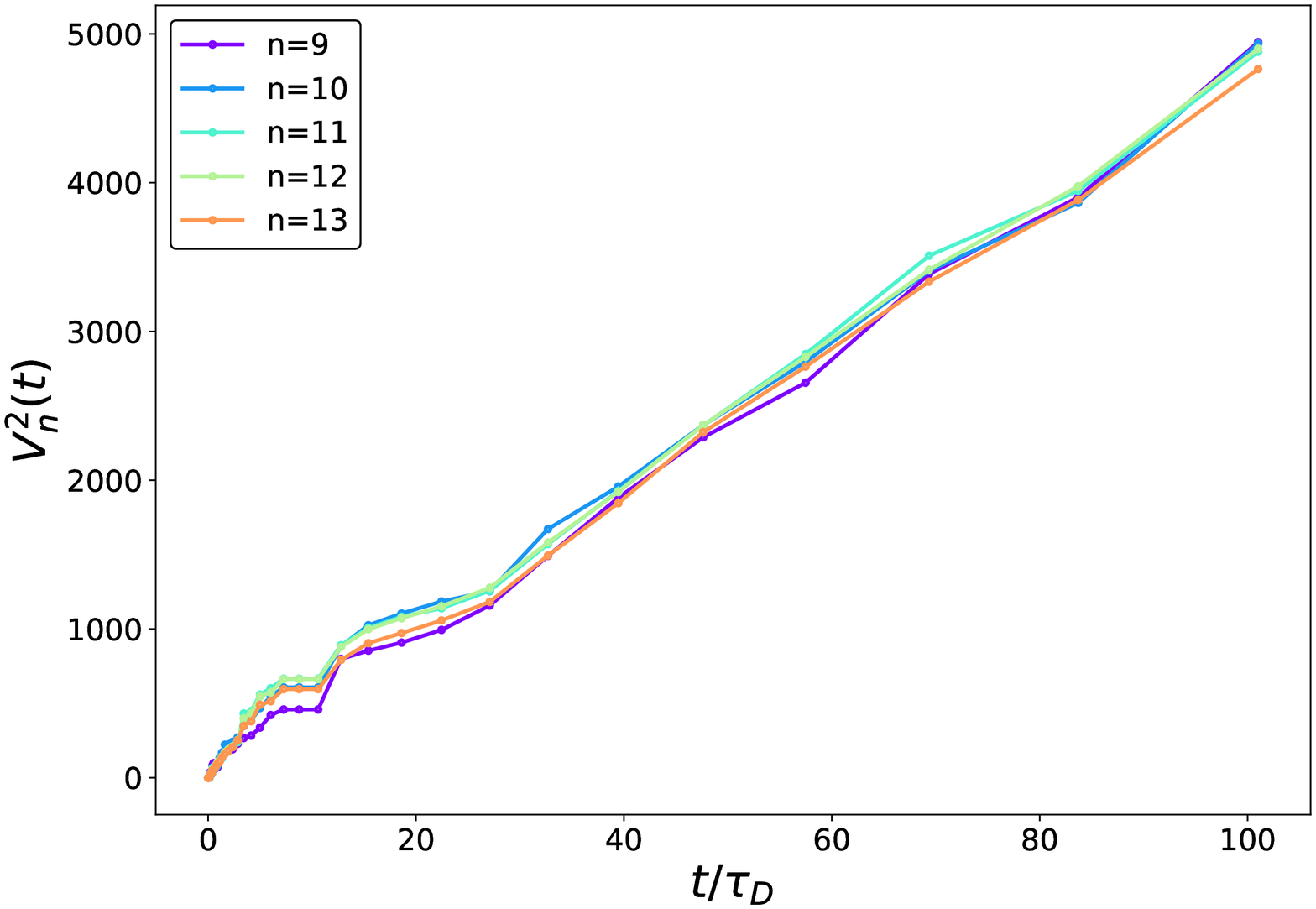}
\\
(c)
\includegraphics[height=4cm,width=7cm]{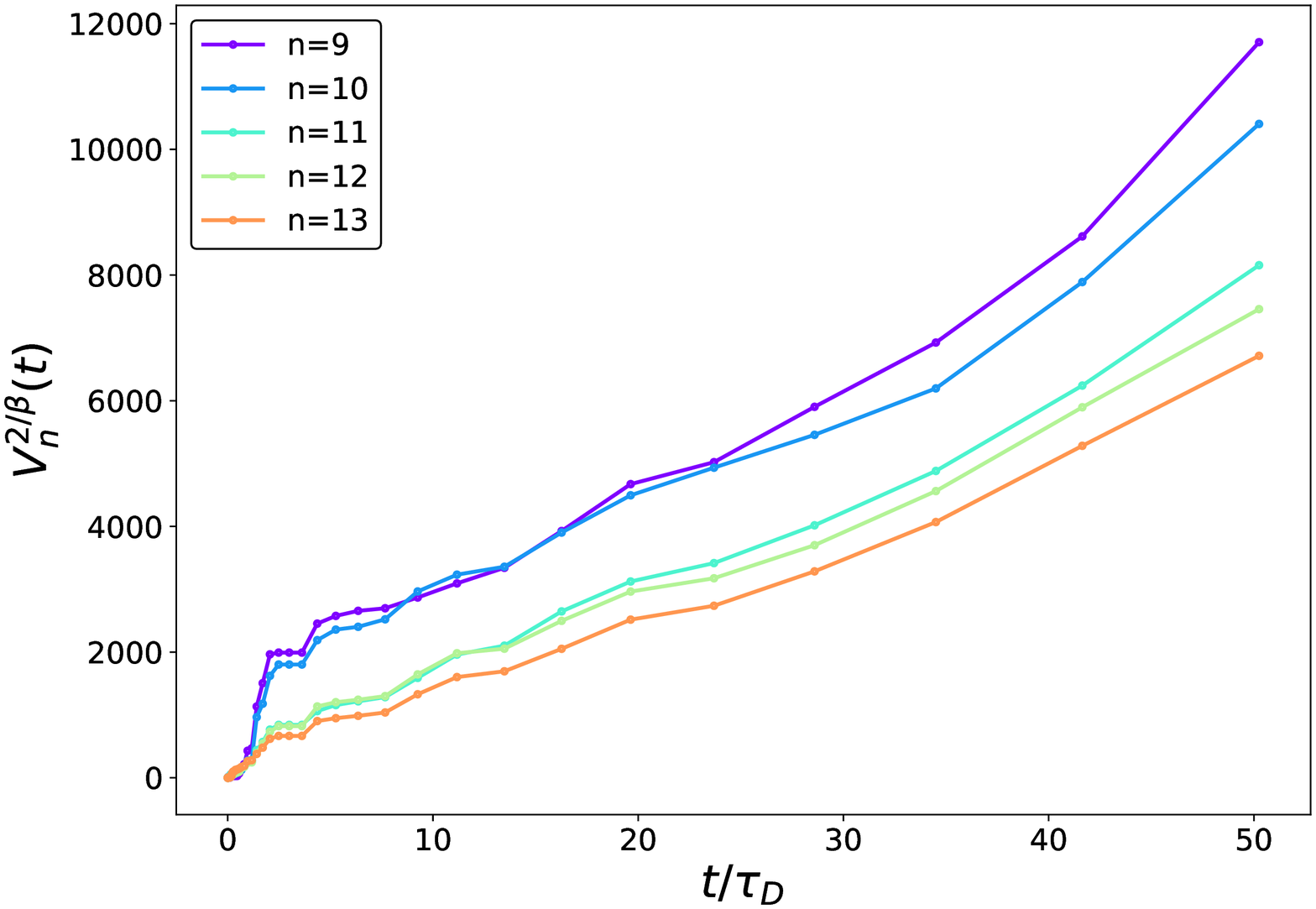}
\includegraphics[height=4cm,width=7cm]{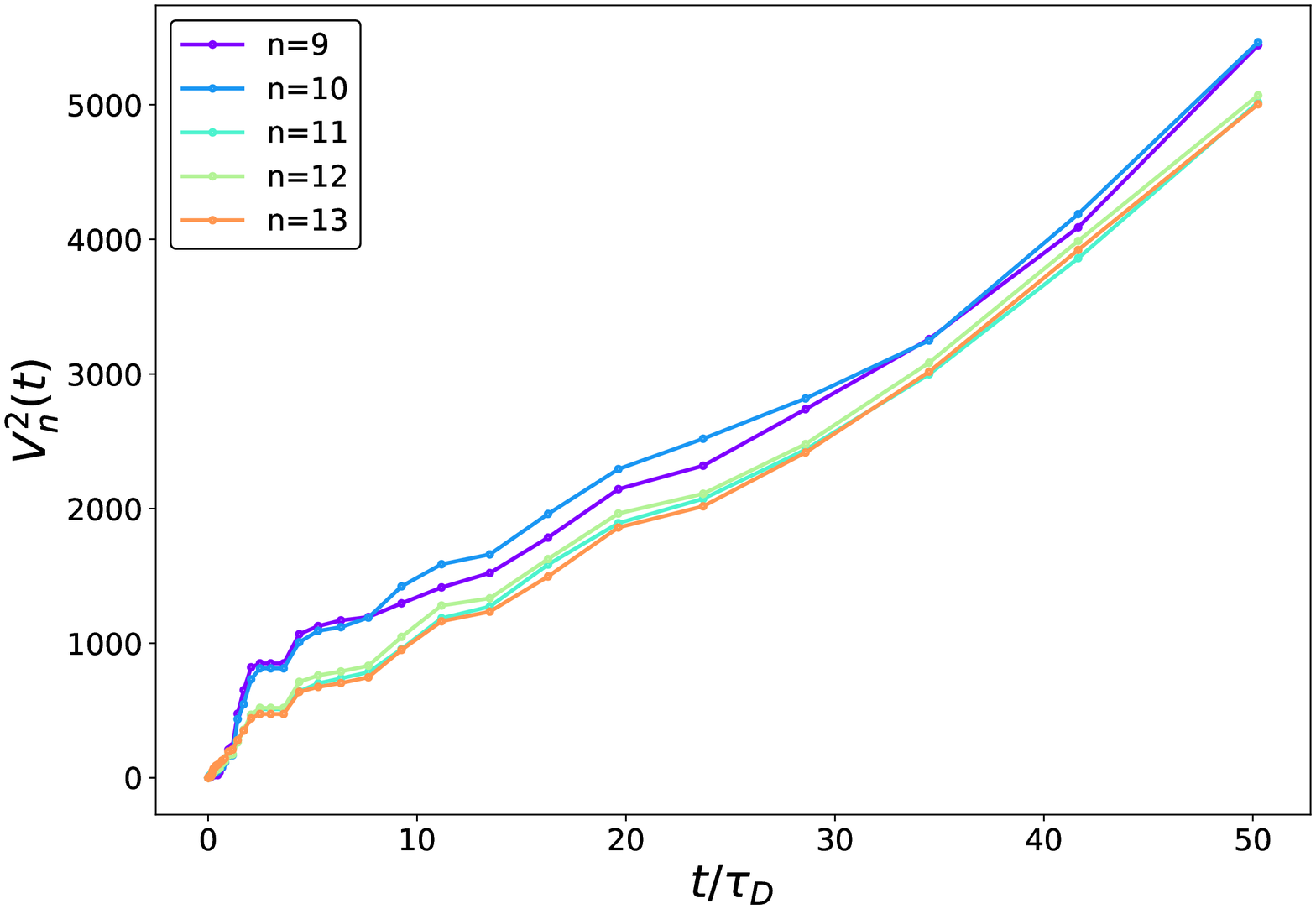}
\\
(d)
\includegraphics[height=4cm,width=7cm]{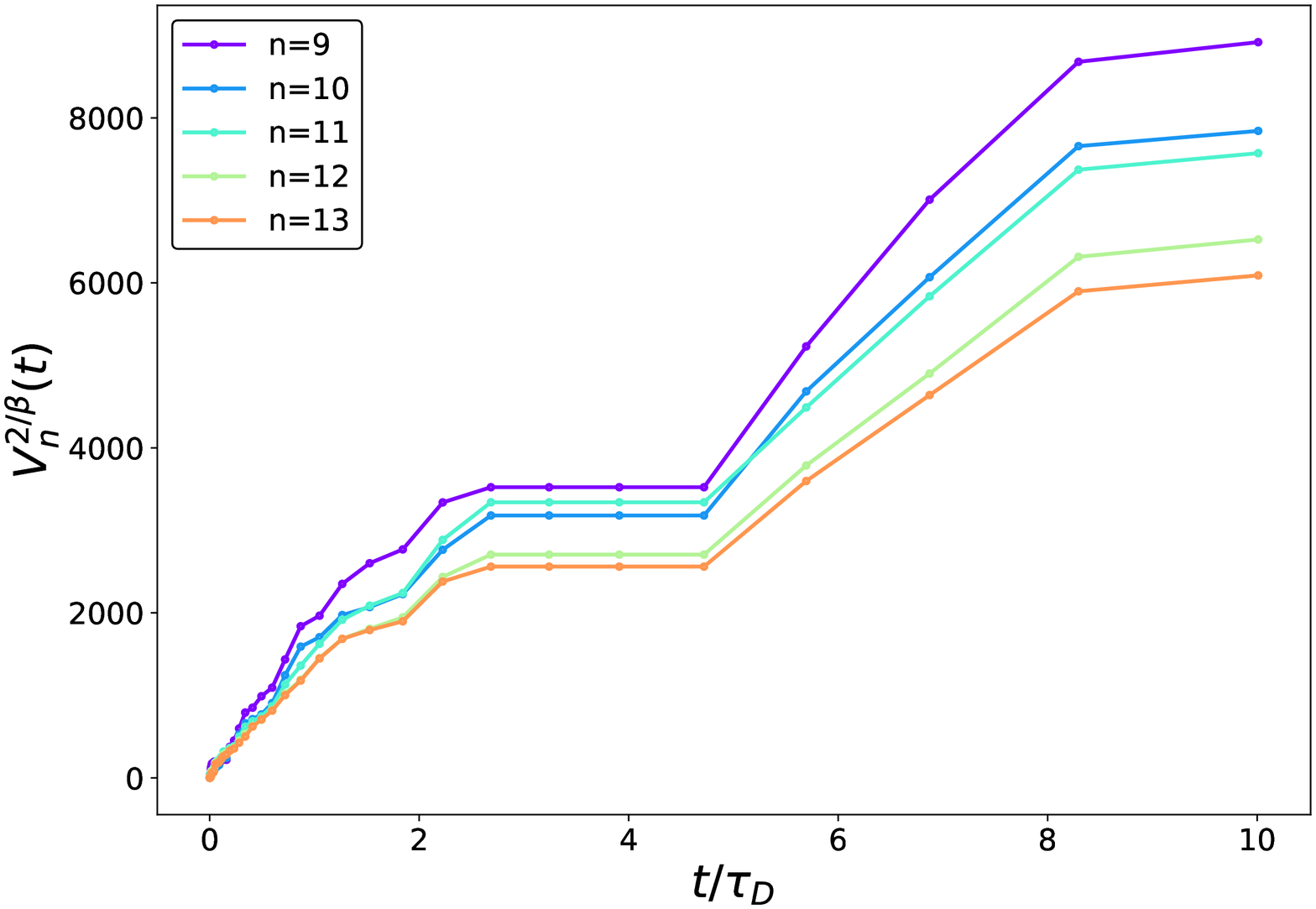}
\includegraphics[height=4cm,width=7cm]{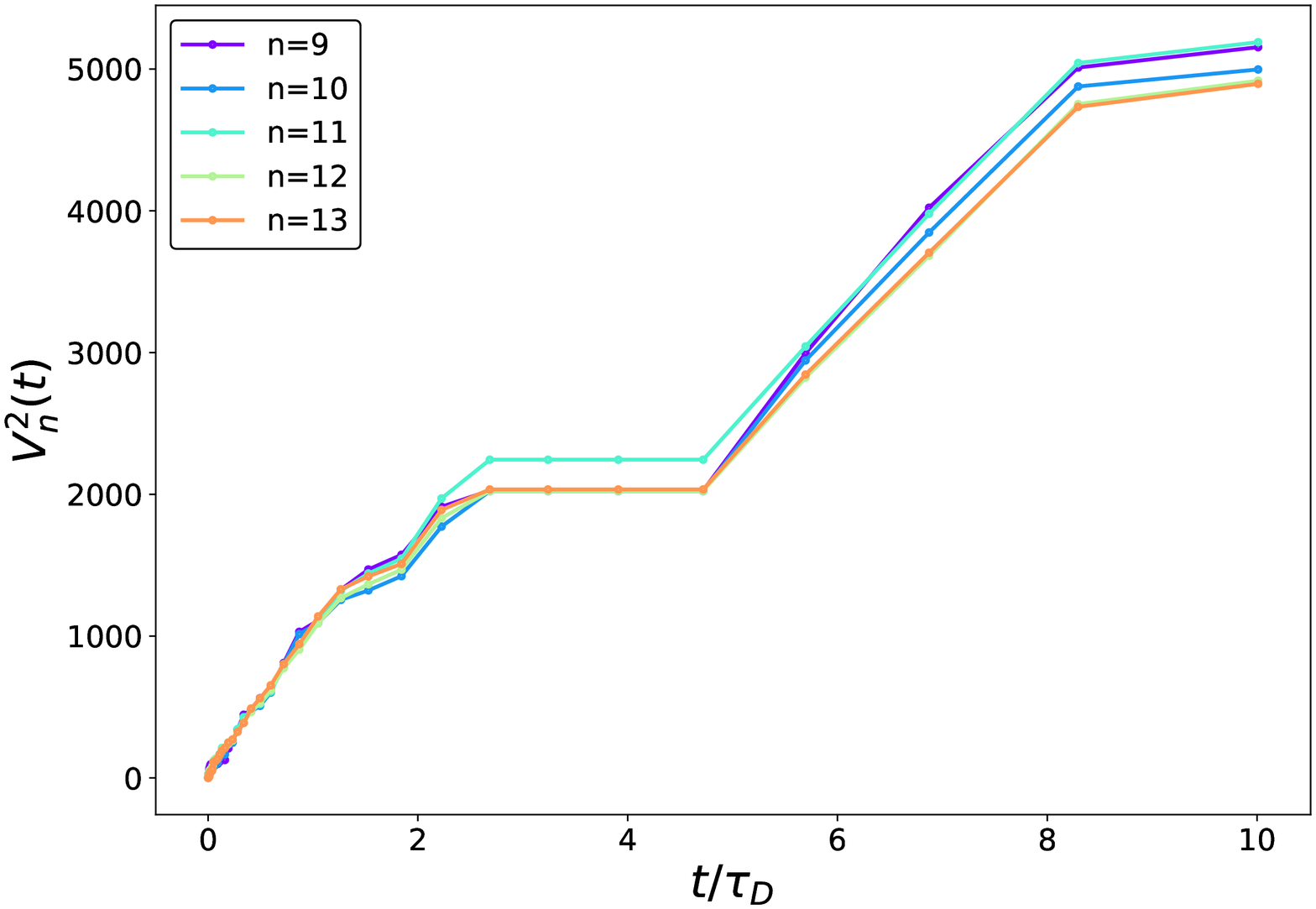}
\caption{
Plots of p-variation tests of variation orders $2/\beta$ (left) and $2$ (right)
for different values of $p\in(0,1/2)$, from (a) to (d):
$p= 0.05 \,, 0.01 \,, 0.005 \,, 0.001$ that is
$\beta= 0.7465 \,, 0.8213 \,, 0.8411 \,, 0.8735$.
The monotonic-continuous growing is consistent 
with an underlying Gaussian-like motion,
in particular with the fBm \cite{magdziarz_etal-prl-2009}
and the ggBm \cite{molina_etal-pre-2016}, and then consistent also with
a typical signature of motion inside living cells,
see, e.g., \cite{magdziarz_etal-prl-2009}.
}
\label{fig:p-variation}
\end{center}
\end{figure}

\begin{figure}
\begin{center}
\subfloat[][]{
\includegraphics[height=5cm,width=7cm]{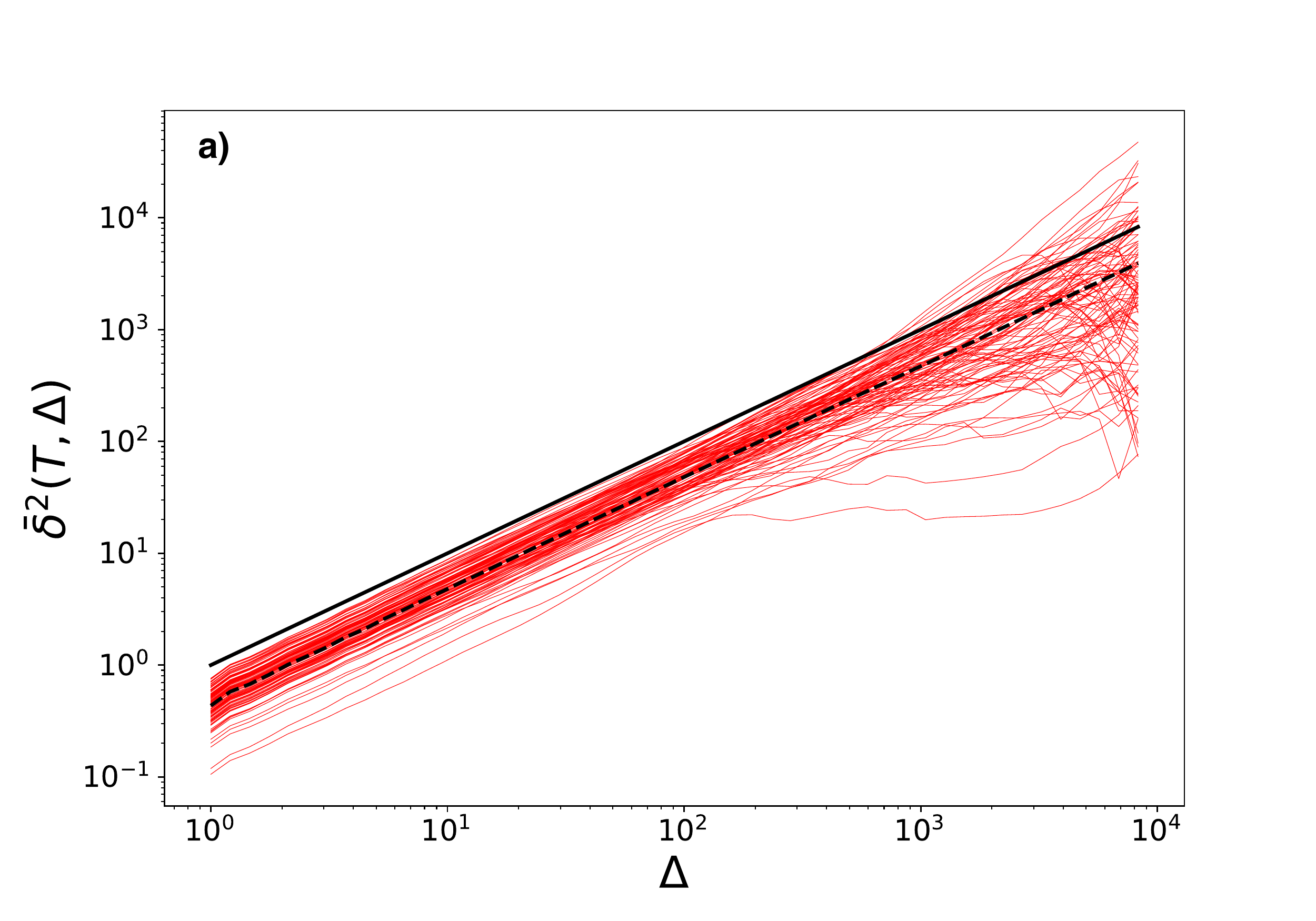}}
\subfloat[][]{
\includegraphics[height=5cm,width=7cm]{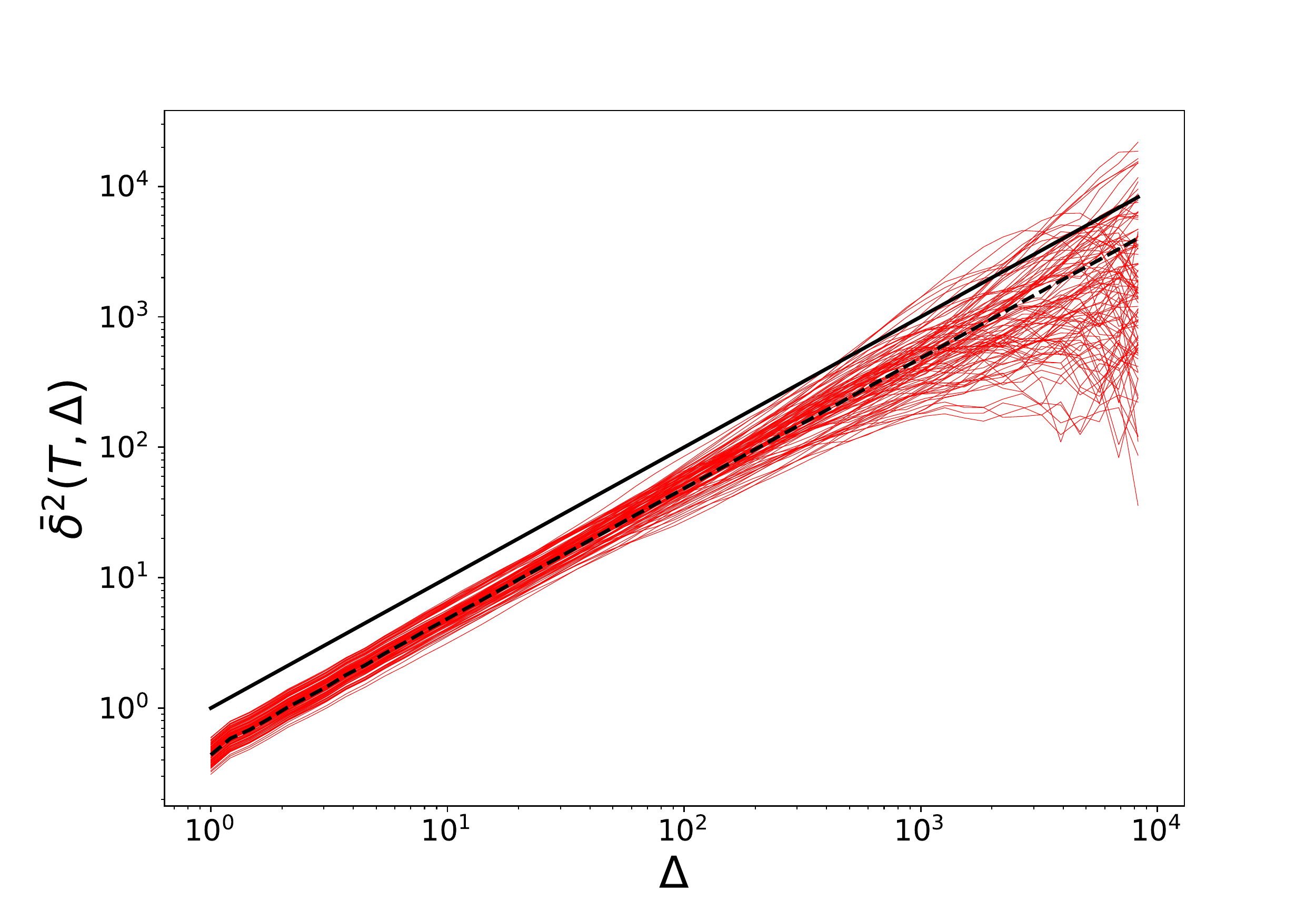}}
\\
\subfloat[][]{
\includegraphics[height=5cm,width=7cm]{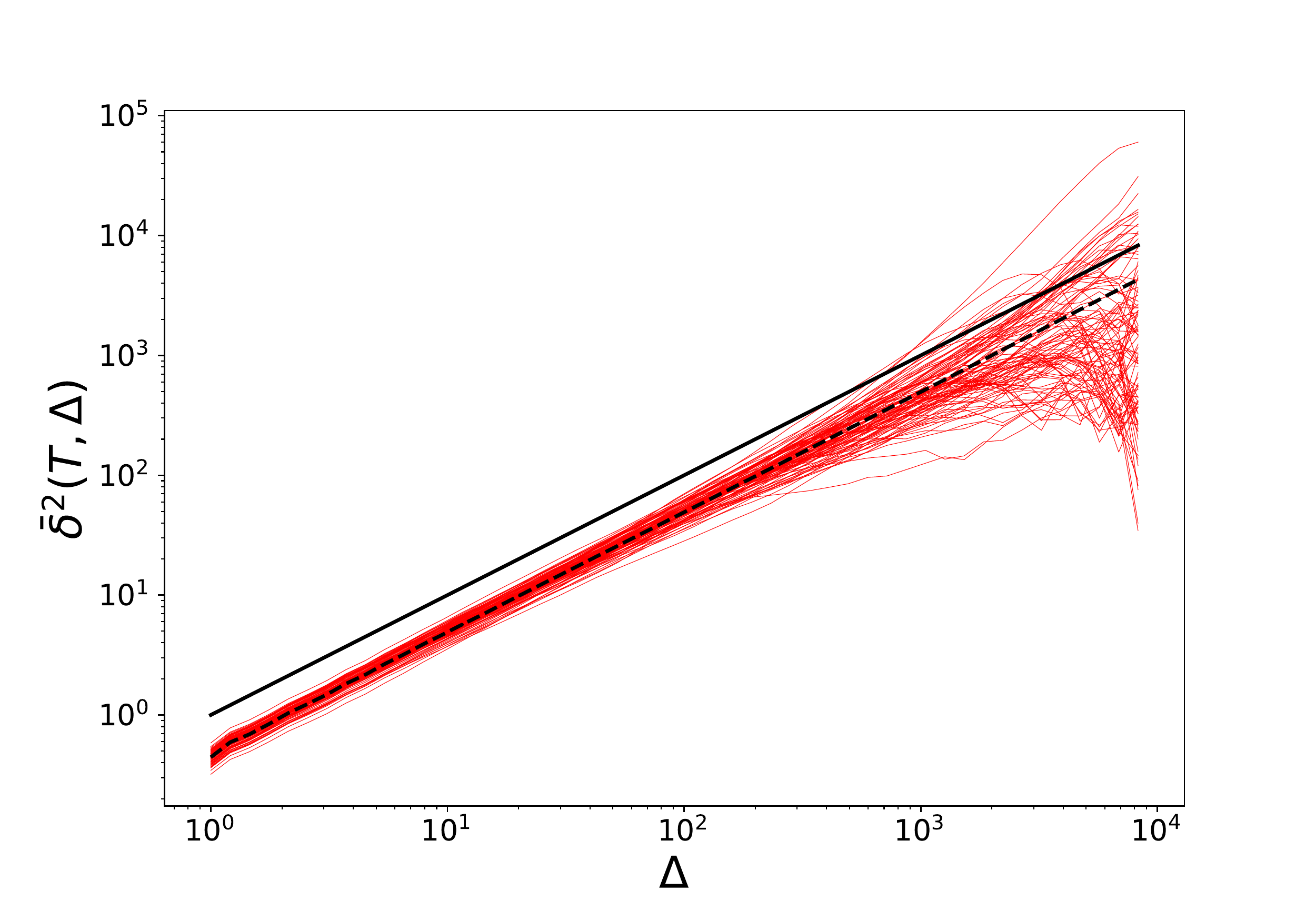}}
\subfloat[][]{
\includegraphics[height=5cm,width=7cm]{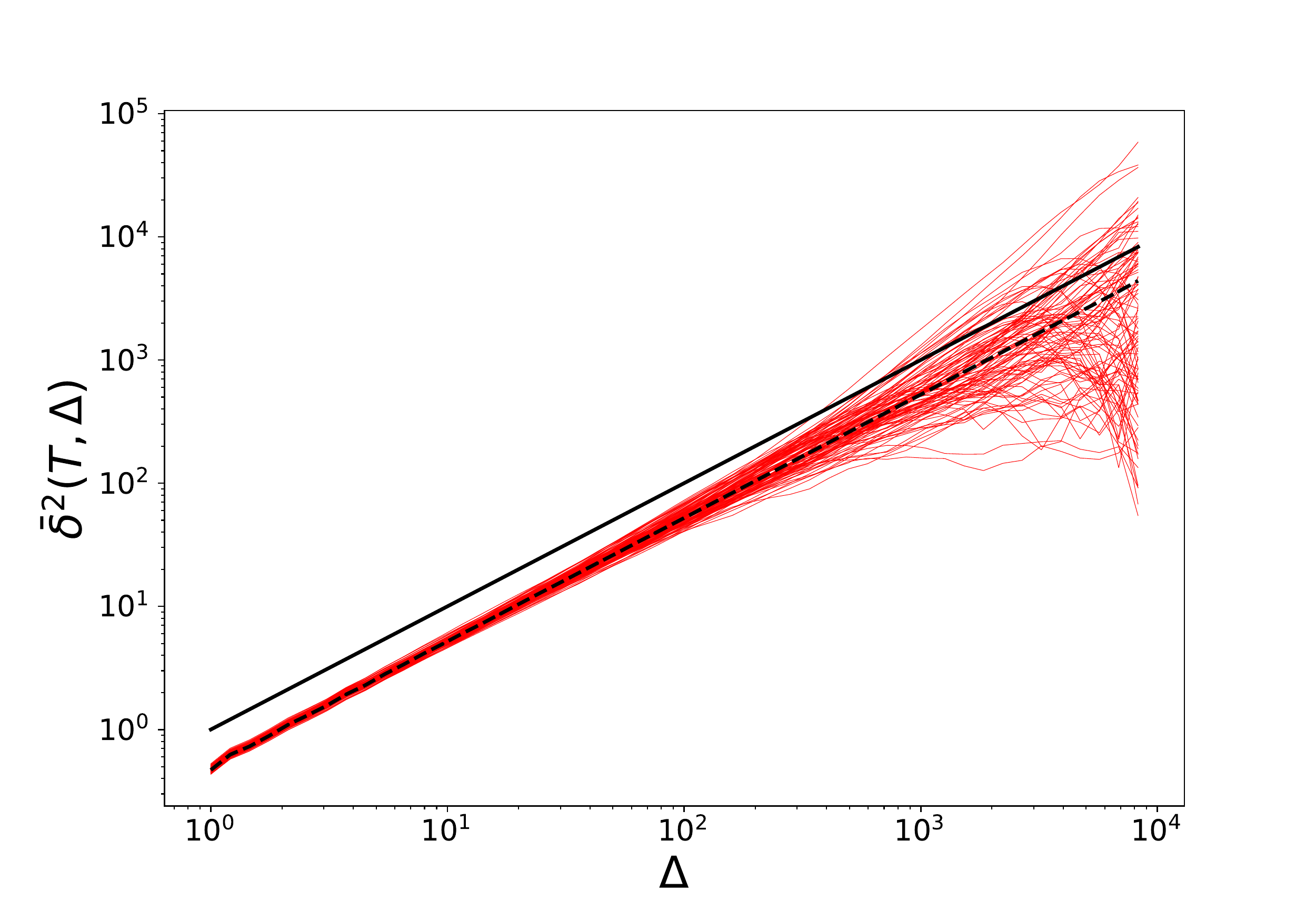}}
\caption{
Plots of the TAMSD for different values of $p\in(0,1/2)$,
from (a) to (d): $p=0.001 \,, 0.005 \,, 0.01 \,, 0.05$ that is
$\beta=0.8735 \,, 0.8411 \,, 0.8213 \,, 0.7465$.
The dashed-line represents the ETAMSD and the
solid-line provides a guide to the eye of the linear scaling.
}
\label{fig:TAMSD}
\end{center}
\end{figure}

\begin{figure}
\begin{center}
\subfloat[][]{
\includegraphics[height=5cm,width=7cm]{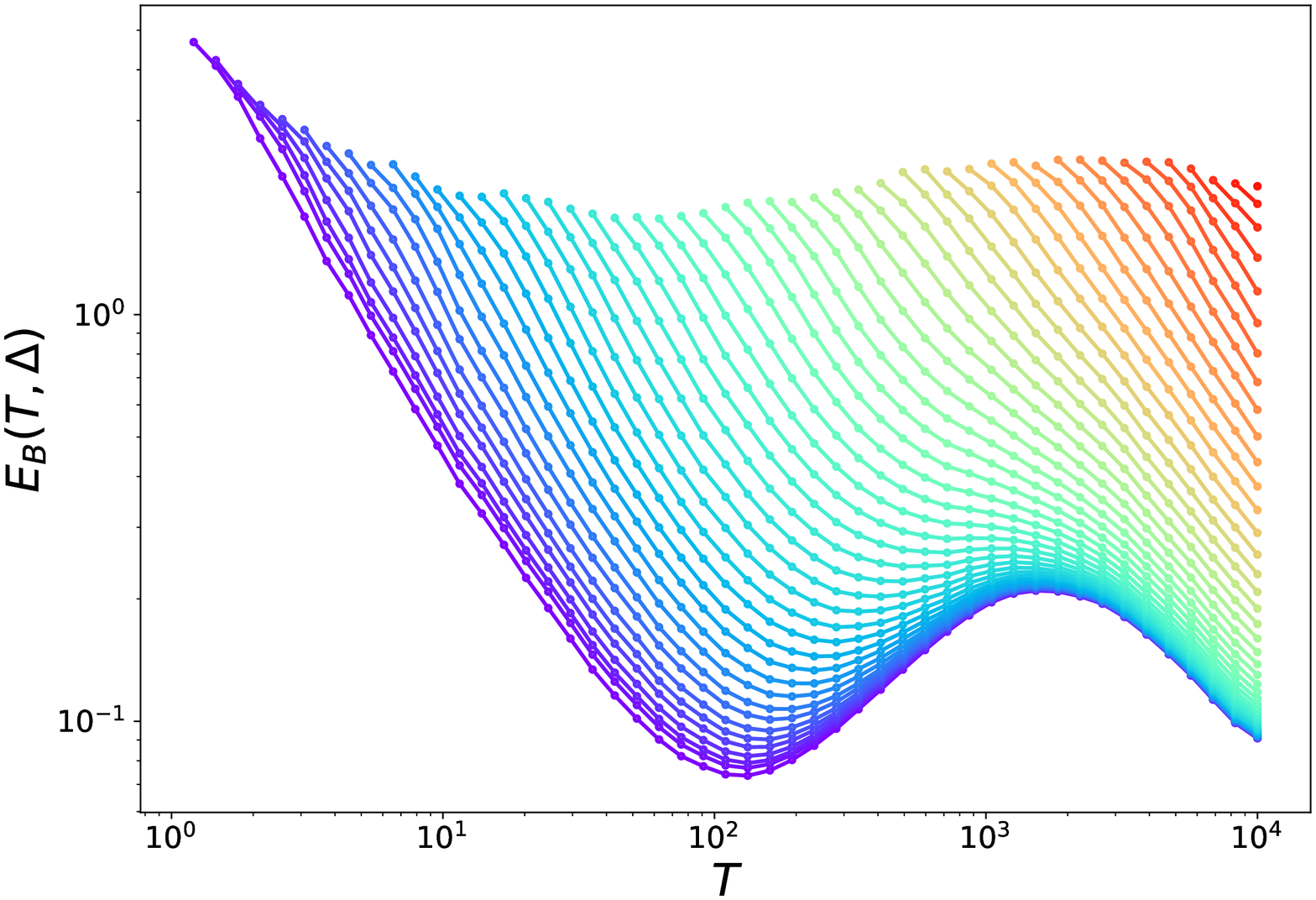}}
\subfloat[][]{
\includegraphics[height=5cm,width=7cm]{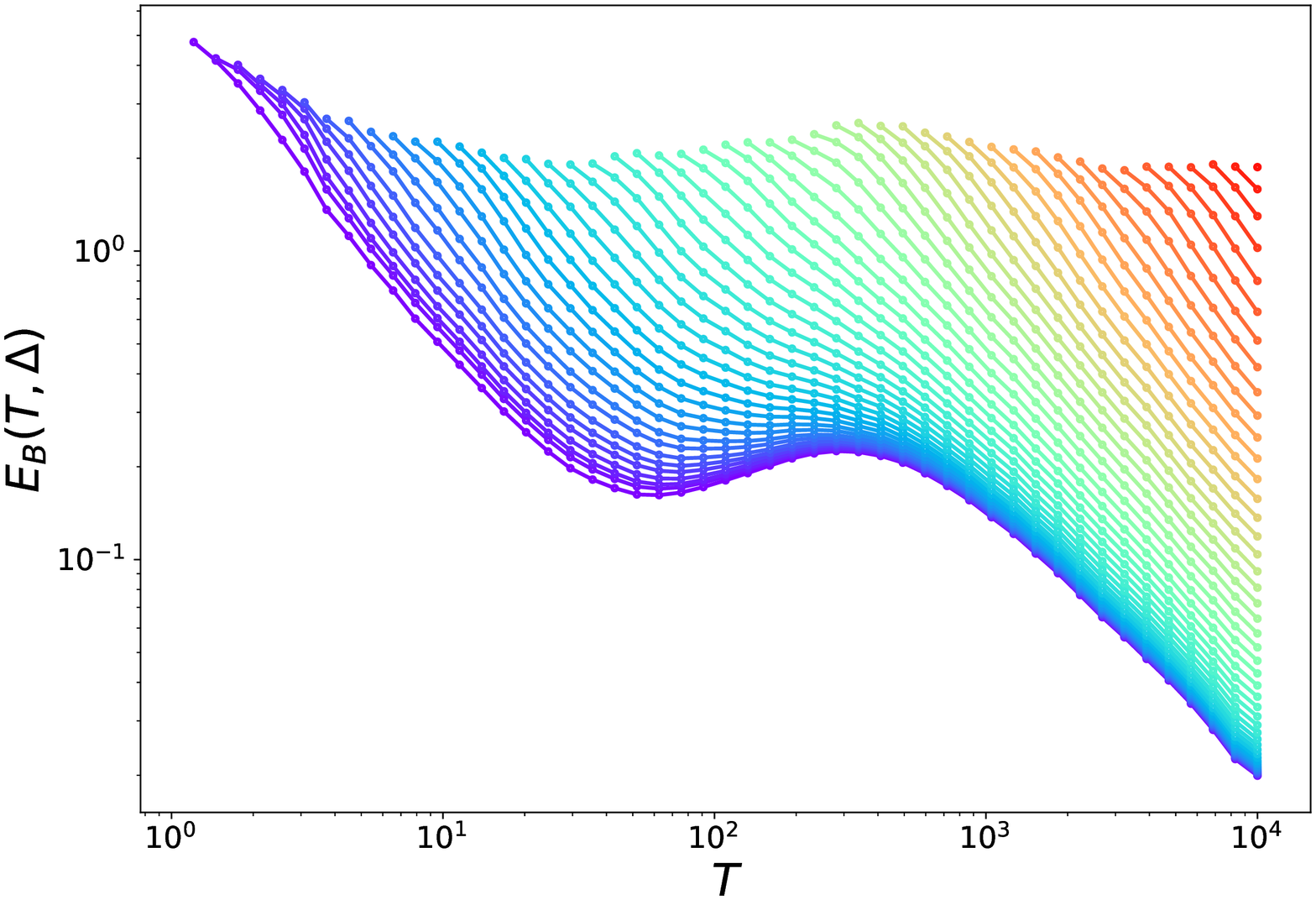}}
\\
\subfloat[][]{
\includegraphics[height=5cm,width=7cm]{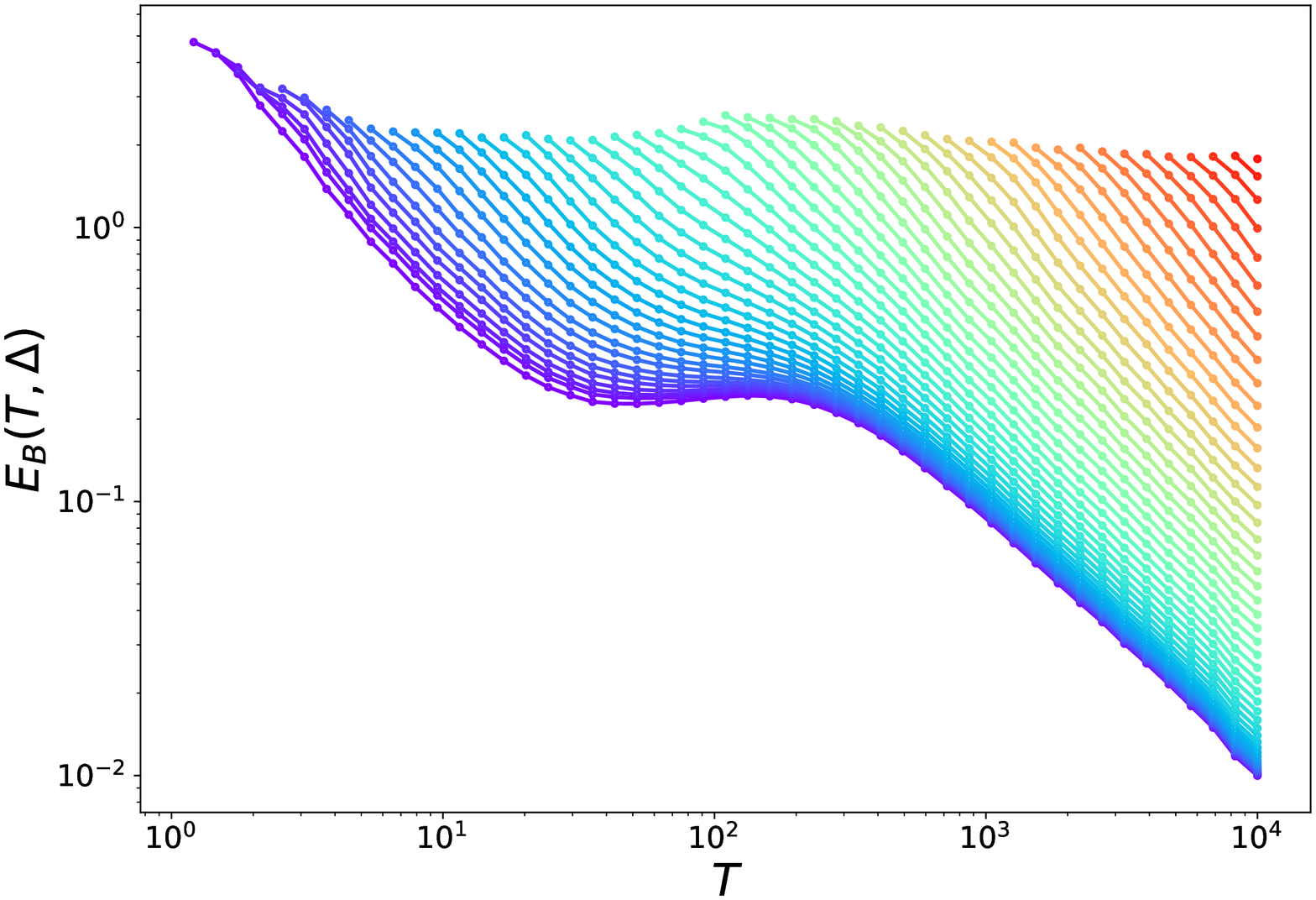}}
\subfloat[][]{
\includegraphics[height=5cm,width=7cm]{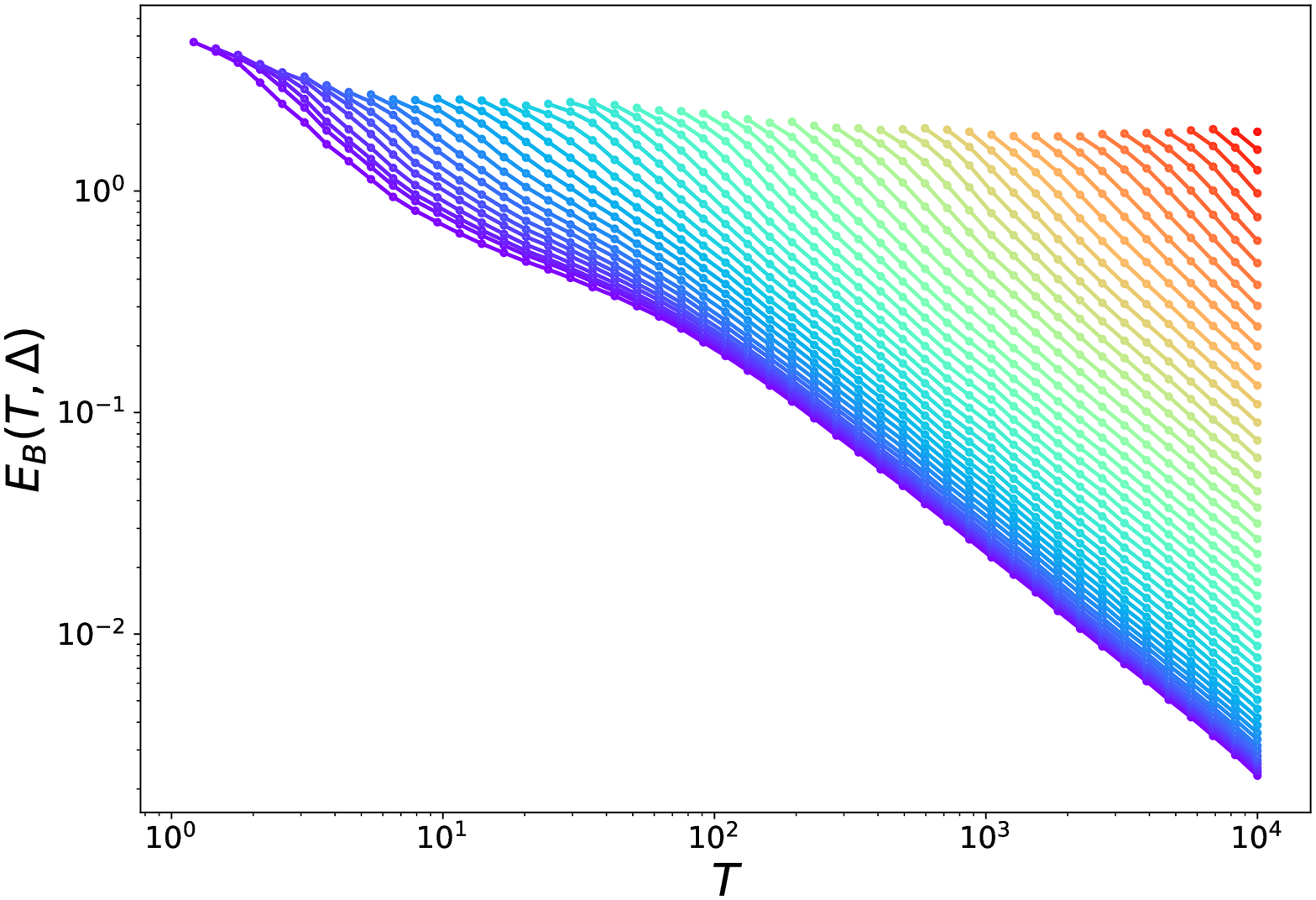}}
\caption{Plot of the ergodicity breaking parameter 
$E_B$ (\ref{eq:EB}) as a function of the measurement time $T$ 
and of the time-lag $\Delta$ for different values of $p \in (0,1/2)$,
from (a) to (d): $p=0.001 \,, 0.005 \,, 0.01 \,, 0.05$
that is $\beta = 0.8735 \,, 0.8411 \,, 0.8213 \,, 0.7465$.
The color map provides the dependence on the parameter $\Delta$: 
from purple (small $\Delta$) to red (large $\Delta$). 
The decreasing of the initial and final regimes is $T^{-1}$. 
}
\label{fig:EB}
\end{center}
\end{figure}

\begin{figure}
\begin{center}
\subfloat[][]{
\includegraphics[height=5cm,width=7cm]{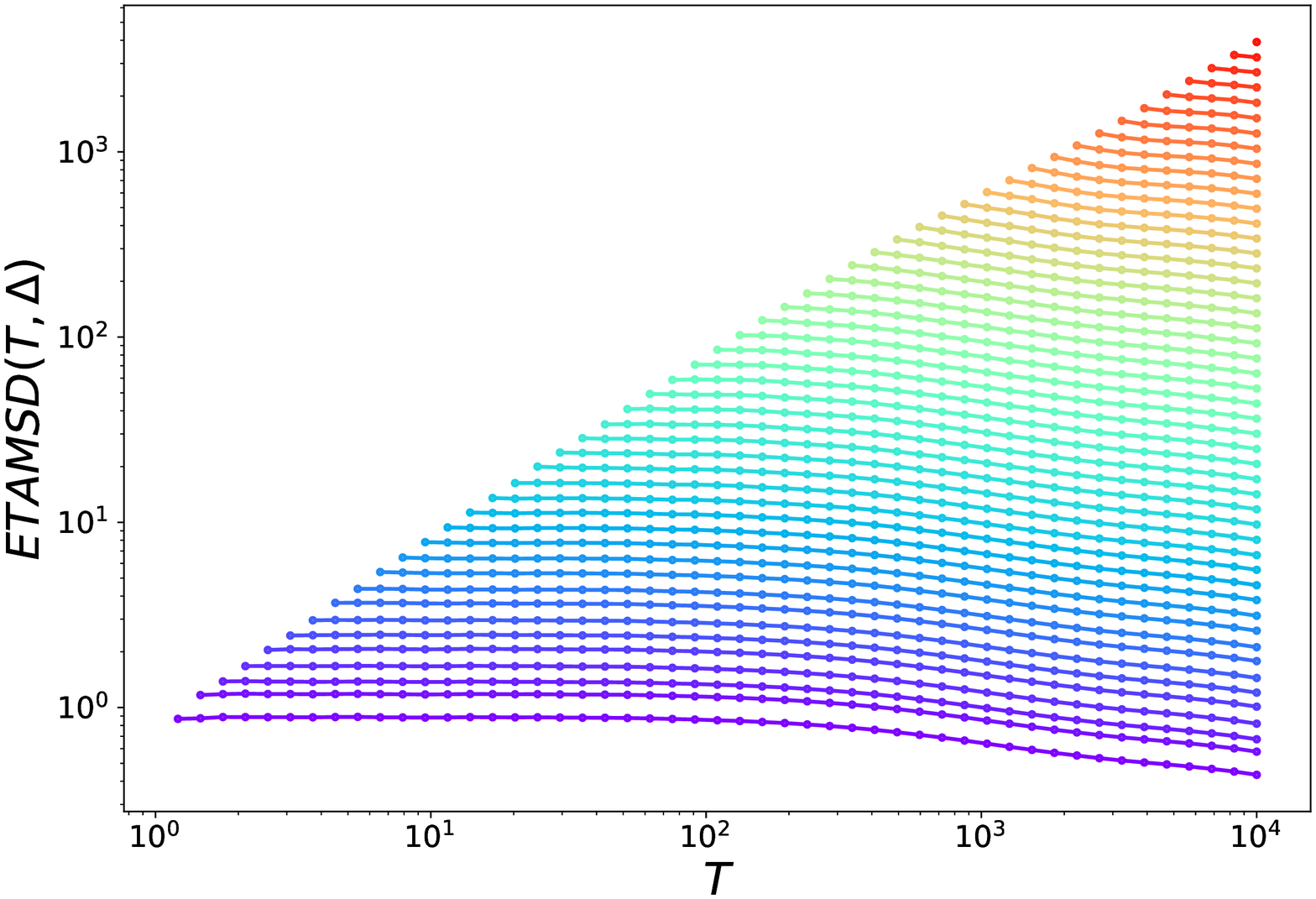}}
\subfloat[][]{
\includegraphics[height=5cm,width=7cm]{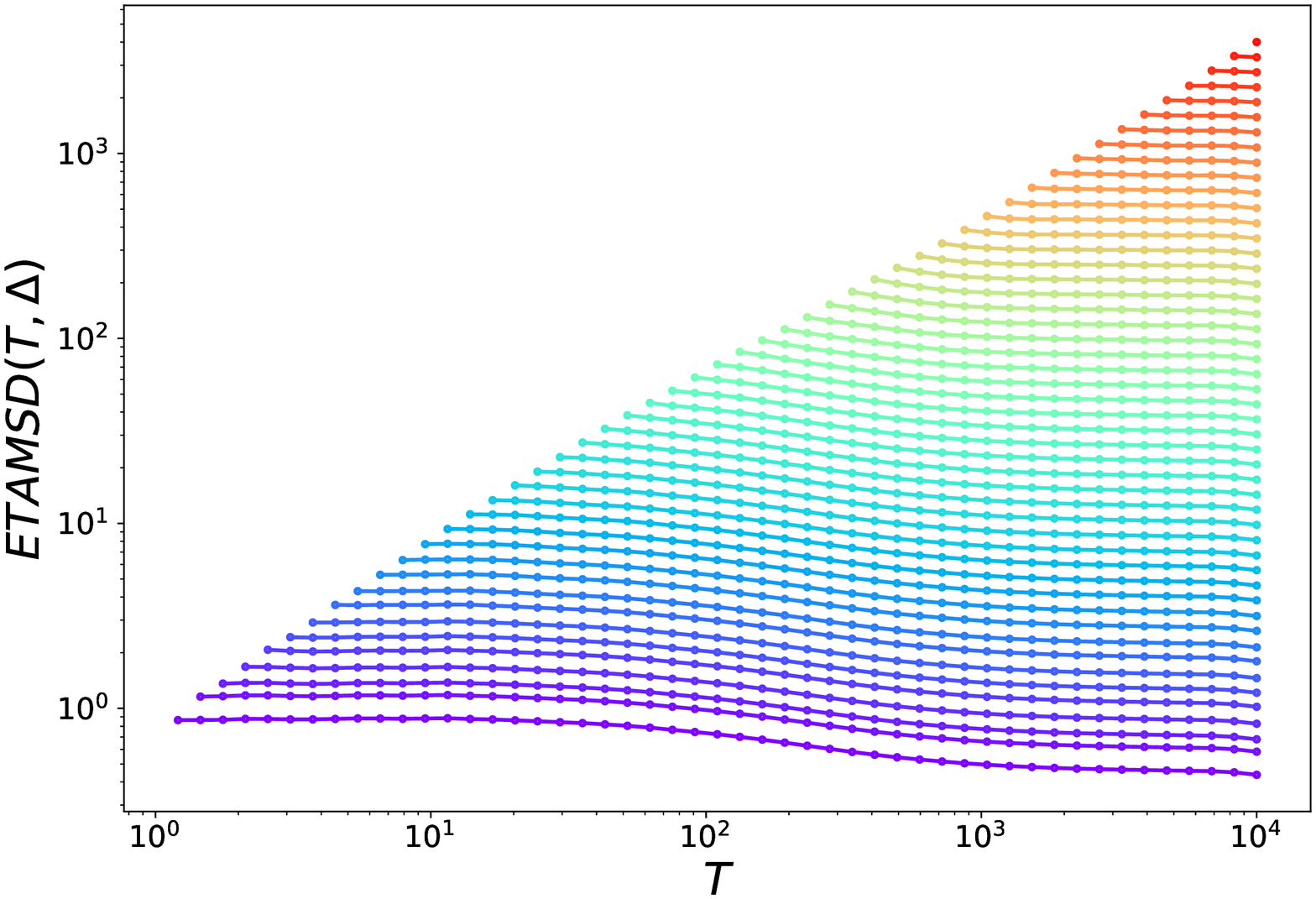}}
\\
\subfloat[][]{
\includegraphics[height=5cm,width=7cm]{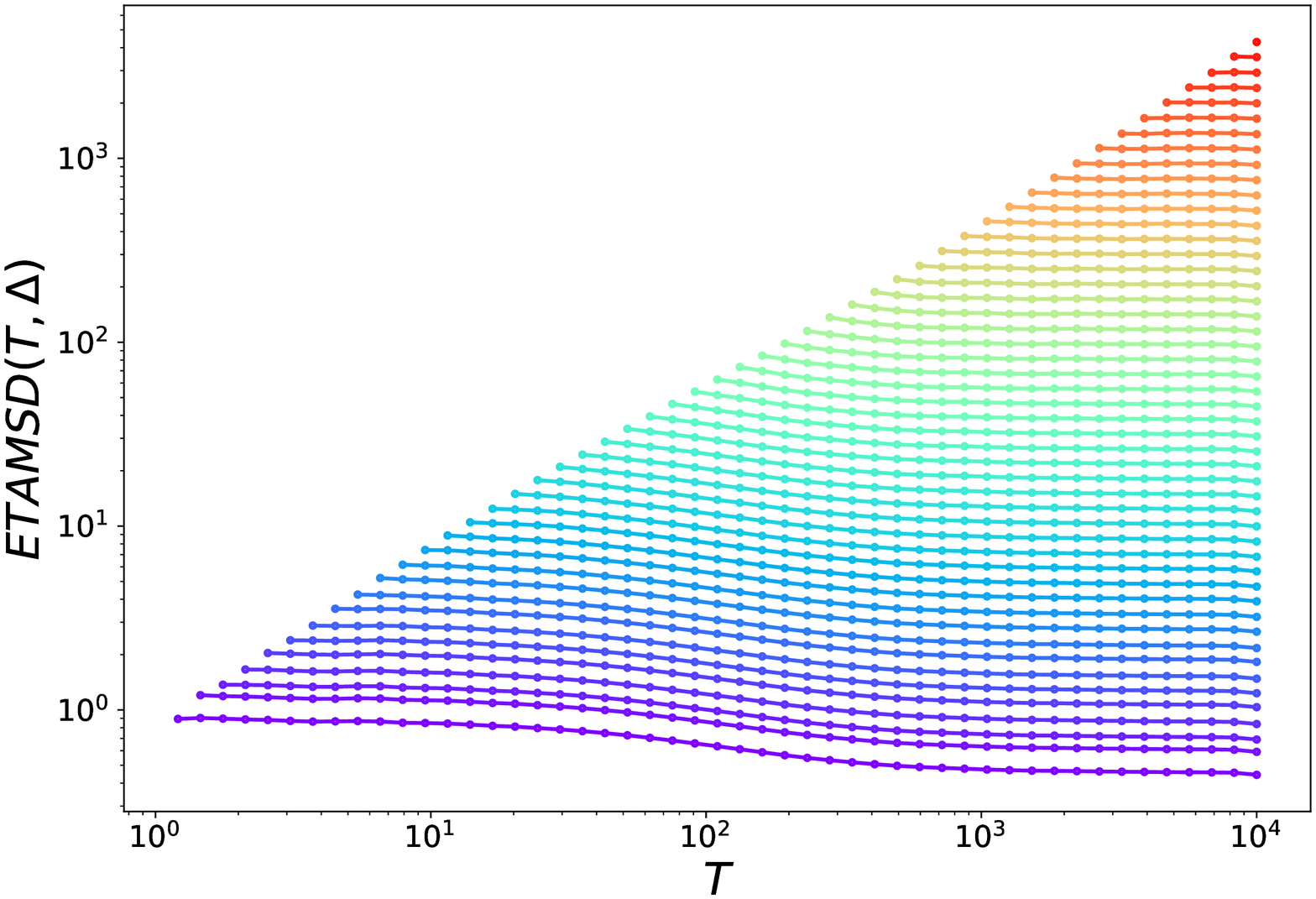}}
\subfloat[][]{
\includegraphics[height=5cm,width=7cm]{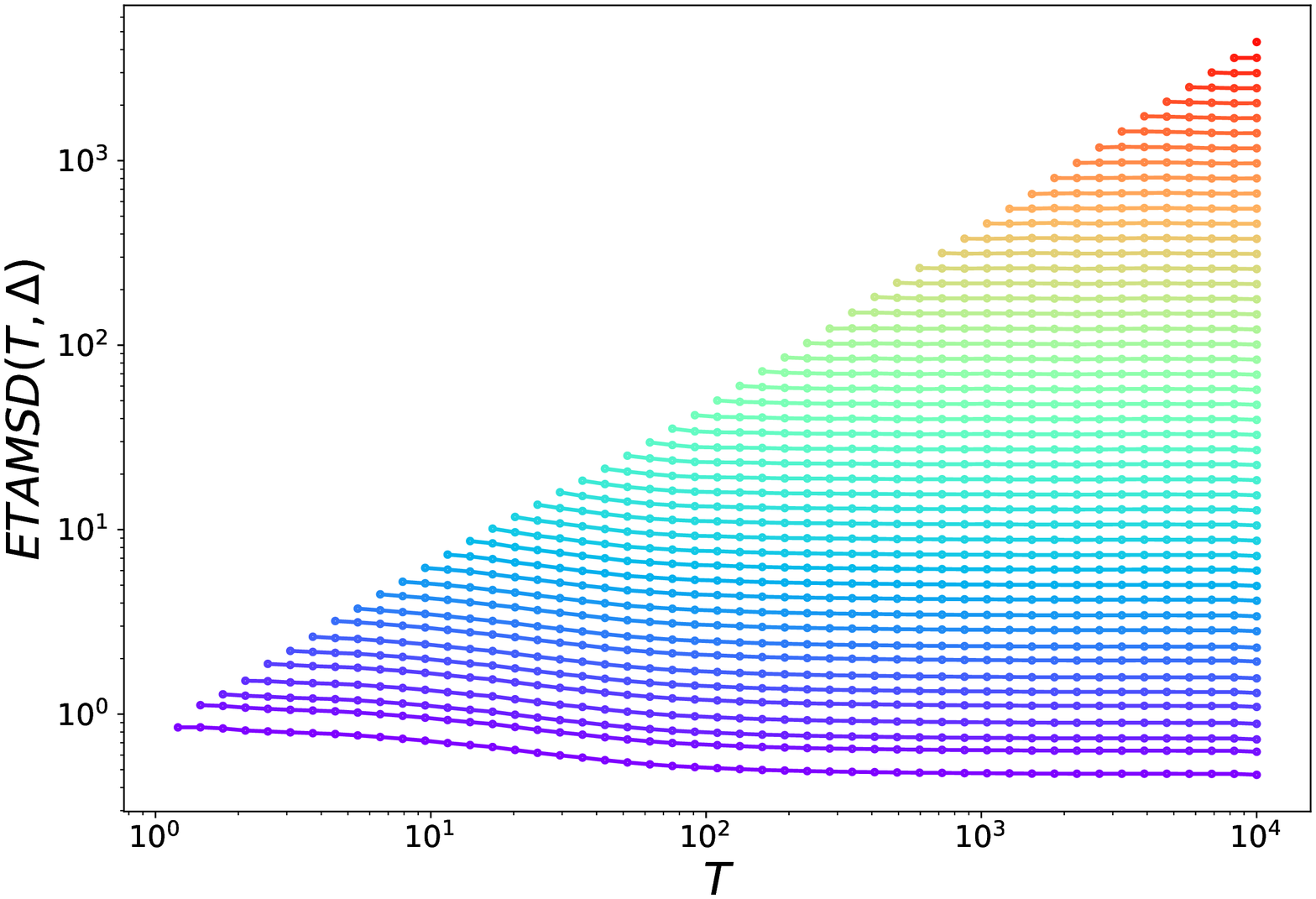}}
\caption{
Plot of the ETAMSD 
as a function of the measurement time $T$ 
and of the time-lag $\Delta$ for different values of $p \in (0,1/2)$,
from (a) to (d): $p=0.001 \,, 0.005 \,, 0.01 \,, 0.05$
that is $\beta = 0.8735 \,, 0.8411 \,, 0.8213 \,, 0.7465$.
The color map provides the dependence on the parameter $\Delta$: 
from purple (small $\Delta$) to red (large $\Delta$). 
When model (\ref{model}, \ref{model2}) displays aging, the 
decreasing of the curve is approximately $T^{-0.2}$, 
see figure \ref{fig:scaling-aging} for details. 
}
\label{fig:aging}
\end{center}
\end{figure}

\begin{figure}
\begin{center}
\subfloat[][]{
\includegraphics[height=5cm,width=7cm]{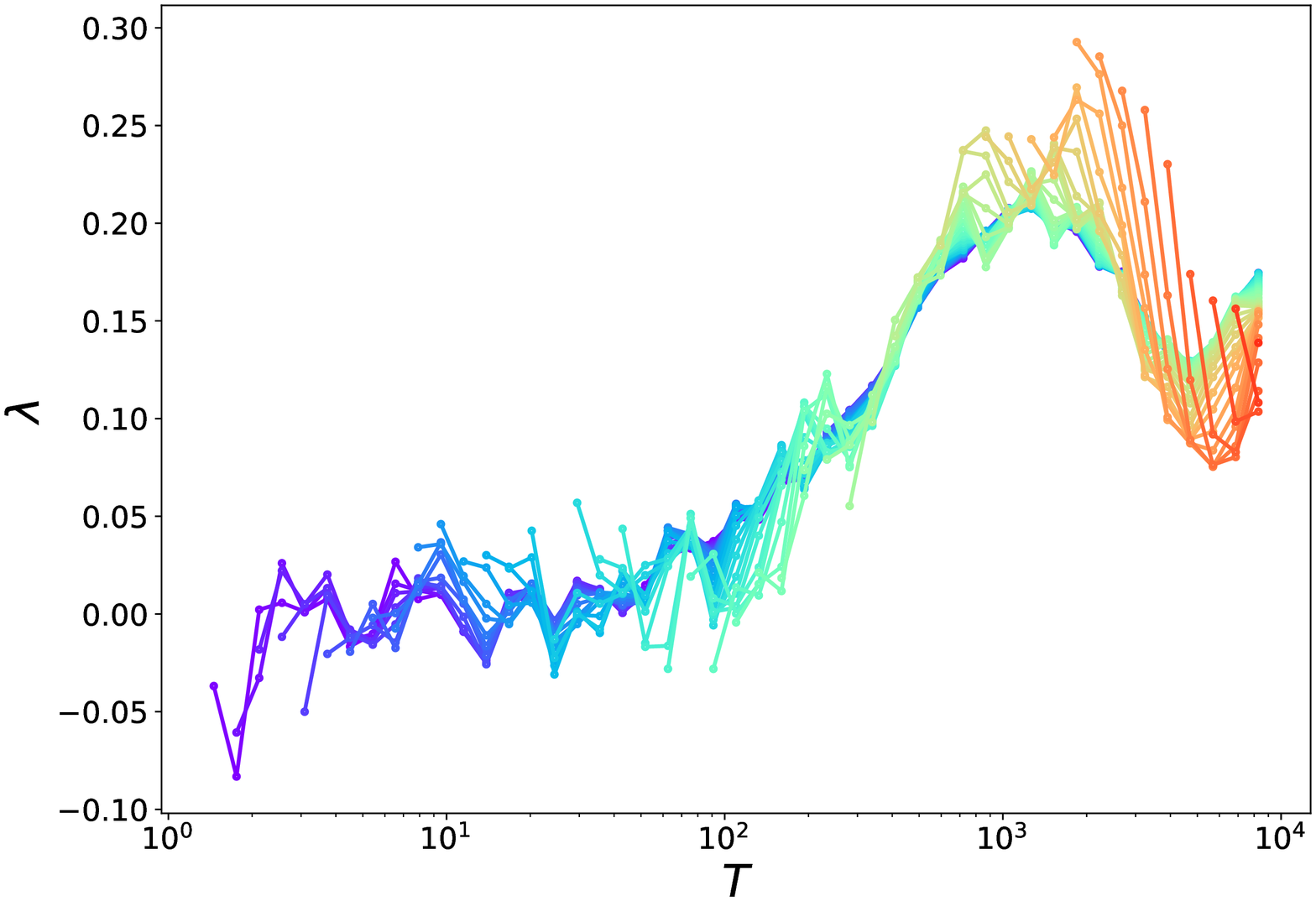}}
\subfloat[][]{
\includegraphics[height=5cm,width=7cm]{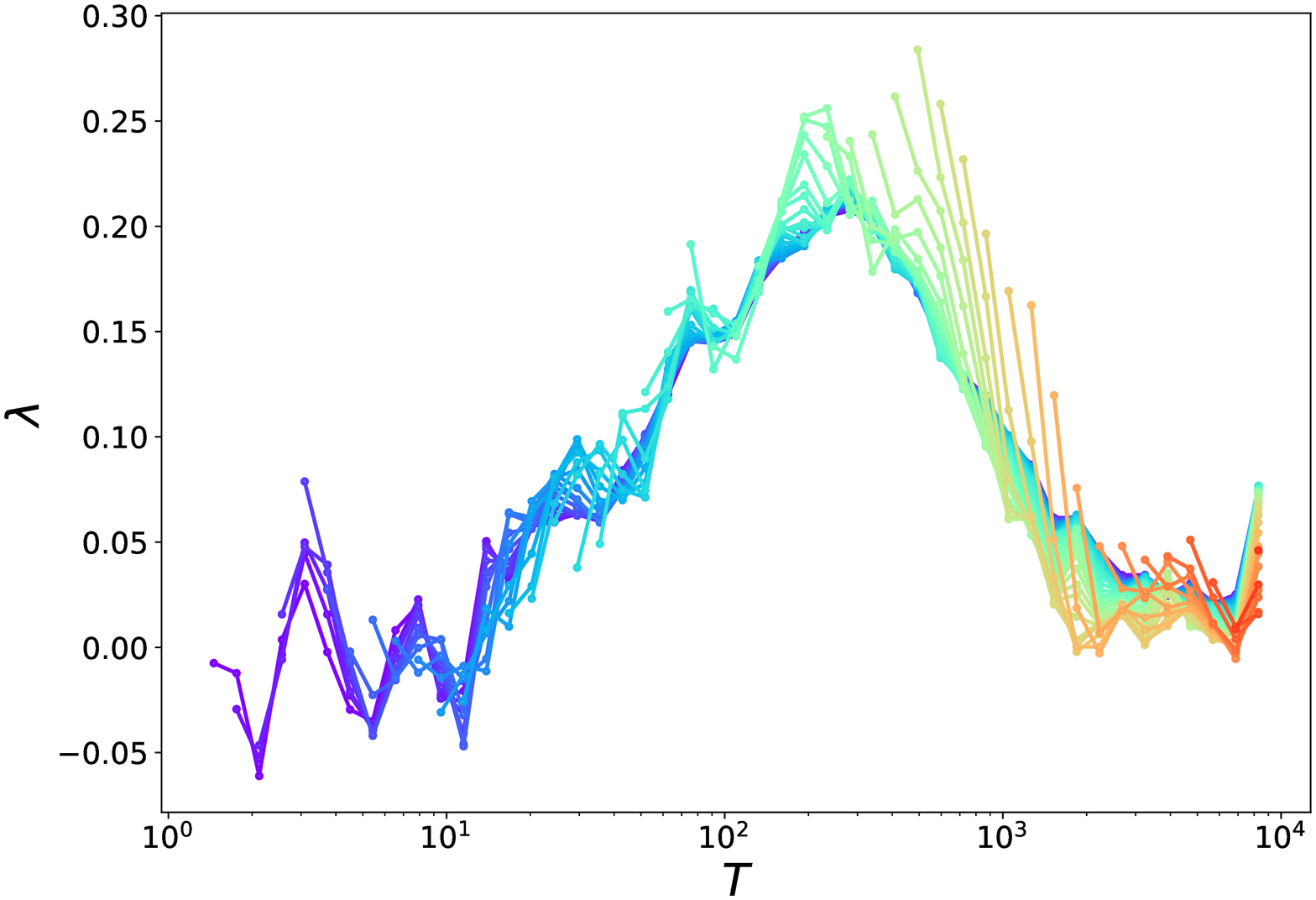}}
\\
\subfloat[][]{
\includegraphics[height=5cm,width=7cm]{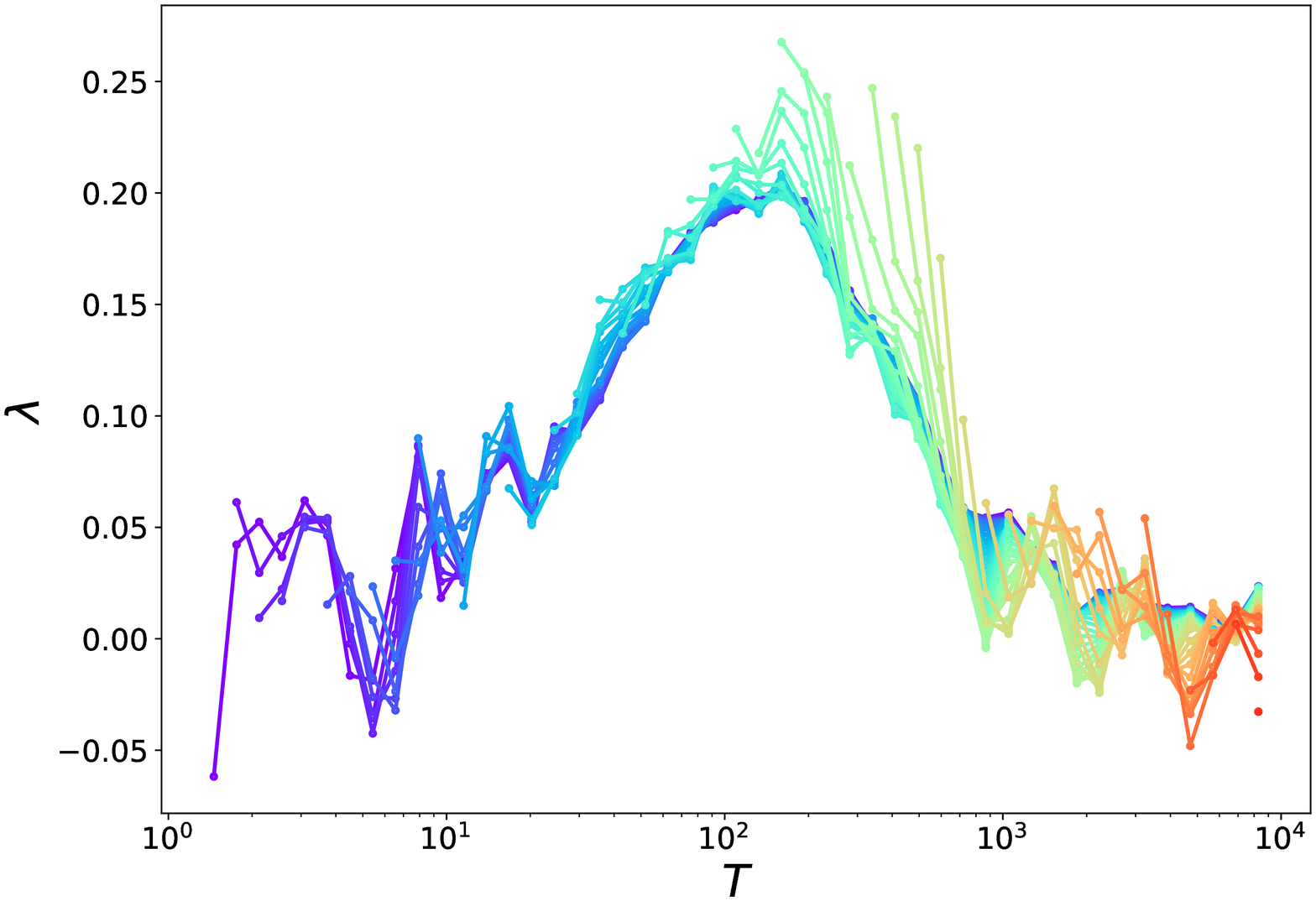}}
\subfloat[][]{
\includegraphics[height=5cm,width=7cm]{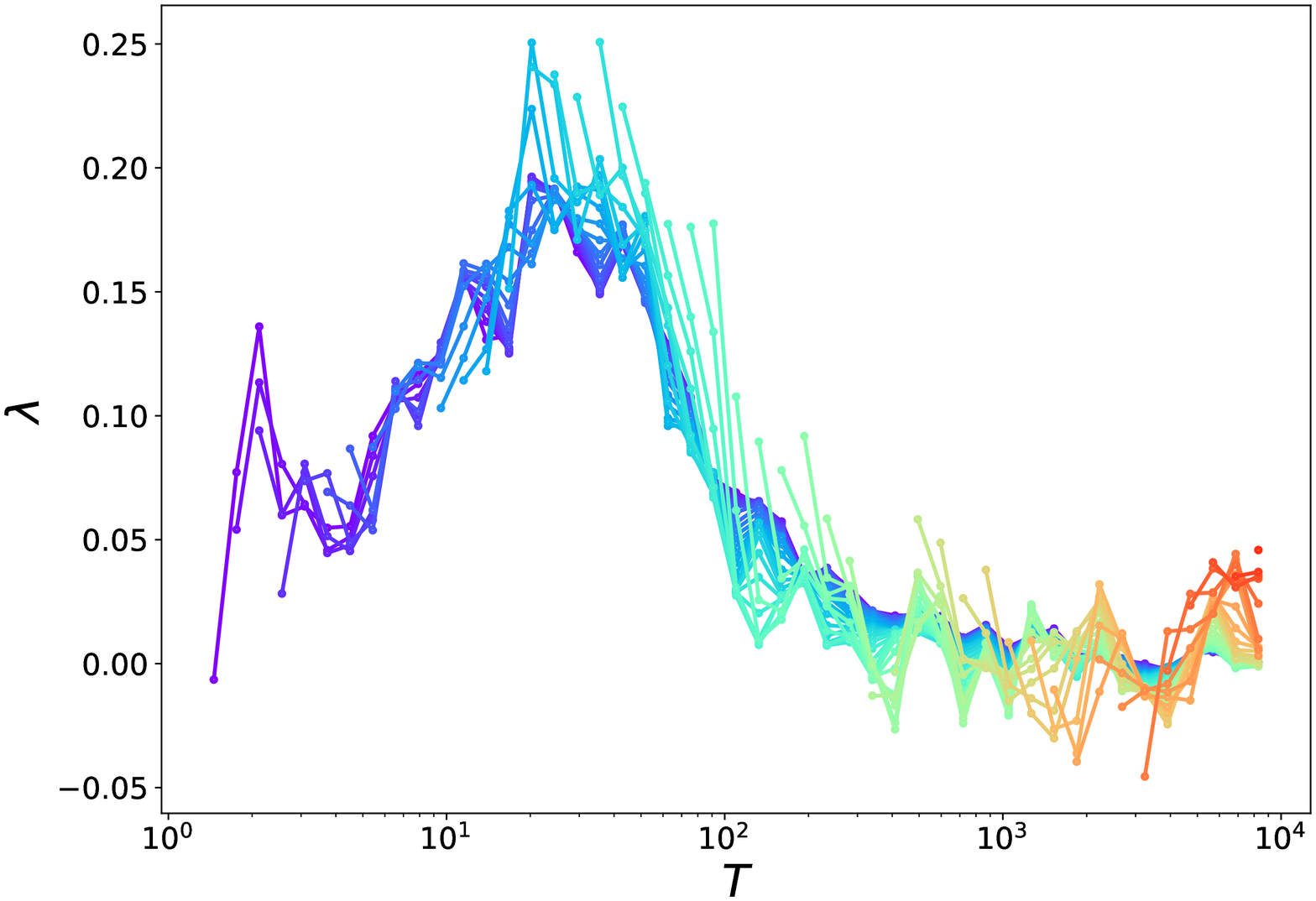}}
\end{center}
\caption{Analysis of the decreasing-law $T^{-\lambda}$ of the ETAMSD. 
The exponent $\lambda$ is plotted as a function of the measurement time $T$
and of the time-lag $\Delta$ for different values of $p \in (0,1/2)$,
from (a) to (d): $p=0.001 \,, 0.005 \,, 0.01 \,, 0.05$
that is $\beta = 0.8735 \,, 0.8411 \,, 0.8213 \,, 0.7465$.
The color map provides the dependence on the parameter $\Delta$: 
from purple (small $\Delta$) to red (large $\Delta$). 
At the peak of maximum aging it holds $\lambda \approx 0.20$ and 
it corresponds to the intermediate anomalous diffusion regime.
}
\label{fig:scaling-aging}
\end{figure}

About statistics of single-trajectory, 
in figure \ref{fig:p-variation} it is shown that
model (\ref{model}, \ref{model2}) displays 
a p-variantion test consistent with an underlying Gaussian-like motion,
e.g., the fBm \cite{magdziarz_etal-prl-2009} 
and ggBm \cite{molina_etal-pre-2016}, and then consistent also with
the typical signature of motion inside living cells,
see, e.g., \cite{magdziarz_etal-prl-2009}. 
Moreover, 
in figure \ref{fig:TAMSD} it is shown that the TAMSD scales
linearly as the Bm but it displays a distribution
of diffusion coefficients among the trajectories.
This behaviour is also observed in data of molecular motion
in living cells, see, e.g., \cite{manzo_etal-prx-2015}.
Together with the p-variation trend,
the TAMSD is another characteristic consistent with the 
ggBm formalism \cite{mura_etal-jpa-2008,molina_etal-pre-2016} that, 
for a distribution of the diffusion coefficients
which is characteristic of each data set,
was succefully applied 
to anomalous diffusion observed in live Escherichia coli bacteria
by tracking mRNA molecules,
see reference \cite{golding_etal-prl-2006} for the data  
and \cite{mackala_etal-pre-2019} for the ggBm-like model, 
and by tracking DNA-binding proteins,
see reference \cite{sadoon_etal-pre-2018} for the data
and \cite{itto_etal-jrsi-2021} for the ggBm-like model.
The distribution of the diffusion coefficient
causes weak ergodicity-breaking
in the anomalous diffusion regime $\tauB < t < \tauD$, 
see figure \ref{fig:EB}, and also aging, see figure \ref{fig:aging}. 
In particular, figure \ref{fig:EB} shows that
the $E_B$ parameter (\ref{eq:EB}) displays  
an initial and a final linear decreasing-law $\sim T^{-1}$ towards
ergodicity that is broken during the intermediate anomalous regime. 
The strength of the break reduces when $p$ grows. 
This behaviour confirms that weak ergodicity breaking 
is the cause of the emerging of fractional diffusion \cite{molina_etal-pre-2016}
and it is the ruler of its extension. 
In figure \ref{fig:aging} the aging of model (\ref{model}, \ref{model2})
is shown through the plot of the ETAMSD.
Again during the intermediate anomalous regime,
we observe that the ETAMSD is not constant and a decreasing-law 
$\sim T^{-\lambda}$, with $\lambda > 0$, manifests 
a transition between two aged regimes that are independent of $T$. 
The aging exponent $\lambda$
is analysed in figure \ref{fig:scaling-aging} and it emerges to be
$\lambda \simeq 0.20$ in the interval with maximum slope.

We conclude that model (\ref{model}, \ref{model2}) meets
all the paradigmatic features that belong to the anomalous diffusion
as it is observed in living systems. 
Moreover, the plots show a clear characterization of the 
intermediate anomalous regime as due and driven by the 
ratio between the time-scales of two Markovian mechanisms (\ref{model2}) and
this ratio determines the anomalousness parameter $\beta$ (\ref{beta}). 

\section{Conclusions}
\label{sec:conclusions}
In this study we analysed a simple CTRW model with a
waiting-time distribution defined as the weigthed sum of two
exponential distributions with different time-scales 
$\tauB$ and $\tauD$ (\ref{model}):
$\tauB$ is the Barkai--Burov time-scale related to the 
universal exponential tails of walker's PDF and 
$\tauD$ is the time-scale of the diffusive limit.
The weight parameter $p \in (0,1/2)$ is a free-parameter
that rules non-Markovianity of model (\ref{model}) and 
relates the two time-scales according to formula (\ref{model2}).
We tested this model against paradigmatic features of anomalous
diffusion. In particular,
ensemble features 
as the sub-linear MSD growing with a power-law of degree $0 < \beta < 1$,
a stretched-exponential walker's PDF and the occurrence of the BynG interval,
and single-particle features, as the TAMSD,
p-variation test, weak ergodicity breaking and aging.
Remarkably, model (\ref{model}, \ref{model2}) 
meets all these features that therefore are caused by a 
process characterized by solely two time-scales.
This allows to avoid the introduction of 
a wide spectrum of time-scales \cite{shlesinger-arpc-1988,pagnini-pa-2014}
as adopted in superposition of fBm 
\cite{molina_etal-pre-2016,mackala_etal-pre-2019} 
or in CTRW models with trapping mechanism with infinite-mean waiting-times,
e.g., \cite{shlesinger-jsp-1974}.
Moreover, 
the model dynamically provides the anomalousness parameters $\beta$
as a function of $p$ (\ref{beta}).

Model (\ref{model}, \ref{model2}) describes a diffusive
hopping-trap mechanism in a disordered medium 
where two families of Markovian sites characterize its structure. 
These two families of sites are not compartmented in separated zones,
but they are uniformly mixed togheter with different 
percentage and randomly shuffled in time.
On a phenomenological level, this is equivalent to saying that 
the two families of Markovian sites reflect the fact that  
the energy barrier may fluctuate within each 
equilibrium state 
(because of an exponential distribution of waiting-times)
and also between two equilibrium states.
The statistical occurence of such state-fluctuations 
along walker's trajectory causes 
the emergence of an intermediate regime where walker moves according to
features of anomalous diffusion.

In a more dynamical sense,
we argue here that a CTRW model with a waiting-time distribution 
given by the weighted sum of two waiting-time distributions
is indeed a model for walker's trajectory going under the action
of {\it two} hopping-trap mechanisms:
namely a trajectory that goes under the action of two 
{\it co-existing} random forcings 
each one responsible for each hopping-trap mechanism. 
This observation brings to our mind the case of  
the motion of a material-particle in a fluid, 
that moves under the {\it co-existing} effects of the velocity of the 
fluid-particle hosting the material-particle 
and of the molecular diffusion, this last allowing the material-particles
to shift between fluid-particles 
\cite{saffman-jfm-1960,sawford_etal-jfm-1986,borgas_etal-jfm-1996}. 
Actually, in this case, the motion of a material-particle in a fluid
is described by a SDE equipped with a delta-correlated Wiener process
that takes the form
$\rmd Z^\omega_t = U^\omega_t \, \rmd t + \sqrt{2 \, \kappa} \, \rmd W^\omega_t$,
where $Z^\omega_t:[0,\infty) \times \Omega
\to \R$, with $\omega \in \Omega$, 
is the position of the material-particle at time $t$,
$U^\omega_t$ is the random velocity of the fluid-particle containing 
the material-particle,
and $\kappa$ is the molecular diffusivity.
Since turbulent velocity $U_t$ is correlated,
it results that the process $Z_t$ is non-Markovian.
By reminding that in homogeneous, stationary and isotropic turbulence
the fluid-particle velocity is Gaussian \cite{thomson-jfm-1987}
as well as the Wiener process,
we have that the above scheme composed by 
a transport flow plus molecular diffusion can indeed be applied to model
(\ref{model}, \ref{model2}) at least as interpretative scheme. 

As a matter of fact, in the framework of the CTRW,
the Markovian hopping-trap mechanism,
i.e., a CTRW with an exponential waiting-time distribution, 
is the one that fulfills the Onsager principle \cite{allegrini_etal-pre-2003}
analogously to the turbulent motion and the molecular diffusion. 
Therefore model (\ref{model}) describes the motion of a walker
under two co-existing forcings that are properly set, separately, 
for out-of-equilibrium systems.
Hence, model (\ref{model}) is a model for the motion of
a diffusive particle under the action of a co-existing
large-scale process, understandable as the mixing 
by an underlying hydrodynamical forcing,
and a small-scale process, understandable as the molecular diffusion,
provided that (\ref{model}) is non-Markovian (\ref{model2}). 

We want to conclude by remarking that,
even if it is not new that anomalous diffusion is just an intermediate
regime in a row of three \cite[see figure 1]{hofling_etal-rpp-2013}
and that it is known that
its extension is limited by thermodynamic uncertainty relation 
\cite{hartich_etal-prl-2021},
this intermediate regime 
follows indeed the time-fractional diffusion law (\ref{TFDE}) and this
provides an argoument for its modelling through
fractional diffusion.
This bridging modelling-role of fractional diffusion 
supports previous "physical mathematics"\footnote{Physical mathematics
is here used in the spirit of Sommerfeld:
{\it The topic with which I regularly conclude my six-term series of 
lectures in Munich is the partial differential equations of physics. 
We do not really deal with mathematical physics, 
but with physical mathematics;
not with the mathematical formulation of physical facts, but
with the physical motivation of mathematical methods.} 
Foreword in: Sommerfeld A 1949 Partial Differential Equations in Physics  
(Academic Press Inc.).}
interpretations of fractional kinetics 
by Grigolini, Rocco and West \cite{grigolini_etal-pre-1999} 
and by Zaslavsky \cite{zaslavsky-pr-2002}. 
Grigolini \etal argue that fractional diffusion describes systems
where there is no separation of time-scales between 
the microscopic and the macroscopic level of the process, 
such that the randomness of the microscopic level is, at least partially, 
transmitted to the macroscopic level and the macroscopic dynamics 
is described by means of fractional calculus operators
\cite{grigolini_etal-pre-1999}. 
Zaslavsky argues that, 
since chaotic dynamics is a physical phenomenon 
whose evolution bridges between a completely regular integrable system 
and a completely random process, 
kinetic equations and statistical tools arise as modelling methods
\cite{zaslavsky-pr-2002}. 
In the present approach, fractional diffusion emerges
as a mathematical method for bridging two 
co-existing equilibrium states in a disordered medium.

\ack 
The authors acknowledge an anonymous Referee for highlighting the 
e-prints \cite{doerries_etal-arxiv-2022,metzler_etal-arxiv-2022} 
that appeared on the arXiv during the peer-review process of the manuscript. 
This research is supported by the Basque Government through the 
BERC  2018--2021 and 2022-2025 programs and 
by the Ministry of Science, Innovation and Universities: 
BCAM Severo Ochoa accreditation SEV-2017-0718.
The research was carried out under the auspices of INDAM-GNFM
(the National Group of Mathematical Physics of the 
Italian National Institute of High Mathematics).

\section*{References}

\providecommand{\newblock}{}

\end{document}